\documentclass[]{aa} 
\usepackage{graphicx}
\usepackage{hyperref}
\usepackage{multirow}
\usepackage{float}	
\usepackage{textcomp}
\usepackage{subfigure}
\usepackage{gensymb}
\usepackage{natbib}
\usepackage{xcolor}
\usepackage[justification=centering]{caption}
\usepackage{ulem}

\definecolor{katia}{rgb}{0.5, 0.0, 0.5}

\begin{document}
	\title{{\Large } A novel analytical model of the magnetic field configuration in the Galactic center}
	\author{M. Guenduez\inst{1} \and J. Becker Tjus\inst{1} \and K. Ferrière\inst{3} \and R.-J. Dettmar\inst{2}}
	\institute{
		Ruhr University Bochum, Faculty of Physics and Astronomy, RAPP Center, TP IV, 44780, Bochum Germany\\
		\email{Mehmet.Guenduez@rub.de}
		\and
		Ruhr University, Faculty of Physics and Astronomy, Astronomical Institute (AIRUB), 44780 Bochum, Germany
		\and 
		University of Toulouse, IRAP, CNRS, 31028 Toulouse Cedex 4, France
	}
	\date{Received / Accepted }
	\abstract
	{Cosmic-ray propagation is strongly dependent on the large-scale configuration of the  Galactic magnetic field. In particular, the Galactic center (GC) region provides highly interesting cosmic-ray data from 
	gamma-ray maps and it is clear that a large fraction of 
	the cosmic rays detected at Earth originate in this region of the Galaxy. 
Yet because of confusion from line-of-sight integration,
	the magnetic field structure in the Galactic center is not well known and no large-scale magnetic field model exists 
	at present.}
{In this paper, we develop a magnetic field model, derived from observational data 
on the diffuse gas, nonthermal radio filaments, and molecular clouds.}
{We derive an analytical description of the magnetic field structure in the central molecular zone by combining observational data with the theoretical modeling of the basic properties of magnetic fields.}
{
We provide a first description of the large-scale magnetic field in the Galactic center region.
We present first test simulations of cosmic-ray propagation and the 
 impact of the magnetic field structure on the cosmic-ray distribution in 
the three dimensions.}
{Our 
 magnetic field model is able to describe the main features of polarization maps; it is particularly important to note that they are significantly better than standard global Galactic magnetic field models. 
It can also be used to model cosmic-ray propagation in the Galactic Center region more accurately.}
	\titlerunning{Galactic center magnetic field}
	\authorrunning{M. Guenduez et al.}
	\maketitle
\section{Introduction}
\label{Introduction}
Cosmic-ray 
(CR) propagation is highly influenced by the ambient conditions. Especially the structure and strength of the magnetic field are often modeled in a simplified way. Global models, such as \cite{Farrar2012} (in the following abbreviated as {JF12}), \cite {Unger2019}, \cite {Kleimann2019}, \cite{MagneticFieldModelSun2008}, \cite{MagneticFieldModelJaffe2013}, or the recently published model by \cite{ShukorovGMF}, do not consider the Galactic center (GC) region due to its complexity.  However, a large fraction of the CR flux is supposed to originate from this region. In particular, for the interpretation of spatially resolved gamma-ray data, it is therefore essential to model the transport out of this region correctly.
To date no magnetic field model has described both the large-scale structure and strength in this inner region of the Galaxy. Nevertheless, many individual works have been published on the observation of structures and objects in the GC that are known to have a well-defined field structure. This concerns, in particular, the nonthermal radio filaments \citep{Yusef-ZadehArc1987,FilamentSpectralIndexSgrA,Anantharamaiah1991,AnantharamaiahPelican1999,Cornelia1999,Lang1999,LaRosaFilaments,LaRosaFilaments2001,FilamentsYusufZadeh,LaRosa330MHz,Law2008} and molecular clouds \citep{CNRMagneticField,G0253,CMZCloud2}. From now on, the localized nonthermal filaments are abbreviated as NTF and the localized molecular clouds MC.
Moreover, \cite{Chuss2003} published polarization data within the GC for a longitudinal extent of $|y|\leq 50\,$pc, complemented by \cite{Nishiyama} for $|y|\leq 250\,$pc, as well as \cite{Mangilli2019} and  \cite{Polarization2} for  $|y|\leq 220\,$pc.

In this paper, we offer a description of the magnetic field in the central 200\,pc around the GC that is as detailed as possible with current data and with the existing information. We derived a total magnetic field via the superposition of the contributions from four structures: the diffuse inter-cloud medium, localized NFT regions, localized MC regions, and the supermassive black hole at Sgr~A$^\ast$.

This list of chosen objects is based on the relevance of the structures for the large-scale field and the possibility to deduce the directionality and intensity of the field from  their parameters. In doing so, we separated the total magnetic field model into poloidal and  azimuthal components.

These two components are described by a number of free parameters (four for the poloidal component and five for the azimuthal component).\par
The results of this work can be used for any  CR propagation purpose in the GC region.
They are particularly important to extend the spatially resolved modeling of CR interactions. In addition, it also allows for a more detailed modeling of the interaction with the gas as the magnetic field model is based on the gas structure, which also becomes available in this paper. One of the most prominent features detected in recent years concerns the PeVatron seen with H.E.S.S. The detected outflows at $\gamma$-ray \citep{FermiBubble2,FermiBubble3}, microwave \citep{WmapHaze,WmapHazeDobler}, and radio wavelengths \citep{RadioHalo} are also in need of proper modeling including the GC magnetic field structure. 
Further, the $\gamma$-ray excess detected with Fermi in the GC region \citep{FermiGCGammaRayExcess2017} and its interpretation as a signature of the annihilation of weakly interacting massive dark matter particles
\citep{DMAnnihilationGC,FermiGCGammaRayExcess2017} can be investigated in more detail. A regular classification of these high-energy features cannot be done without a realistic background magnetic field or gas distribution as many compelling and powerful sources are omnipresent.

This paper is structured as follows. This introduction is followed by a summary to the state-of-the-art research on the environmental conditions in the GC (Section \ref{gc:sec}). Section \ref{MagneticField} represents the central part of this work, where we derive a total magnetic field for the GC region.  We go through all components individually and explain our reasoning systematically for deriving the field structure or using pre-existing templates for the choice of parameters. Section \ref{Application} then presents the application of our model using the CR propagation tool CRPropa, which is introduced as well.
Here, the results are compared with the propagation in the magnetic field of  JF12. The last chapter provides a summary, conclusion, as well as an outlook for the future.

\section{The Galactic Center \label{gc:sec}}
The extreme conditions of the GC make it one of the most exciting regions in our Galaxy. 
The densest region, also known as the Central Molecular Zone (CMZ), is dominated by molecular gas. The region within a Galactocentric radius of $|r|\simeq430\,$ pc contains about 3\% - 10\% of the total gas and star formation in the Milky Way \citep{StarFormationGC}.

Recent observations by H.E.S.S.\ reveal nonthermal emission from the GC  \citep{AbramowskiNature, HessGC2017} indicating a hadronic accelerator, a so-called PeVatron that can accelerate particles up to energies of the first break in the CR energy spectrum, the so-called CR knee.
Many exciting and powerful sources are present in the GC region that could be responsible for causing such a high-energy flux, like SgrA$^*$ itself or pulsars and supernova remnants (SNRs) in the very central region. The explanation of the large-scale diffuse $\gamma$-ray detection by H.E.S.S. inspired this work, since a correct modeling of the spatial dependence is only possible if the propagation of cosmic rays is modeled in a realistic large field configuration.

\subsection{Mass distribution}

As the local magnetic field strength in the GC cannot be determined by direct measurements except for a few Zeeman measurements, such as by \cite{Lin2000}, a correlation between the magnetic field and the gas density and gas dynamics $n$ based on equipartition with gas pressure and kinetic (turbulent) energy will be considered.

This is one reason why we present a detailed state-of-the-art inventory of the mass distribution in this section. The mass distribution is furthermore required for the interpretation of $\gamma$-ray observations in terms of hadronic processes. In hadronic processes, $\gamma$-rays are produced via CR interaction with the ambient gas. The resulting neutral pions decay into two $\gamma$-rays. Thus, even for this interaction process, knowledge on the gas distribution is crucial.

Field lines that pass through MCs are sheared out by the relative motions between MCs and the intercloud medium. Since the Alfv{\'e}n speed is higher in the intercloud medium than directly in the MCs, where the density is much lower, the shearing is expected to be stronger within the MCs.
Although \cite{MassGalacticCenter2} presented the average gas density distribution for the CMZ, this model does not resolve local structures such as MCs or the inner 10\,pc. Therefore, this model will only be considered for the diffuse intercloud (IC) medium (ICM).
This results in the following concept for deriving an overall density profile of the GC region, which consists of three individual components: firstly, the diffuse component as the diffuse ICM,
secondly, localized dense regions in the form of identified MCs, and finally the inner 10\,pc, including the known substructures surrounding the supermassive black hole SgrA$^*$.
In the following we describe the observations of these three components and how we model them. 

\subsubsection{Diffuse component}
 \cite{MassGalacticCenter2} built an analytical 3D model describing the gas distribution in the CMZ. Therefore, in order to determine the diffuse component, we subtract the mass of all MCs that are known to us, as well as the inner 10\,pc structure masses from the total mass obtained in \cite{MassGalacticCenter2}. The subtracted components will be modeled individually and are therefore not part of the diffuse component. 
The total mass in the GC is supposed to be $\sim2.05\cdot10^7\, M_{\odot}$ \citep{MassGalacticCenter2} which is in the range of recent observation of $\sim6\cdot10^6\, M_{\odot}$ for the inner 150\,pc \citep{Oka2019}.
\begin{equation}
\begin{split}
n_{\text{diffuse}}=&2\cdot n_{0,\rm H2} \exp\left(-f(X,Y)^4\right) \cdot \exp\left( -( \frac{z}{18\,\text{pc}})^2 \right) \\
+ &n_{0,\rm H} \exp\left(-f(X,Y)^4\right) \cdot \exp\left( -( \frac{z}{54\,\text{pc}})^2 \right)
\end{split}
\end{equation}
with the definitions 
\begin{equation}
\begin{split}
f(X,Y)=&\frac{\sqrt{X^2+(2.5\,Y)^2}-125\, \text{pc}}{137\, \text{pc}}\\
X:=&(x+50\, \text{pc})\cos(70\degree)+ (y-50\, \text{pc})\sin(70\degree)\\
Y:=&-(x+50\, \text{pc})\sin(70\degree)+ (y-50\, \text{pc})\cos(70\degree)\\
\end{split}
\end{equation}
In this work, we use Galactocentric Cartesian coordinates $(x, y, z)$ where the $x$-axis points from the observer at Earth (or equivalently from the Sun) toward the Galactic Center, the $y$-axis corresponds to the direction $l = +90\degree$, and the $z$-axis points toward the North Galactic Pole. 
The normalization factor after subtracting the substructures, which amounts for $\sim40\%$ of the total mass, yields $n_{0,\rm H2}:=91.1\, \text{cm}^{-3}$ and 
$n_{0,\rm H}:=5.3\ \text{cm}^{-3}$.
\subsubsection{Molecular clouds}
 \cite{CMZMC} derive the masses and radii from Herschel data, where the target region is obtained with the Submillimetre Common - User Bolometer Array (SCUBA) 850\,$\mu$m dust emission maps. Here, we consider all MC mentioned in \cite{CMZMC}. Due to the comparatively larger extension of Sgr B2, however, we have decided on an even more detailed and inhomogeneous description given in \cite{SgrB2}. Additionally, we treated all MC in the dust ridge region separately. Thus, six different dust ridge MCs are taken into account from \cite{Immer2012,Pillai2015,Walker2015,Walker2018}. It turns out that the source Sgr B1-off has been identified as dust ridge E and F. We assume that the mass of Sgr B1-off is included in the dust ridges and thus neglect this source in our modeling. The MC Sgr D was first identified as an \ion{H}{ii} region \citep{Wink1982}.  The Sgr D molecular cloud is  included due to its significant mass. The size and position are taken from \cite{LaRosaFilaments} and the density is considered as given in \cite{MassSgrD}. 
The resulting parameters of the above-described clouds are presented in Table \ref{table1}.
\begin{sidewaystable*}
	\begin{table}[H]
		\centering
				\caption{
		Table of the  molecular cloud parameters.}
		
		\begin{tabular}{||c|c| c| c|c|c|c|c||} 
			\hline
			i&Name & ($l$,$b$)& Radius &  $n_{\text{H}_2}$ &$\delta_n$&$M$ &Ref.\\
			&&  ($\deg$,$\deg$)& pc& $10^4$ cm$^{-3}$&&$\cdot10^4 \, M_\odot$&\\ [0.5ex] 
			\hline\hline
			1&Sgr C &( 359.45, -0.11)&1.7$\pm$0.1&1.8&2.0&2.5 &1, 1, 1, 1, 1\\ 
			\hline
			2&SgrA* & (359.944, -0.046) & $R_s$&-& 1.04&4$\cdot10^6$& 4, -, -, 8, 8\\
			\hline
			3&Inner 5 pc & (359.944, -0.046) & 5 $\pm$1.5 &s. Table \ref{table2} &5.0&s. Table \ref{table2}& 4, -, 4, 4, 4\\
			\hline 
			4&20 km s$^{-1}$ & (359.87, -0.08) & 9.4$\pm$1.5&1& 5.0&22& 1, 4, 4, 4, 4\\
			\hline 
			5&50 km s$^{-1}$ & (359.98, -0.07) & 4.5$\pm$1.5&1& 5.0&19&  1, 4, 4, 4, 4\\
			\hline 
			6&Dust Ridge A & (0.253,0.016) &2.4$\pm$0.1&1.3&2.0&7.2 &5, 5, 5, 5, 5\\
			&(G.0253+0.016)&&&&&&\\
			\hline
			7&Dust Ridge B & (0.34, 0.055) &1.9$\pm$0.1&0.7&2.0&1.3&6, 6, 6, 6, 6\\
			\hline
			8&Dust Ridge C & (0.38, 0.050) &1.9$\pm$0.1&0.9&5.0&1.8 &6, 6, 6, 6, 6\\
			\hline
			9&Dust Ridge D & (0.412, 0.052) &3.3$\pm$0.1&0.8&2.0&7.4&6, 6, 6, 6, 6\\
			\hline
			10&Dust Ridge E & (0.478, 0.005) &3.5$\pm$0.1&2.8&2.0&13.3&6, 6, 6, 6, 6\\
			\hline
			11&Dust Ridge F & (0.496, 0.020) &2.4$\pm$0.1&3.2&2.0&7.3&6, 6, 6, 6, 6\\
			\hline
			12&Sgr B2 & (0.66,-0.04)&(1.25-22.5)$\pm$0.4& 0.3&1.3&700&1, 7, 7, 7, 7\\
			\hline
			13&Sgr D&(1.13,-0.11)& 1.8$\pm$0.1&0.6& 3.0&1.2&2, 2, 3, 3, 3\\
			[1ex] 
			\hline\hline
		\end{tabular}
		\tablefoot{Entries are: Number (first column), name (2nd Position (3rd), radius (4th), $H_2$ particle density (5th), density uncertainty factor (6th) and mass (7th), references (8th). The parameters of the magnetic field strength $B$ and $\delta_B$ the uncertainty factor of the magnetic field are based on the results of this work (see a detailed description in Section \ref{MagneticField}).
 
Further, $R_s$ is the Schwarzschild radius of SgrA* $\simeq1.12\cdot10^{12}$ cm.}
		\tablebib{
		(1) \cite{CMZMC}, (2) \cite{LaRosaFilaments}, (3) \cite{MassSgrD}, (4) \cite{Inner10pc}, (5) \cite{G0253}, (6) \cite{Walker2018}, (7) \cite{SgrB2}, (8) \cite{SgrA*Mass}.}
		\label{table1}
	\end{table}
\end{sidewaystable*}
\subsubsection{Inner 10\,pc \label{inner10_gas:sec}}
\begin{figure}[H]
	\centering
	\subfigure{\includegraphics[width=0.8\linewidth]{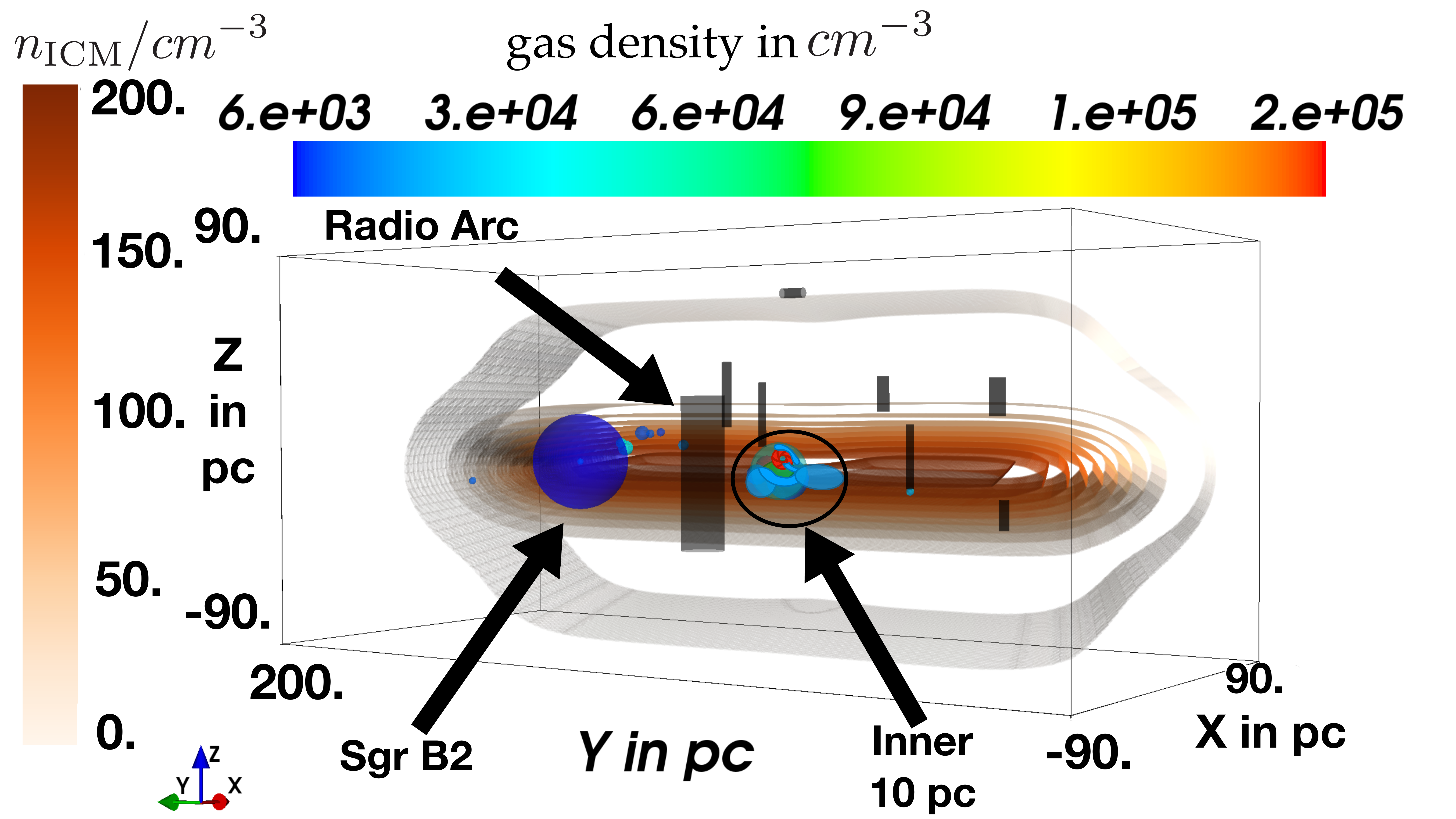}}
	\caption[]{Representation of the combined gas density of the CMZ region in units of cm$^{-3}$ (color scales).  Here and in the figures that follow, the x-axis points from the Sun to the Galactic Center. The diffuse ICM is shown as red contours. The non-thermal filament objects are presented as cylindrical black colored objects.}
	\label{MassDistr1}
\end{figure}

\begin{figure}[H]
	\vspace*{-0cm}
	\centering
	\subfigure{\includegraphics[width=0.9\linewidth]{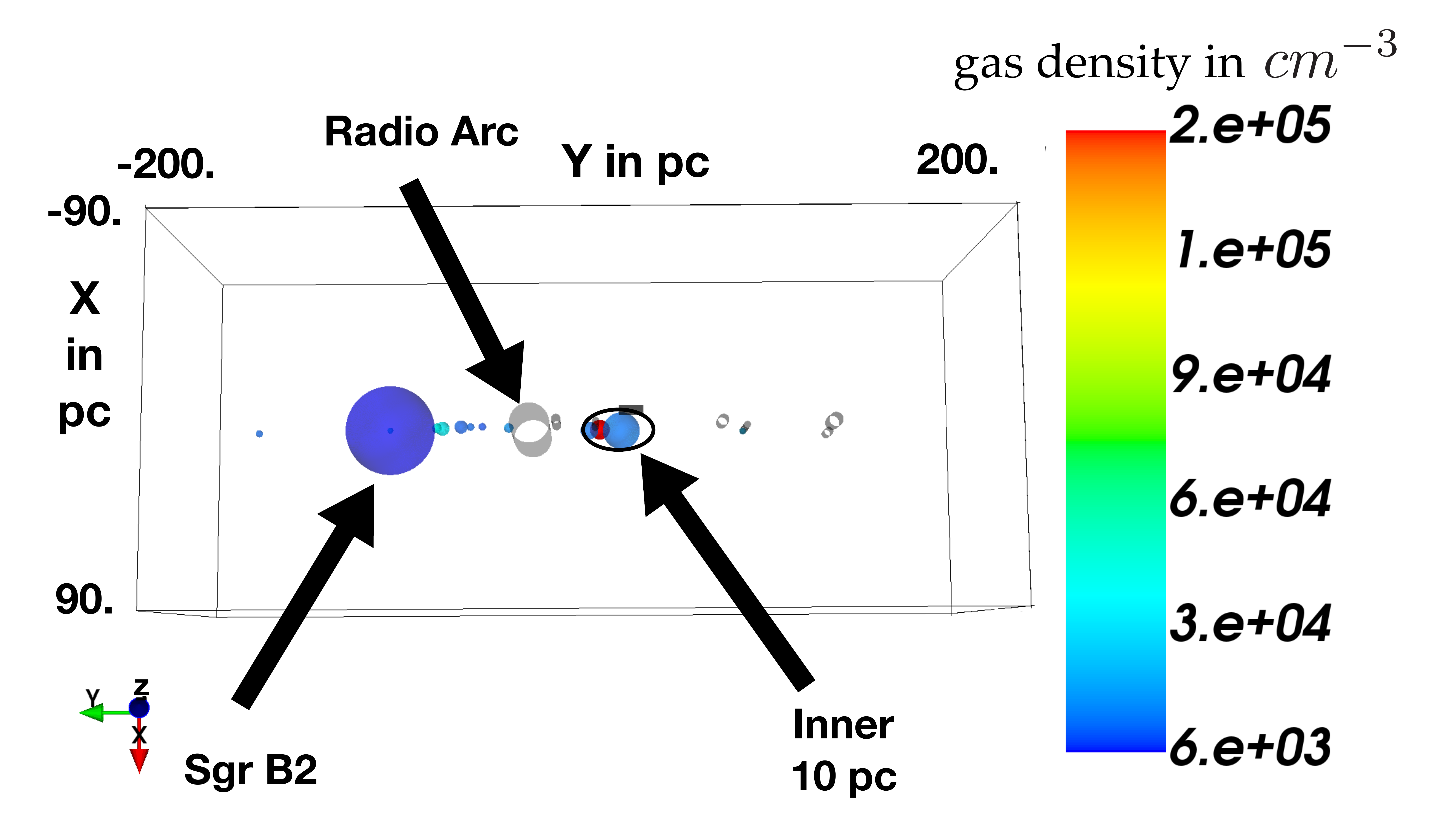}}
	\caption[]{Same as Fig.\ \ref{MassDistr1}, but with the projection along the $z$-axis and without considering the ICM.}
	\label{MassDistr2}
\end{figure}

The central 10\,pc around SgrA$^*$ are of particular importance as SgrA$^*$ is one of the most interesting candidates for CR acceleration \citep{HESS2018GC,AbramowskiNature}. Thus, the morphology of the mass distribution and magnetic field is highly relevant for CR propagation models. Even if the sources of CRs would mainly lie outside of this inner core, diffusion would lead to the propagation of particles into the central $10$\,pc and again make a detailed modeling of both the gas distribution and magnetic field configuration important.  \cite{Inner10pc} has constructed a realistic three-dimensional picture of this region by reviewing existing observational studies. 
Table \ref{table2} lists ten components by order of decreasing density that are adopted according to the 3D morphology and density derived in \cite{Inner10pc}.
\begin{sidewaystable}
\begin{table}[H]
	\captionsetup{width=.75\textwidth}
	\caption{Summary of the gas components of the interstellar gas in the inner 10\,pc around SgrA*, ordered by decreasing density.  }
	\centering
	\begin{tabular}{||c| c| c| c||} 
		\hline
		\#Component & Name  & $n_{\text{H}}$ & Shape \\ [0.5ex] 
		\hline\hline
		1 & Circumnuclear disk (CND) & 4.4$\times 10^5$ cm$^{-3}$ & trapezoidal ring \\ 
		\hline
		2 & Molecular ridge (MR) & 3$\times 10^4$ cm$^{-3}$ & curved cylinder \\
		\hline
		3 & Southern streamer (SS) & 3$\times 10^4$ cm$^{-3}$ & curved cylinder \\
		\hline
		4 & Western streamer (WS) & 3$\times 10^4$ cm$^{-3}$ & curved cylinder \\
		\hline
		5 & Northern ridge (NR) & 3$\times 10^4$ cm$^{-3}$ & curved cylinder \\
		\hline
		6 & M-0.13-0.08/ 20 km s$^{-1}$ (SC) & 2$\times 10^4$ cm$^{-3}$ & ellipsoid\\
		\hline
		7 & M$-$0.02$-$0.07/ 50 km s$^{-1}$ (EC) & 2$\times 10^4$ cm$^{-3}$ & indented sphere \\
		\hline
		8 & Central cavity (CC) & 1.6$\times 10^3$ cm$^{-3}$ & ellipsoid \\
		\hline
		9 & Radio halo & 210 cm$^{-3}$& sphere \\
		\hline
		10 & Sgr A East SNR & 3 cm$^{-3}$& ellipsoid \\ [1ex] 
		\hline\hline
	\end{tabular}
	\tablefoot{Positions and  shapes are taken from \cite{Inner10pc}.}
	\label{table2}
\end{table}
\end{sidewaystable}
\begin{sidewaystable}
	\begin{table}[H]
		\caption{Summary of the properties of the filaments in the CMZ.}
		\centering{
		\begin{tabular}{||c|c| c|c|c|c|c|c|c||} 
			\hline\hline
			&&($l$,$b$) &  $\Delta b \times \Delta l$   &$I_1$ &  &  &$\overline{B}_{\rm NFT}$& \\ 
			i&Name & (deg,deg)   &pc$\times$pc&Jy/beam&$\alpha_{\text{0.33GHz}}^{\text{1.4GHz}}$ &$\alpha_{\text{1.4GHz}}^{\text{4.8GHz}}$&($\mu$G)&Ref.\\[0.5ex] 
			\hline\hline
			1&G359.15-0.2 & (359.15, -0.17 )& 12.9$\pm$1.8$\times$1.5$\pm$0.5 & 0.1 & -0.5 & -0.5 &88 &1; 1; 1; 2; 2; 2 \\
			&(The Snake) & & & & & &&\\
			\hline
			2&Sgr C-NTF& (359.45,-0.01)& 27.4$\pm$1.8$\times$1.7$\pm$0.5 & 0.51& -0.5 & -&100& 1; 1; 1; 1; -; 1\\
			\hline
			3&G359.54+0.18& (359.55,0.17) & 15$\pm$1.8$\times$2.7$\pm$0.5&0.19&-& -0.8  &1000&1; 1; 1; -; 3; 1\\
			&(Ripple) & & & & && &\\
			\hline
			4&G359.79+0.17 & (359.80,0.16) & 16.2$\pm$1.8$\times$3.5$\pm$0.5&0.3&-0.6  &-0.6  &1000 &1; 1; 1; 3; 3; 1\\
			\hline
			5&G359.85+0.47&(359.85,0.47)&13.4$\pm$1.8$\times$2.2$\pm$0.5& 0.12&-0.6  &-0.8 &70 &1; 1; 1;  1; 4,5; 6\\
			&(Pelican) & & & & &&& \\
			\hline
			6&G359.96+0.09 &(359.96,0.11)&28.7$\pm$1.8$\times$1.7$\pm$0.5& 0.15 & -0.6&- &100& 1; 1; 1; 1; -; 1\\
			&(Southern Thread)& & &  & &&& \\
			\hline
			7&G0.09+0.17& (0.09,0.17 )& 29.4$\pm$1.8$\times$2.0$\pm$0.5&0.28 & -0.6 & -0.5 &140 &1; 1; 1; 1; 4; 1,4\\
			&(Northern Thread)& & & & &&&\\
			\hline
			8&The radio arc & (0.18,-0.07) & 70.5$\pm$1.8$\times$9.9$\pm$0.5&0.54&-0.4 &0.3 &1000 &1; 1; 1; 7; 8; 9\\ [1ex] 
			\hline\hline
		\end{tabular}
		\tablefoot{The position ($l$,\,$b$), intensity density $I_1$ at 330 MHz  and $\Delta l$ as longitudinal and $\Delta b$ as latitudinal extension of each object are taken from \cite{LaRosaFilaments}. $\alpha_{\text{0.33GHz}}^{\text{1.4GHz}}$ and $\alpha_{\text{1.4GHz}}^{\text{4.8GHz}}$ represent the non-thermal radio emission spectral index. The flux is connected to a uncertainty of 15\% which leads to a $\sim$ 5\% uncertainty for the equipartition magnetic field.}
		\tablebib{(1) \cite{LaRosaFilaments}, (2) \cite{FilamentsYusufZadeh}, (3) \cite{Law2008}, 4=\cite{Lang1999}, (5) \cite{Cornelia1999}, (6) \cite{AnantharamaiahPelican1999}, (7) \cite{Anantharamaiah1991}, (8) \cite{FilamentSpectralIndexSgrA}, (9) \cite{Yusef-ZadehArc1987}.}}
		\label{table3}
	\end{table}
\end{sidewaystable}
\noindent 
The locations and sizes are taken from \cite{Inner10pc}.
The combined gas density distribution is visualized in Figs.\ \ref{MassDistr1} and \ref{MassDistr2} with and without contours of the diffuse medium, respectively.
\subsection{Nonthermal filaments}
\label{filaments}
While the origin of the NTFs is not well understood yet, their presence in the GC region indicates the existence of strong magnetic fields and relativistic electrons. Several reasonable alternatives are describing the formation process. The first scenario (more detailed in \cite{Benford1987},  \cite{Lesch1992}, \cite{Serabyn1994}, \cite{Uchida1996} and \cite{Staguhn1998}) suggests that the relativistic electrons originate from fast-moving clouds. Hence, the interaction between the background magnetic field and these clouds fulfills both ingredients for generating synchrotron radiation.\\ The alternative explanation, for example, presented in  \cite{YusufZadehOriginNTF} suggests a link to star-formation regions where collective winds of massive WR and OB stars generate shock waves. Thus, within a dense stellar region, particles can be accelerated to relativistic energies.\\
The third explanation is based on recent MeerKAT observations. \cite{Heywood2019} observed a huge radio bubble (140\,pc$\times$430\,pc) extending above and below the GC. These lobes can be seen in both the radio \citep{Heywood2019} and X-ray \citep{Nakashima2019, Ponti2019} parts of the spectrum. 
The observations indicate an outbursts of relativistic electrons from the central black hole (BH) or from its surroundings.
We are interested in NTFs because of their strong non-thermal radio emission and, thus, the connection to the magnetic field. Their expected strong fields, in combination with the known structures, represent an influential contribution to the modeling of the large magnetic field in the GC region.

\cite{LaRosaFilaments} presented a wide-field 90\,cm-VLA image of the GC region.
They cataloged the 90\,cm flux density, position, and size of NTFs among other objects. Some of the sources were also investigated concerning their 20\,cm/90\,cm spectral index. Furthermore, isolated NTFs show a constant spectral index of $\alpha\approx -0.6$, consistent with a relatively flat, first-order-Fermi accelerated electron population.
We summarize the properties of these filaments as found in \cite{LaRosaFilaments} in Table \ref{table3}.
Figure \ref{MassDistr1} visualizes the NTFs relative to the MC distribution.

\section{Magnetic field in the CMZ}
\label{MagneticField}
The main application of the global magnetic field models of the Milky Way is the propagation of CR. Divergence-freeness of the \textit{JF12} field has been implemented in \cite{Kleimann2019,Unger2019}, which makes the model useful for MHD simulations as well, in which the lack of sources and sinks is of high importance.
The \textit{JF12} model and its updates, however, do not explicitly take into account the central 1\,kpc of the disk component. 
The use of this field for CR propagation throughout the Galaxy leads to magnetic mirror effects, resulting in the trapping of CRs within 1\,kpc \citep{MertenMaster}. Interpolating the disk field from further out has been used for CR propagation modeling by \cite{CRpropa2017}. This is a good first approximation for global CR models, but it is doubtful that CR propagation and its signatures in the GC can be studied properly with this simplified approach. Global propagation models can also rely on the field configuration in the GC range - the CR source density is expected to be highest in the central part of the galaxy and local diffusion in the first kpc could also change the global result. 
The aim of this paper is to answer these open questions by first creating a more sophisticated magnetic field model in the CMZ region (from now on \textit{GBFD20}) and performing a first comparison with the simplified \textit{JF12} extrapolation in the GC region.
 
This section focuses on building the model for the inner radius of 200\,pc, while Section \ref{Application} presents first simulations that compares the two approaches.
\subsection{Magnetic field observations in the CMZ}
\label{Observation}
The magnetic field of the diffuse ISM in the GC region has been investigated with different methods, among those are polarization measurements from far-infrared to near-infrared or radio wavelength  \citep{Chuss2003,Polarization2,Nishiyama}. Still, the magnetic field strength and orientation in some regions in the CMZ remain rather uncertain. Nevertheless, the measurements provide us with pieces of information on a general structure and intensity of the field as summarized below for the three different density regions - the MCs, NTFs and the diffuse ICM. Pieces of evidence for the directionality in the CMZ are the following:
\begin{enumerate}
	\item $\vec{B}\sim$ poloidal in the diffuse ICM and NTF region:
	\begin{itemize}
		\item  \cite{Nishiyama} present a polarization map deduced from near-infrared wavelength of nearly the whole CMZ that exhibits the existence of a poloidal field for $|b|>0.4$ deg.
		\item \cite{LaRosa330MHz} refer to an approximately poloidal component in the diffuse IC region.
		\item The NTFs can best be described by a poloidal component \citep{LaRosaFilaments2001}. 
		One exception is the \textit{Pelican} region, which reveals a dominantly horizontal orientation \citep{LaRosaFilaments}, with \textit{horizontal} being defined as the $B_r-B_\phi$ plane.
	\end{itemize}
	\item $\vec{B}\sim$ horizontal in the MC region:
	\begin{itemize}
		\item The dense clouds, on the other hand, require a dominant horizontal field component \citep{FerriereMagneticField2009,Nishiyama}.
		\item Far-infrared polarization measurements from dust indicate a field which is predominantly parallel to the Galactic Plane for $|b|<0.4$ \citep{Nishiyama} where most of the cloud are located. This indicates that the X-field component modeled for the large-scale field in the \textit{JF12} paper is subdominant in the CMZ for small Galactic longitudes.
		\item  \cite{Chuss2003} (submillimeter observation), \cite{Nishiyama2009} (from far-infrared (FIR) and submillimeter observations), and several other groups suggest a dominant horizontal field component with respect to the Galactic Plane in dense regions. 
		
	\end{itemize}
\end{enumerate}
Evidence for the magnetic field strength are summarized here:
\begin{itemize}
	\item Although the measurement of the field strength at the event horizon of SgrA* is still inaccessible, in simulations, it requires a field strength of about 30-100\,G at the event horizon in order to explain the synchrotron radiation \citep{Eatough2013,Moscibrodzka2009,Dexter2010}. Therefore, we apply this constraint in our modeling. The field is supposed to be azimuthally sheared by the differential rotation as discussed by \cite{Johnson2015}. These authors also found evidence for a partially ordered fields near the event horizon ($\sim$6 Schwarzschild radii) as well as an associated variability on the time-scale of an hour. 
	\item 
Simulation results from \cite{Peratt1984} suggest a field strength up to 10\,mG inside the inner 10\,pc.  On average, however, the field might be weaker in order to ensure the stability of the dominant cloud in the Circumnuclear disk (CND) \citep{Morris1990}. The highest direct detection of the field strength inside 10\,pc was $\sim$ 3\,mG \citep{Plante1995} based on the detection of \ion{H}{i} Zeeman splitting.
	\item A field strength of 1\,mG has been measured in Sgr B2 based on \ion{H}{i} Zeeman measurements \citep{Crutcher1996}.
	\item The dense clouds are thought to have a typical field strength of 1\,mG \citep{FerriereMagneticField2009} (see also Table \ref{table1}).
	\item An average value of $\sim$10 $\mu$G with an uncertainty factor of 10 is supposed to be present in the ICM \citep{LaRosa330MHz,FerriereMagneticField2009}. However, since the field strength in the Galactic disk amounts to $\sim3~\mu$G a smaller value than $\sim$10 $\mu$G seems to be not plausible.
This value is from now on defined as $\overline{B}_{\rm IC}$.
	\item  The NTFs also show a field strength up to 1\,mG obtained from the related equipartition magnetic field and the radio luminosity (see Table \ref{table3}).
\end{itemize}
In summary, the background  field in the ICM and the NTF regions\footnote{Except for the \textit{Pelican} region, which particularly has a dominant horizontal component \citep{LaRosaFilaments}} is predominantly poloidal, whereas the horizontal field dominates in the MC regions. Additionally, each region has a specific field strength with the evidence as discussed above. These pieces of information together with the constraints from electrodynamics, that is $\vec{\nabla}\cdot\mathbf{B} = 0$ and a zero net magnetic flux, can be used to develop a first full description of the magnetic field in the CMZ.
\subsection{Determination of the magnetic field strength}
\label{field_strength:sec}
The highly polarized emission observed \citep{LaRosaFilaments2001} in the NTF regions has an average strength of the order $\sim$ 1\,mG (see Table \ref{table3}) which is derived from radio luminosities.\\ 
An average value of $\overline{B}_{\rm IC}\approx$10 $\mu$G determines the field strength in the ICM \citep{LaRosa330MHz,FerriereMagneticField2009}. Concerning the MCs, only two clouds out of 12 (CND \citep{Plante1995} and Sgr B2 \citep{Crutcher1996}) have direct magnetic field measurements via \ion{H}{i} Zeeman splitting. However, it is possible to estimate the field strength in MCs based on a theoretical argument, which assume equipartition between magnetic pressure and  turbulent pressure.
\paragraph{Equipartition with turbulent pressure}\mbox{}\\
The second method builds on the assumption that the turbulent and magnetic energy are in equilibrium, that is,\ 
\begin{equation}
\frac{1}{2} \rho \cdot\Delta v^2=\frac{\mathbf{B}_{\rm eq}^2}{8\pi}
\label{B-v-Relation}
\end{equation}
Here, $\Delta v$ represents the turbulent velocity, which is estimated considering cloud line width and presented in Table \ref{BFieldTurbulent}. The magnetic field strength results are also given in Table \ref{BFieldTurbulent}. Here, we interpret $\mathbf{B}_{\rm eq}$ as the average field strength $\overline{B}_{\rm MC}$ of the MCs.
\\
	\begin{table}[H]
	\centering
		\caption{Magnetic field strength for the different objects, derived from equipartition with turbulent pressure is listed. }		
		\begin{tabular}{||c|c||c|c||c|} 
			\hline
			i&Name &$\Delta v$/&$\mathbf{B}_{\rm eq}$/&Ref.\\
			&&km~s$^{-1}$& mG&\\ [0.5ex] 
			\hline\hline
			1&Sgr C&6.5 &0.6&1\\ 
			\hline
			2&SgrA* &-&$6.5\cdot10^3$&5\\
			\hline
			3&inner 5~pc&- &8.4&2\\
			\hline 
			4&20 km s$^{-1}$&10.2 &0.9&2\\
			\hline 
			5&50 km s$^{-1}$&13.9 &1.3&2\\
			\hline 
			6&Dust Ridge A &6.5&0.5&3 \\
			\hline
			7&Dust Ridge B &6.8& 0.4&-\\
			\hline
			8&Dust Ridge C &6.8& 0.4&-\\
			\hline
			9&Dust Ridge D &7.0& 0.4&3\\
			\hline
			10&Dust Ridge E &6.7& 0.7&3\\
			\hline
			11&Dust Ridge F &7.2& 0.9&3\\
			\hline
			12&Sgr B2 & -&1.3&4\\
			\hline
			13&Sgr D&2.5&0.2&1\\
			[1ex] 
			\hline\hline
		\end{tabular}
		\tablefoot{Magnetic field results for the inner 5\,pc and Sgr B2 are adopted from observations. For dust ridge B and C, the average velocity dispersion in the dust ridge is used. 1=\cite{CMZMC}, 2=\cite{CNDMF2}, 3=\cite{Walker2015}, 4=\cite{Crutcher1996}, 5=\cite{Eatough2013}.}
		\label{BFieldTurbulent}
	\end{table}
\subsection{Modeling the magnetic field structure}
In this section, we make use of analytical and divergence-free magnetic field components contributing to a total field model and constrain the resulting free parameters by the geometric structure, location and size of the considered regions and objects. The details of this mathematical approach are described below.

Concerning the structure of the field, we start with the introduction of the poloidal field component in Section \ref{PFM}, based on the model presented in \cite{X-ShapeModel}, as this can be used to describe both, the IC and NTF regions. A purely poloidal configuration has radial and vertical (vertical=$z$ in cylindrical coordinates) unit vectors. In contrast, a purely horizontal configuration depends on a radial and azimuthal unit vectors.
The poloidal field does not apply to the MC regions, as the field is predominantly horizontally oriented. Thus, a horizontal field component will be derived for the MCs in the CMZ. The relation between these components will be based on the ambient conditions in the MCs. Therefore, Section \ref{MC} will introduce the horizontal component and constrain the parameters by the boundary conditions also described in the same section. 
The total magnetic field model is then given by a superposition of all poloidal and horizontal components. 
\subsubsection{Poloidal field component}
\label{PFM}
\cite{X-ShapeModel} published an analytical expression of poloidal and x-shaped divergence-free magnetic fields models. A good description of a poloidal field is given by \textit{Model C} described in equations (80) \& (81) in \cite{X-ShapeModel}
\begin{equation}
\begin{split}
B_r&=&\frac{2\, a\, r_1^3\, z}{r^2} \cdot B_s(r, \phi, z)\\
B_{\phi}&=&0\\
B_z&=&\frac{r_1^2}{r^2}\cdot B_s(r, \phi, z)
\end{split}
\label{xshape:equ}
\end{equation}
with
\begin{equation}
r_1=\frac{r}{1+a\, z^2} \text{ and } B_s=B_1\, e^{-r_1/L}\, \cos(m_0(\phi-g(r,z) )) .\nonumber
\end{equation}
Since we are interested in a purely poloidal field, we set the azimuthal component 
to zero.
From now on the magnetic field given by {Ferri{\`e}re \& Terral - Model C (FT14-C)} is called $\mathbf{B}^C=(B_r,0,B_z)$. This model
has four free parameters: $L$, $m_0$, $B_1$ and $a$. 
Here, $L$ denotes the radial exponential scale length, 
$m$ is the azimuthal wavenumber ($m_0=0$ for axisymmetric, $m_0$=1 for bi-symmetric and $m_0$=2 quadri-symmetric, ... magnetic field). For simplicity and due to the lack of information on the azimuthal wavenumber, we use $m_0=0$ for all poloidal fields.
$B_1$ is the magnetic field strength normalization factor and 
$a$ is a strictly positive parameter governing the opening of field lines away from the z-axis. 
Finally, $g(r,z)$ is a smoothly varying function describing the spiraling of field lines.
Those interested in a single global model can use \textit{Model C} with its azimuthal field component given in
\cite{X-ShapeModel}. This type of model is easier to handle, but cannot reproduce local but influential configurations, such as those in the NTFs or dense regions in the GC.
\subsubsection*{Magnetic field in the diffuse inter-cloud medium}
\begin{figure}[H]
	\centering
	\subfigure{\includegraphics[width=1.0\linewidth]{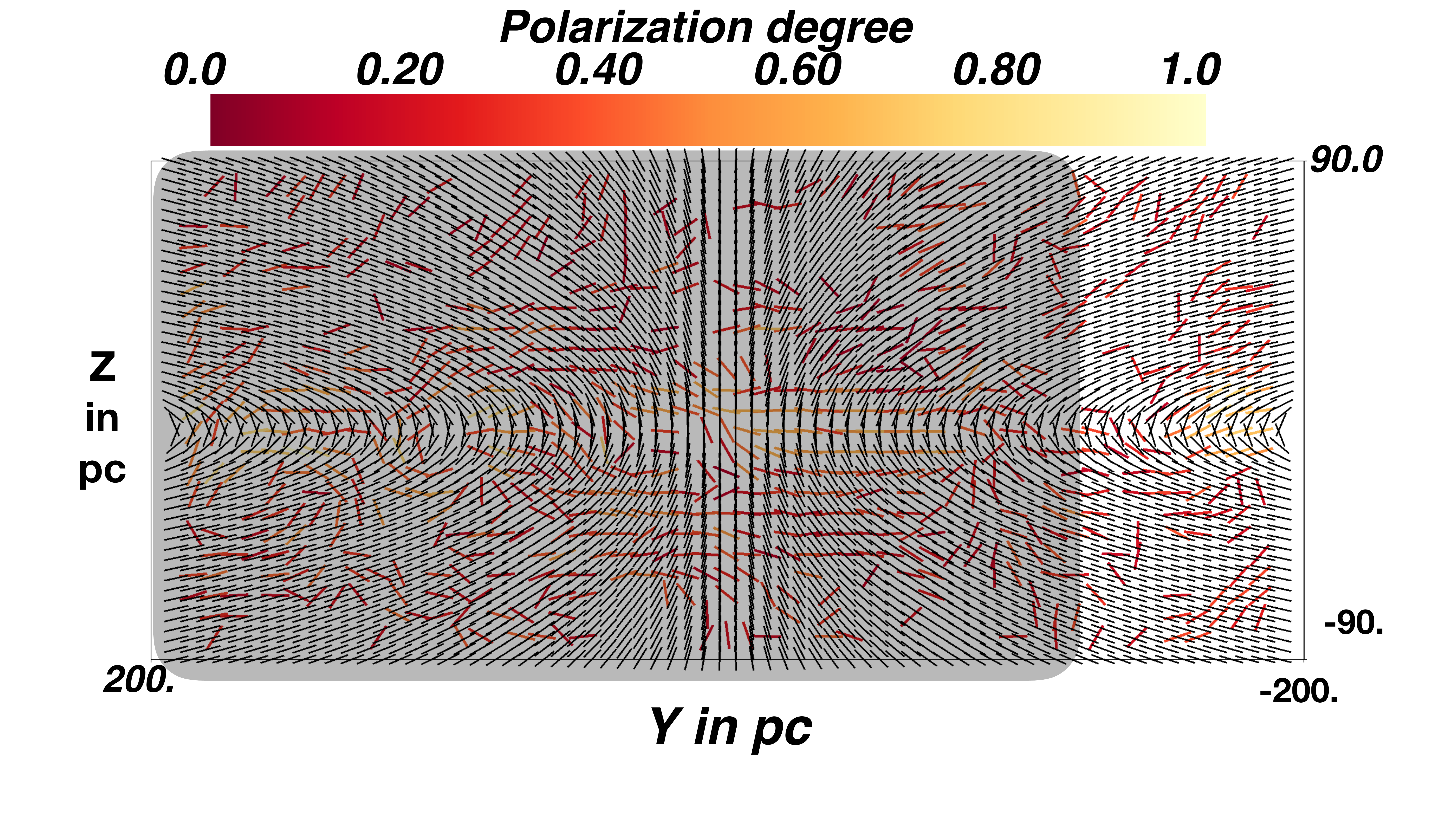}}
	\caption[]{Magnetic field configuration as derived for $\mathbf{B}_{\rm IC}$ is visualized by black dashes lines and by colored dashed lines as measured by \cite{Nishiyama2}.}
	\label{BICPol}
\end{figure}

\begin{figure}[H]
	\centering
	\subfigure{\includegraphics[width=1.0\linewidth]{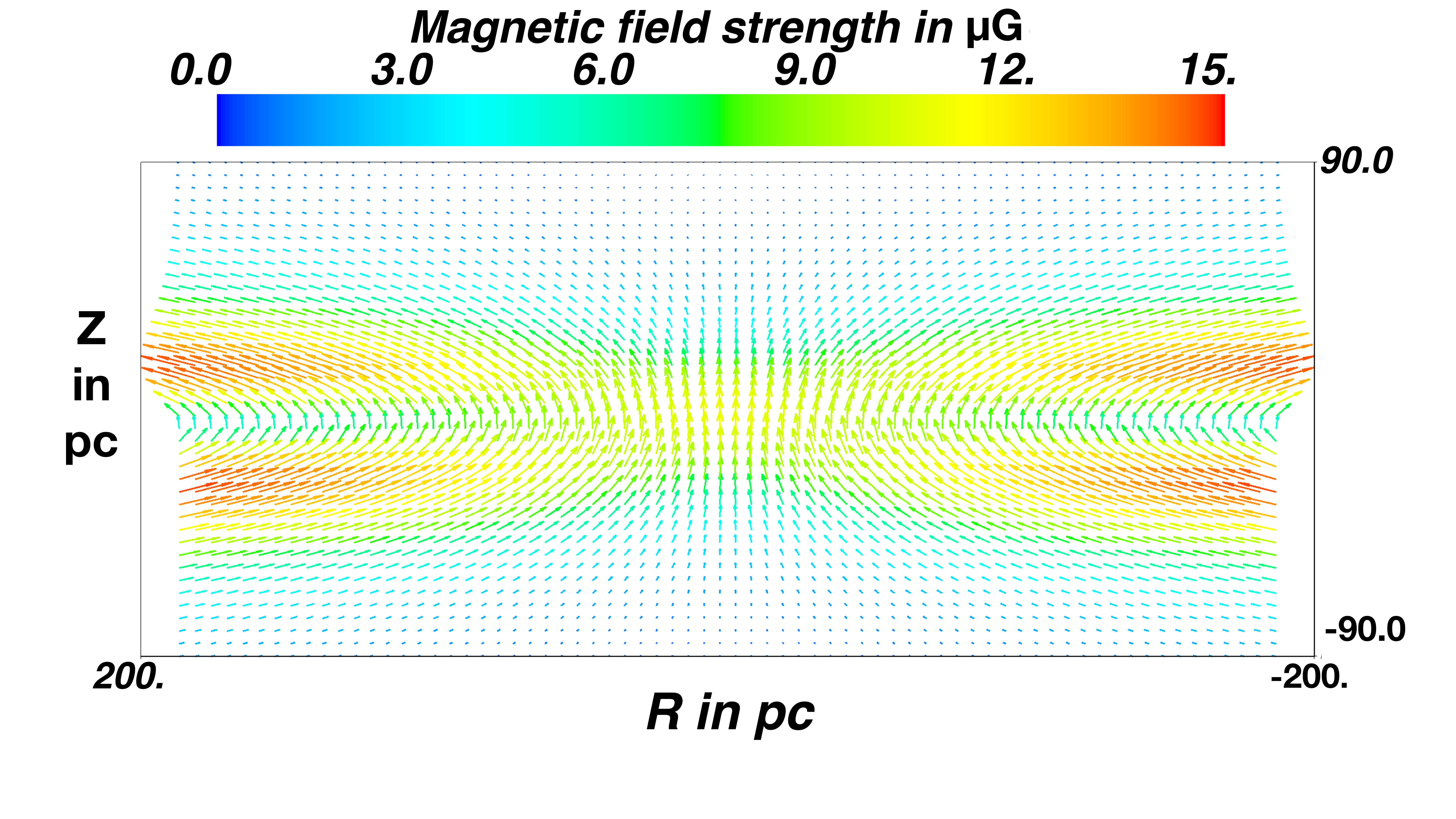}}
	\caption[]{
	Magnetic field strength in units of $\mu$G in the IC region. The relative length and color of the arrows represent the relative magnetic field strength.}
	\label{FieldIntercloud}
\end{figure}

 As the background field in the ICM is predominantly poloidally oriented, we  make use of the poloidal field component described in Equ.\ (\ref{xshape:equ}). The background field in the ICM, from now on $\mathbf{B}_{\mathrm{IC}}^{C}$, is assumed to follow the mass distribution in the CMZ as described in \cite{MassGalacticCenter2}. There, the semi-major axis corresponds to 250\,pc with an axis ratio of 2.5. Transforming this property to an axisymmetric description delivers a radius of $R_{\text{IC}}=$158\,pc. The radial exponential scale length is then obtained by assuming half-max value of $\mathbf{B}$ at $R_{\text{IC}}$.
\begin{equation}
L=\frac{R_{\text{IC}}}{\ln(2)}=114\,\text{ pc}
\label{ExpScaleL}
\end{equation}
The parameter $a$ governing the opening of field lines away from the z-axis will be determined by comparing the configuration obtained by this model and the polarization map of \cite{Nishiyama2}. Here, we consider a region which does not contain any significant NTFs or MCs. In doing so,  we assume that we can neglect the field contributed by the NTF and MC regions and just have a dominant background field of the ICM. Therefore, we take the region from $-200$\,pc$<y<-125$\,pc into account. Varying $a$ from $1/(1\,\rm pc)^2$ to $1/(100\,\rm pc)^2$, a best fitting procedure delivers $a=1/(42\,\rm pc)^2$ which leads to similar disk thickness as used in \cite{MassGalacticCenter2}. Figure \ref{BICPol} displays observed and expected configuration for our best fit value.
The map derived from this work is deduced from the direction of the field lines and calculated by the poloidal field model. Thus, it does not correspond to a polarization map in its entirety. Rather, it should only be used as a qualitative comparison with the observation to explain features.
Since a single and ordered magnetic field model describes the ICM region, the configuration in the direction of the line of sight does not change.
The fitting parameter $B_1$ is determined by assuming that the average value of the model needs to match the observed value, here,  $\overline{B}_{\rm IC}$
\\
In the IC region, $\overline{B}_{\rm IC}=10$~$\mu$G is assumed to correspond to the true average value, following the conclusion of \cite{FerriereMagneticField2009} and \cite{LaRosa330MHz} as already discussed above. Within the model, we can determine the average value as
\begin{equation}
\begin{split}
\overline{B}=&(\frac{1}{V} \int_{V}\sqrt{B_r^2+ B_z^2} \,  \text{d}V)\\
=& \alpha\,B_1 = \overline{B}_{\rm IC}=10\,\mu{\rm G}\,.
\label{BAverage}
\end{split}
\end{equation}
with $\alpha= V^{-1}\,\left(\int_{V}\exp(-\frac{r_{1}}{L})\cdot\sqrt{\frac{4\, a^2\, r_{1}^{6}\, z^2}{r^{4}}+ \frac{r_{1}^{4}}{r^{4}}}\text{d}V\right)$.
Here, the integrand in the bracket denotes the average field strength obtained from the model, and $V$ the total volume of the CMZ. Thus, $B_1$ can be determined as
\begin{equation}
B_1=\frac{\overline{B}_{\rm IC}}{\alpha}\,.
\label{B1}
\end{equation}
Solving this equation delivers $\alpha\approx 0.85$ and therefore
\begin{equation}
B_1
\approx 12\,\mu{\rm G}\, .
\end{equation}
Figure \ref{FieldIntercloud} illustrates the application of the poloidal magnetic field model in the IC region, projected onto the $Y-Z$-plane. Due to the symmetry of the equation in $Y$ and $X$, this graph represents the $X-Z$-plane as well.
\subsubsection*{Magnetic field in the NTF regions}
\begin{figure}[H]
	\hspace*{-0.5cm}
	\centering
	\subfigure{\includegraphics[width=1.1\linewidth]{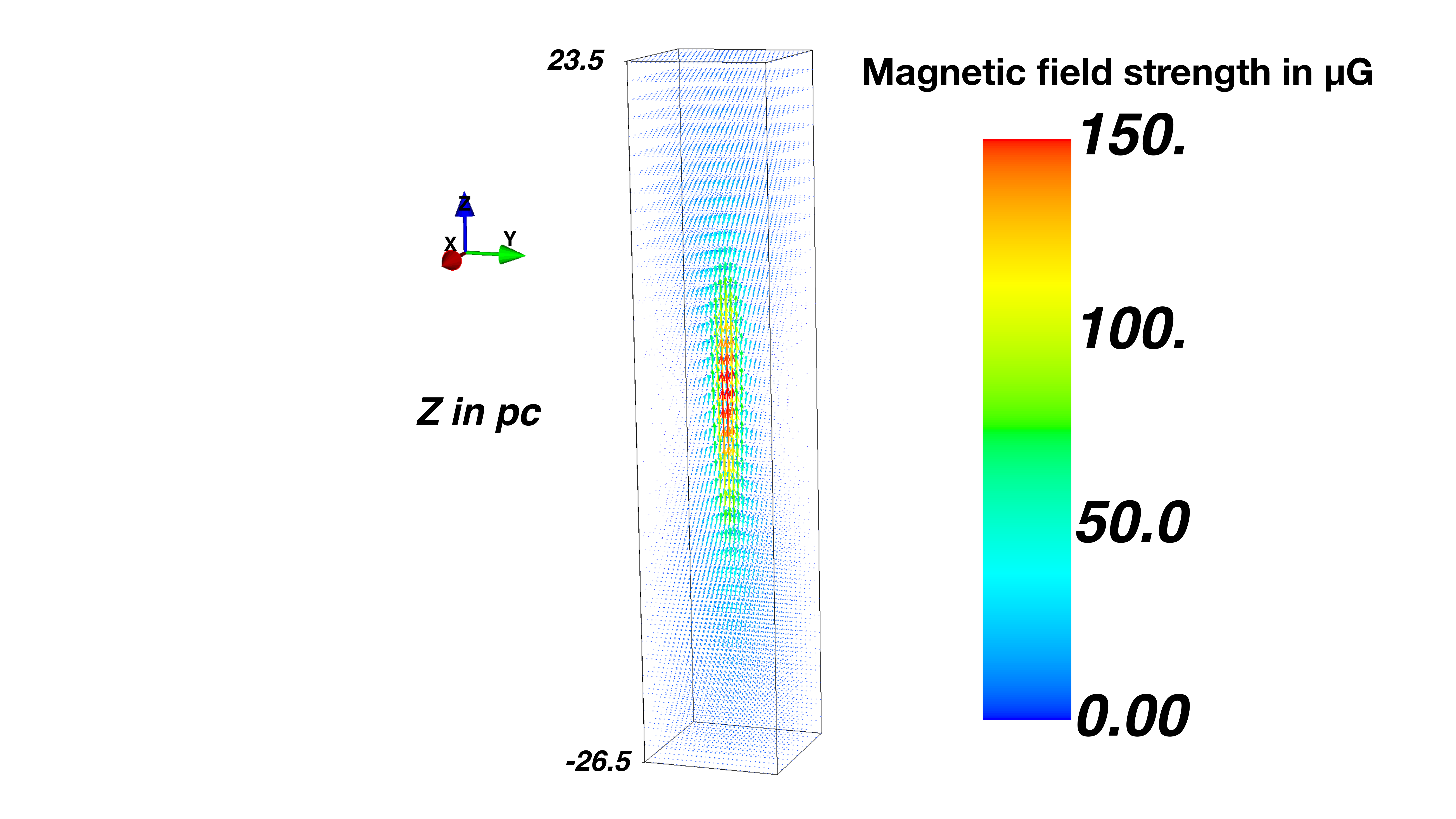}}
	\subfigure{\includegraphics[width=1.0\linewidth]{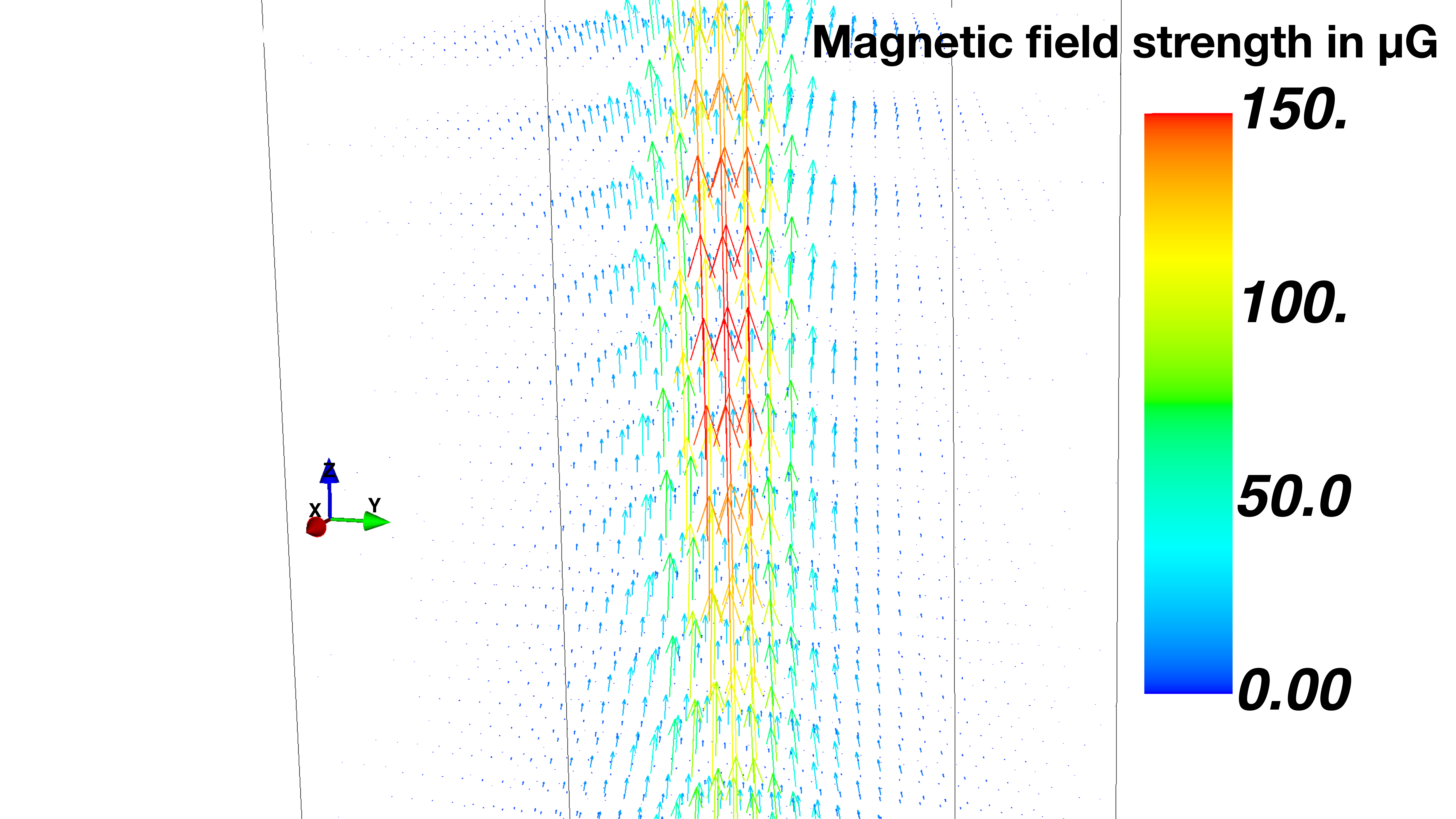}}
	\caption[]{Magnetic field strength in Sgr C visualized by the related 3D magnetic field configuration. The lower panel displays a zoomed view.}
	\label{SgrCBF}
\end{figure}

There are observations that prove that NTFs have an almost vertical magnetic field  \cite{FilamentsYusufZadeh,Morris2015,Mangilli2019}. Using the poloidal field component given by Equ.\ (\ref{xshape:equ}), the opening angle of the field lines (away from the $z$-axis) is governed by the parameter $a$. If $a$ is very small, the poloidal field goes to a vertical field, which can then be used for the NTF regions. Outside the filaments, the contribution from the NTF magnetic field is much weaker than the contribution from the ICM magnetic field due to the rapid decrease of the NFT field strength. This expectation is  exemplified by Sgr C-NTF in Fig.\ \ref{SgrCBF}, where the field strength outside the filaments decreases rapidly.

\begin{figure}[H]
	\centering
	\subfigure{\includegraphics[width=1.1\linewidth]{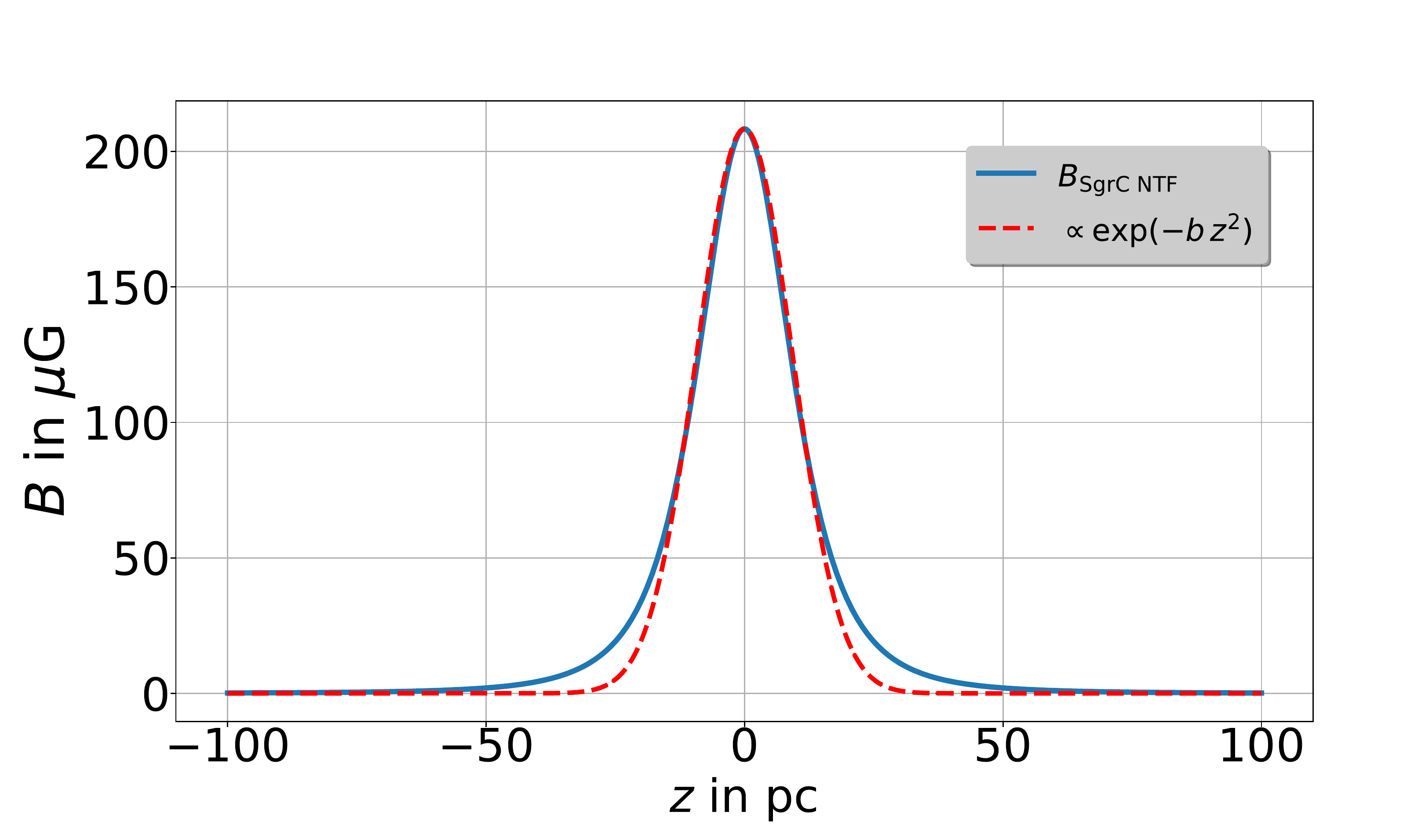}}
	\caption[]{$z-$profile of the poloidal field strength in units of $\mu$G (blue, solid line) in comparison to a Gaussian distribution (red, dashed line).}
	\label{SgrCZProfile}
\end{figure}

\begin{figure}[H]
	\vspace*{+0.5cm}
	\centering
	\subfigure{\includegraphics[width=1.0\linewidth]{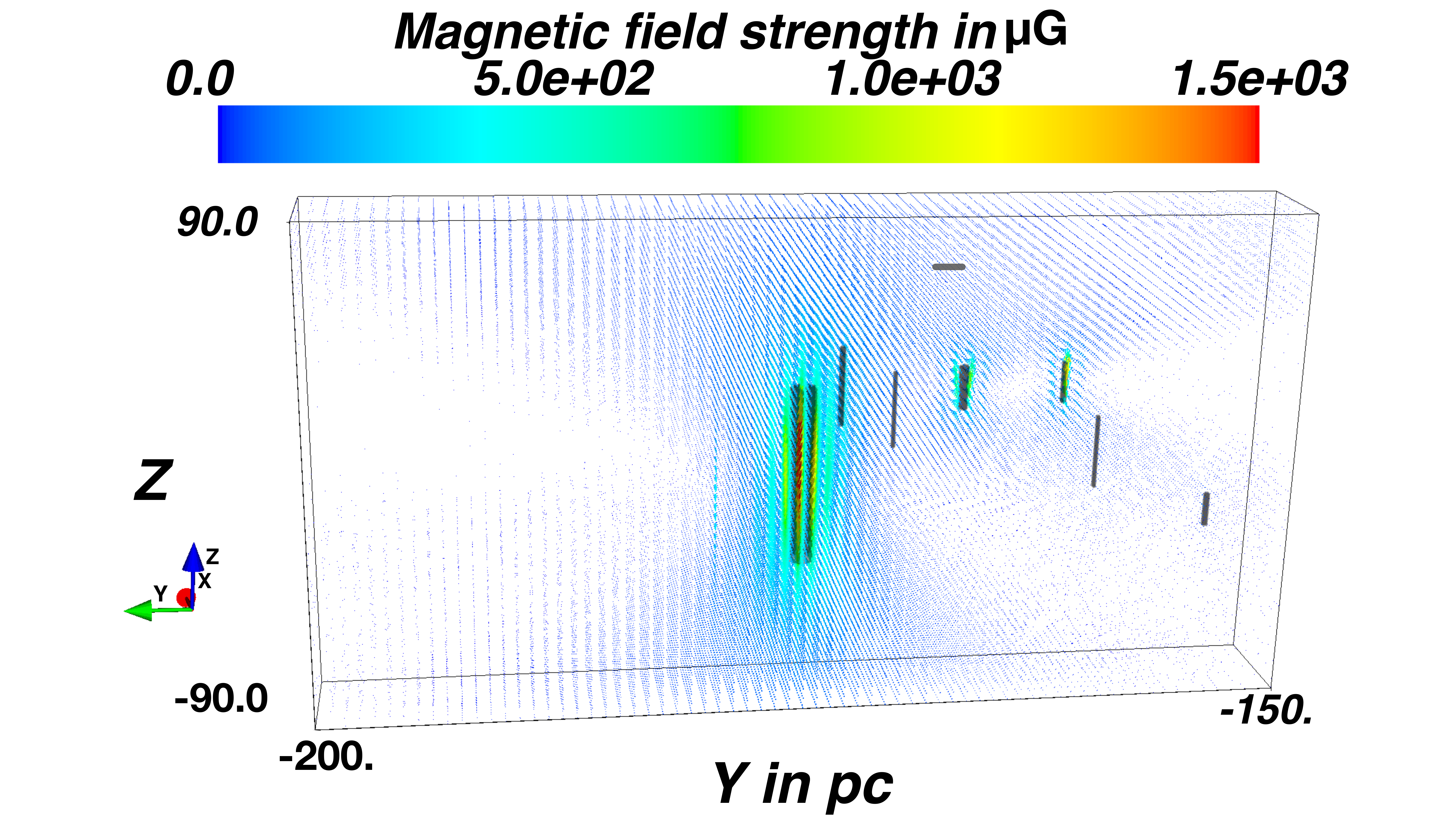}}
	\caption[]{
	Magnetic field strength in $\mu$G is presented considering the NTF components in Table \ref{table3}. The relative length and color of the arrows represent the relative magnetic field strength.}
	\label{FieldNTF}
\end{figure}

An external confining force must be present, otherwise the strong magnetic pressure inside filaments would tend to blow them apart.
Considering that the field strength in the NTF regions lies between $\sim0.1-1$\,mG, the  corresponding magnetic pressure lies in the range
\begin{equation}
    P_m=\frac{B^2}{8\,\pi}\approx (4\cdot10^{-10}\ - \ 4\cdot10^{-8}) 
    \ \mathrm{g\,cm^{-1}\,s^{-2}} \ .
\end{equation}
A first possibility is confinement by gas pressure,
\begin{equation}
    P_g=n_{\rm tot}\,k_B\,T \approx 3 \cdot 10^{-9}\,\left(\frac{n_{\rm e}}{0.1\,{\rm cm}^{-3}}\right)\,\left(\frac{T}{10^8\,\rm{K}}\right) 
    \ \mathrm{g\,cm^{-1}\,s^{-2}} \ ,
\end{equation} 
where $n_{\rm tot}\approx 2 n_{\rm e}$ is the total number density of particles (protons, helium nuclei, and electrons) and $n_{\rm e}$ is the density of free electrons in the ionized gas. The pressure from the observed hot ICM plasma \citep{bland_hawthorn2000} at a temperature $T\sim10^{6}$\,K and with an electron density $n_{\rm e}\sim0.002$\,cm$^{-3}$ is too low to counterbalance magnetic pressure. However, a gas at higher temperature $T\sim10^{8}$\,K and with an electron density $n_{\rm e}\sim0.01$\,cm$^{-3}$ has been observed with Chandra  \citep{muno2004}; this gas could very well counterbalance magnetic pressure. A second possibility is that the magnetic field has an azimuthal component, which provides an inward magnetic tension force.  A third possibility is that the actual field strength is in fact lower than the assumed strength of $0.1~{\rm mG}$, which is uncertain by a factor $\sim 10$. Finally,  it could also be that the filaments are not in pressure balance, but in a dynamical state.
The only exception is the {Pelican} NTF, which is rotated by about 90$^{\circ}$. \citep{LaRosaFilaments}.
Each of the NTFs requires an individual adaption of the free parameters. 
Similar to Equ. (\ref{ExpScaleL}), the radial exponential scale length $L$ is given by 
\begin{equation}
L=\frac{R_{\rm NTF}}{\ln(2)}=\frac{\Delta l}{2\cdot \ln(2)}
\end{equation}
where $\Delta l$ denotes the horizontal extent of the NTF which is presented in Table \ref{table3}.
The parameter $a$  determines the decrease with respect to the $z$-axis but also the opening of the field lines. Considering the observation, the magnetic field contributed by a NTF should not be dominant outside the NTF region. 
If the top arms and bottom arm of the poloidal field are pressed together, the upper and lower opening angle of the X decreases around the $z$-axis. This leads to a monotonous decrease of the magnetic field with respect to the $z$-coordinate. Since the B-field decreases quadratically with respect to the $z$, an approximation to the Gaussian distribution is thus close.
The vertical Gaussian scale should give us a good value of $a$ for an approximately purely vertical configuration. In this case, the vertical Gaussian scale yields
\begin{equation}
\frac{1}{\sqrt{a}}=\frac{H}{\sqrt{\ln(2)}}=\frac{\Delta b}{2\cdot\sqrt{\ln(2)}}
\end{equation}
where $\Delta b$ denotes the vertical extent of each NTF and is presented in Table \ref{table3}.
The comparison of the $z$-profile of the poloidal field model at $(x,y)=(0,0)$ and the profile of a Gaussian distribution $\propto\exp(-b\cdot z^2)$ is presented in Fig. \ref{SgrCZProfile}. 
Applying the poloidal field model to the \textit{Sgr C}-NTF, the best fit value for the Gaussian parameter $b\approx0.075 \ \rm pc^2$ is approximate the parameter $1/a^2\simeq0.061\ \rm pc^2$.
Figure \ref{SgrCBF} shows the field configuration.
The normalization factor $B_1$ is calculated in the same way as presented in Equ. (\ref{B1}). Again, the normalization factor $B_1$ is determined for each NTF individually from the measured values of the average field strength $\overline{B}_{\rm NTF}$ in each NTF presented in Table \ref{table3}. The volume factor $\alpha$  becomes about the same for all NTFs, i.e.\ $\alpha\approx 0.256$ and thus, the measured value of the average magnetic field in each NTF, $\overline{B}_{\rm NFT}$ scales linearly with $B_1$:

\begin{equation}
B_1\approx\frac{\overline{B}_{\rm NFT}}{0.26}\,.
\end{equation}
The different configuration in the {Pelican} NTF region has to be considered. As this NTF is perpendicular to all other NTFs, we apply a rotation matrix at the x-axis, which yields
\begin{equation}
\widehat{R}=\begin{pmatrix}
1 &   0         & 0           \\
0 & \cos \beta & -\sin \beta \\
0 & \sin \beta &  \cos \beta
\end{pmatrix}
\end{equation}
with $\beta=90\degree$. Thus,
\begin{equation}
\mathbf{B}_{\text{Pelican}}=\widehat{R}\, \mathbf{B}_{\mathrm{NTF}}^C=\begin{pmatrix} B_x\\ B_z\\ B_y \end{pmatrix}
\end{equation}
is obtained.
The total magnetic field in the NTF regions, from now on $\mathbf{B}_{\text{NTF}}^C$, is given by a superposition of the magnetic field in each NTF region.
Figure \ref{FieldNTF} illustrates the application of the poloidal field model in all NTF regions. 
\subsection{Horizontal field model for the MC regions}
\label{MC}
The magnetic field in the dense MCs is predominantly parallel to the Galactic plane, that means,  
it only has radial and azimuthal components, while its
vertical component can be neglected.

Here, we relate the azimuthal field component $B_{\phi}$ to the radial component $B_r$ and assume the same constant ratio for all MCs expect for the MC CND, which is highly influenced by the gravitational dynamics of SgrA*.
\begin{equation}
\left| \frac{B_{r}}{B_{\phi}}\right| =\eta  \text{ and } B_z = 0 \, .
\label{Br-Bphi}
\end{equation}
The comparison of polarization results of \cite{CNDObservation} with the innermost $5$\,pc MC of the CMZ which is called CND delivers $\eta=0.77$ \citep{CNRMagneticField}. 
For all other MCs, $\eta$ is not known and therefore will be kept as a free parameter. However, since $\eta$ determines the azimuthal shearing, a value larger than 1 means a very weak shearing and a value equal zero means a complete azimuthal shearing. Therefore, we show the results for three different values with $\eta\in [0,0.5,1.0]$ as examples. \\
For this purpose, we construct another analytical magnetic field model for the horizontal component considering the Euler potentials, $\alpha$ and $\beta$. In doing so, the magnetic field is naturally divergence-free,
\begin{equation}
\vec{B}=\vec{\nabla}\alpha\times\vec{\nabla}\beta
\, .
\label{EulerPotential}
\end{equation}
Here, we apply $\nabla$ in cylindrical coordinates.
Equation (\ref{EulerPotential}) produces a field with a vanishing $z$-component for instance if we choose $\beta=z$. 
In the plane perpendicular to the $z$-axis, the field lines can be described by
\begin{equation}
\begin{split}
\phi&=f_{\phi}(\alpha,r)\\
\text{or } \\ r&=f_r(\alpha,\phi),
\end{split}
\label{transform:equ}
\end{equation}
where each value of $\alpha$ refers to a field line. Thus, an arbitrary, monotonic function of $\alpha$, which we call $\psi(\alpha)$, represents the azimuthal angle of the field line at a specific radius $\rho$.
Here, $\rho$ represents the reference radius at which the field line is crossing the reference angle $\psi$ and is a free parameter.
This relation will simplify our calculation in order to apply the boundary condition of Equ.\ (\ref{Br-Bphi}) for the determination of the field.

With the transformation described in Equ.\ (\ref{transform:equ}) 
we obtain
\begin{equation}
\vec{B}=
\begin{pmatrix}
\frac{1}{r} \frac{\partial \alpha}{\partial{\psi}}\frac{\partial\psi}{\partial \phi}|_r \\
-\frac{\partial \alpha}{\partial{\psi}}\frac{\partial\psi}{\partial r}|_{\phi} \\
0
\end{pmatrix}
\label{bfield_azim:equ}
\end{equation}
We use $\eta=|B_{r}/B_{\phi}|=const.$ in order to determine $\psi$: 
\begin{equation}
\left| \frac{B_{r}}{B_{\phi}}\right| =\left|\frac{1}{r} \frac{dr}{d\phi} \right|_{\psi,\rho}=\eta=\text{const.}
\label{Br-Bphi-eta}
\end{equation}
\begin{eqnarray}
\mp \frac{dr}{r}|_{\rho} &=&\eta\,(d\phi|_{\psi})\nonumber\\
\mp  \ln\left( \frac{r}{\rho}\right) &=&\eta\,(\phi-\psi)\,. 
\label{r-phi}
\end{eqnarray}
As in astrophysical context $r$ extends to many magnitudes, we can transform $r \longrightarrow r+b$ with $r\gg b$ and still hold a constant ratio of $B_r$ and $B_{\phi}$. With the same assumption we also transform $\rho\longrightarrow \rho+b$. This transformation will be necessary for avoiding singularities in the argument of the cosine function in Equ.(\ref{cos}). The free parameters $b$ needs to fulfill $r\gg b$ and later will be set to a negligible fraction of the MC radius. Thus, the function $\psi$ can be described as
\begin{equation}
\psi=\phi\pm \eta^{-1}\,\ln\left( \frac{r+b}{\rho+b}\right) \,.
\end{equation}
The partial derivates become
\begin{equation}
\frac{\partial\psi}{\partial \phi}|_r=1 \text{ and } \frac{\partial\psi}{\partial r}|_{\phi}=\pm \frac{1}{\eta}\,\frac{1}{r+b}\,.
\end{equation}
Finally, we define $\partial \alpha/\partial \psi:= \rho\cdot \xi(\psi)\cdot h(z)$ and include an arbitrary function of $z$ which will not destroy the solenoidal property of the field. In doing so, Equ.\ (\ref{bfield_azim:equ}) becomes
\begin{equation}
\begin{split}
\begin{pmatrix}
B_r\\
B_{\phi}\\
0
\end{pmatrix}&=
\begin{pmatrix}
\frac{\rho}{r} \xi({\psi})\cdot h(z)  \\
\mp\eta^{-1}\,\frac{\rho}{r+b}\cdot \xi({\psi})\cdot h(z)  \\
0
\end{pmatrix}\\&\approx
\begin{pmatrix}
\frac{\rho}{r}\,\xi(\psi)\cdot h(z)\\
\mp \eta^{-1}\cdot\frac{\rho}{r}\, \xi(\psi)\cdot h(z) \\
0
\end{pmatrix}= \mathbf{B}_{\pm}.
\end{split}
\end{equation}
Here, we obtain two solutions $\vec{B}_-$ and $\vec{B}_+$, as only the absolute value of the ratio between the radial and azimuthal components of the magnetic field is known.
This unknown sign transforms into an unknown rotational direction that needs to be constrained by observations, as we discuss later.

The next step will specify $\xi(\psi)$. For this purpose, we assume a simple cosine which yields the same physical results for $\phi=n\cdot 2\pi$ and $n\in\mathbb{N}$. Moreover, this choice ensures the physical reality of a magnetic field, that is, the net magnetic flux is zero as we show below.
\begin{equation}
\begin{split}
\xi(\psi)=&B_1\cdot\cos(m\cdot{\psi})\\=&B_1\cdot\cos\left(  {\pm m\,\eta^{-1}}\cdot\ln\left( (\frac{r+b}{\rho+b})  \right) +m\cdot{\phi}\right)\, .
\end{split}
\label{cos}
\end{equation}
Here, $m$ denotes the azimuthal wavenumber as defined before.
The field becomes
\begin{equation}
\begin{split}
\begin{pmatrix}
B_r\\
B_{\phi}\\
0
\end{pmatrix}=&B_1\cdot\cos\left(  {\pm m\cdot\eta^{-1}}\cdot\ln\left( (\frac{r+b}{\rho+b})  \right) +m\cdot{\phi}\right)\\
&\cdot h(z)\cdot\begin{pmatrix}
\frac{\rho}{r}  \\
\mp\eta^{-1}\cdot\frac{\rho}{r+b}  \\
0
\end{pmatrix}.
\end{split}
\end{equation}
The $\phi$ component of the magnetic field is a continuous function whereas the $r$ component has a singularity at the origin, that means,  $\mathrm{div}\,B$ is not well-determined at the origin. However, in mathematical terms, $\vec{B}$ is divergence-free as
\begin{equation}
\mathrm{\nabla\cdot\vec{B}= \frac{1}{r}\frac{\partial(r\, B_r)}{\partial r} +\frac{1}{r} \frac{\partial B_{\phi}}{\partial \phi}+ 0=0}
\label{divB=0}
\end{equation}
for all combinations of $r,\,\phi$ and $z$.
In physical terms, the $1/r$ dependency in the $r$ component could lead to a divergence so that the field lines are disturbed and not closed and in particular lead to an infinitely large field component in the center of the clouds.
We first argue for the case of a physically divergence-free magnetic field and later tackle the question of an infinitely large field in the center of the clouds.
A physically divergence-free magnetic field in cylindrical coordinates has a net magnetic flux of zero through the surface of a cylinder with radius R and the length extending from $-a$ to $a$.
The magnetic flux is generally given by
\begin{equation}
\Phi_m=\int_A \widehat{n} \cdot \vec{B}\, \rm{d}A.
\end{equation}
In the case of a cylinder, there are three surfaces: lateral, top and bottom
\begin{equation}
\Phi_{m_1}=\int_{0}^{2\pi} \int_{-a}^a R\cdot B_r(R,\phi,z) \rm{d}\phi\, \rm{d}z
\end{equation}
\begin{equation}
\Phi_{m_2}=-\int_{0}^{2\pi} \int_{0}^R r\cdot B_z(R,\phi,-a)\, \rm{d}\phi\, \rm{d}r
\end{equation}
\begin{equation}
\Phi_{m_3}=\int_{0}^{2\pi} \int_{0}^R r\cdot B_z(R,\phi,a)\, \rm{d}\phi\, \rm{d}r
\end{equation}
Here, $\Phi_{m_1}$ denotes the flux through the side, that is, $\vec{e}_r$.  $\Phi_{m_2}$ and  $\Phi_{m_3}$ denote the fluxes through the bottom and top surfaces, respectively. As ab initio, $B_z(r,\phi)$ is zero the expressions $\Phi_{m_2}$ and  $\Phi_{m_3}$ are immediately zero which solely leaves $\Phi_{m_1}$.
$\Phi_{m_1}$ has two possibilities to vanish: \\
\noindent Firstly,

	\begin{equation}
	\begin{split}
	&\int_{0}^{2\pi}B_r(R,\phi)\, \rm d\phi = \\ &\int_{0}^{2\pi}\cos\left(  {\pm m\cdot\eta^{-1}}\cdot\ln\left( (\frac{r+b}{\rho+b})  \right) +m\cdot{\phi}\right)\, \rm d\phi=0
	\label{Brdphi=0}
	\end{split}
	\end{equation}
	
\noindent Here, Equ. (\ref{Brdphi=0}) requires a wavenumber $m\neq0$ and $m\in\mathbb{N}$, wherefore we  set $m=1$ as there is no clue for an explicit value and this ensures the solenoidal property.

\noindent
And secondly,	

	\begin{equation}
	\int_{-a}^a h(z)\, \rm d z\overset{!}{=}0 \text{ and } \underset{z\rightarrow \pm\infty}{lim} h(z) \overset{!}{=}0
	\label{h(z)condition}
	\end{equation}
	
\noindent	In this case, the constraint for the choice of  $h(z)$ is given by a continuous function and in the simplest case by an odd function, that is
	\begin{equation}
	h(z)=\frac{z}{H}\cdot \exp\left( - (\frac{z}{H})^2\right).
	\end{equation}
In order to constrain $\xi(\psi)$, we rely on the first condition, and thus the choice of $h(z)$ is just restricted by a continuous function. A convenient description which also considers the height of the MCs is given by the Gaussian vertical distribution
\begin{equation}
h(z)=\exp\left(- \frac{z^2}{H_c^2} \right)
\end{equation}
with $H_c=H/\sqrt{\ln(2)}$ and $H$ as the MC height.
Thus, the field in MCs has been constrained to
\begin{equation}
\begin{split}
\vec{B}=&B_1\cdot\cos\left(  {\pm m\cdot\eta^{-1}}\cdot\ln\left( (\frac{r+b}{\rho+b})  \right) + m\cdot{\phi}\right) \\&
\cdot\exp\left(-\frac{z^2}{H_c^2}\right)\cdot\begin{pmatrix}
\frac{\rho}{r} \\
\mp\eta \cdot\frac{\rho}{r+b}   \\
0
\end{pmatrix}\, .
\label{BFieldUnmodified}
\end{split}
\end{equation}

The singularity of $B_r\propto 1/r$  arises from the fact that all field lines meet at $r=0$. In order to prevent this unphysical 
behavior, we modify the expression of $\vec{B}$ in the innermost region, $r<r'$, in such a way that all field lines (except for one at each height) avoid the Galactic center. 

\begin{figure}[H]
	\vspace*{-0cm}
	\hspace*{-0cm}
	\subfigure{\includegraphics[width=0.7\linewidth]{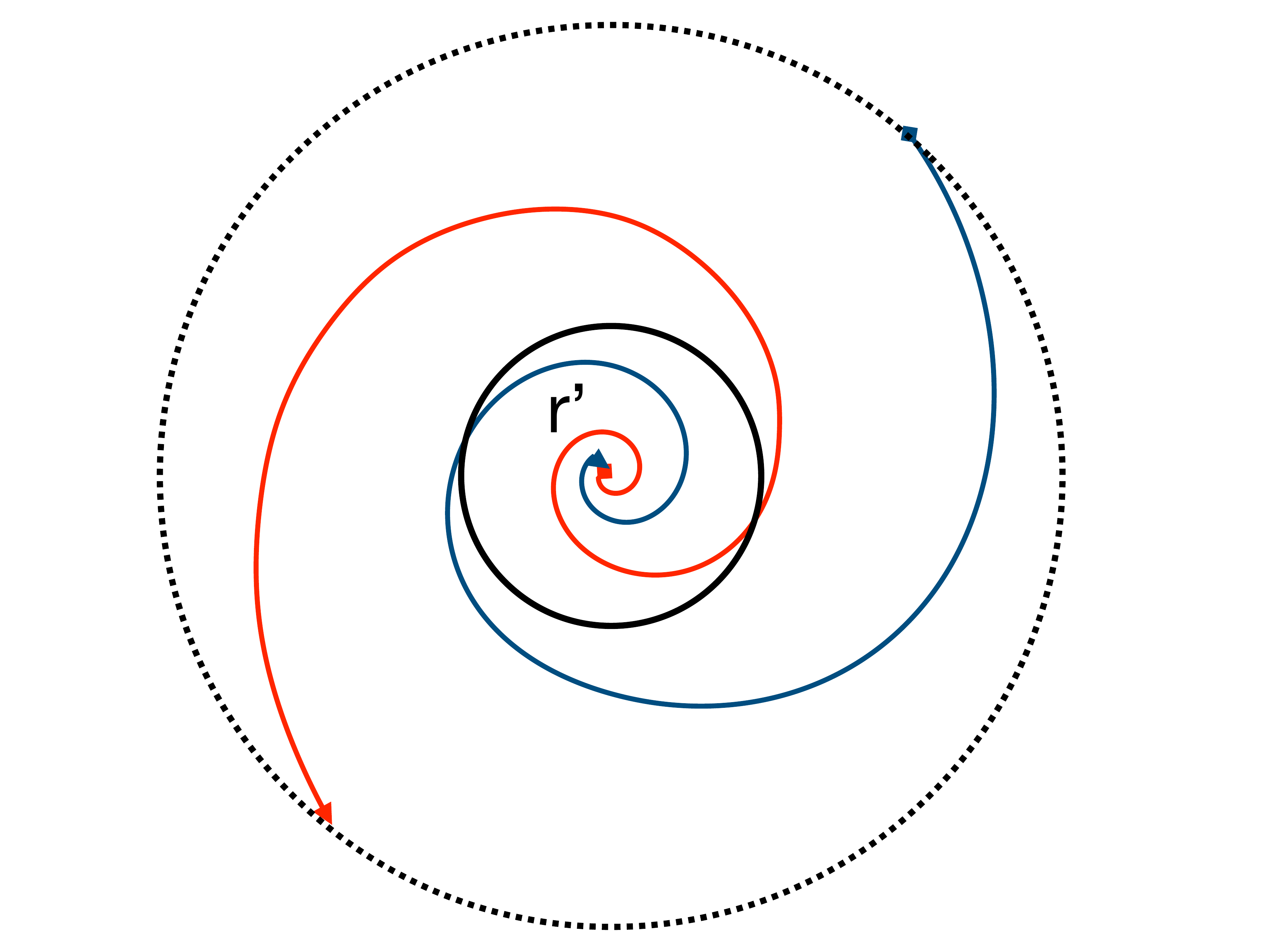}}
	\centering
	\subfigure{\includegraphics[width=0.7\linewidth]{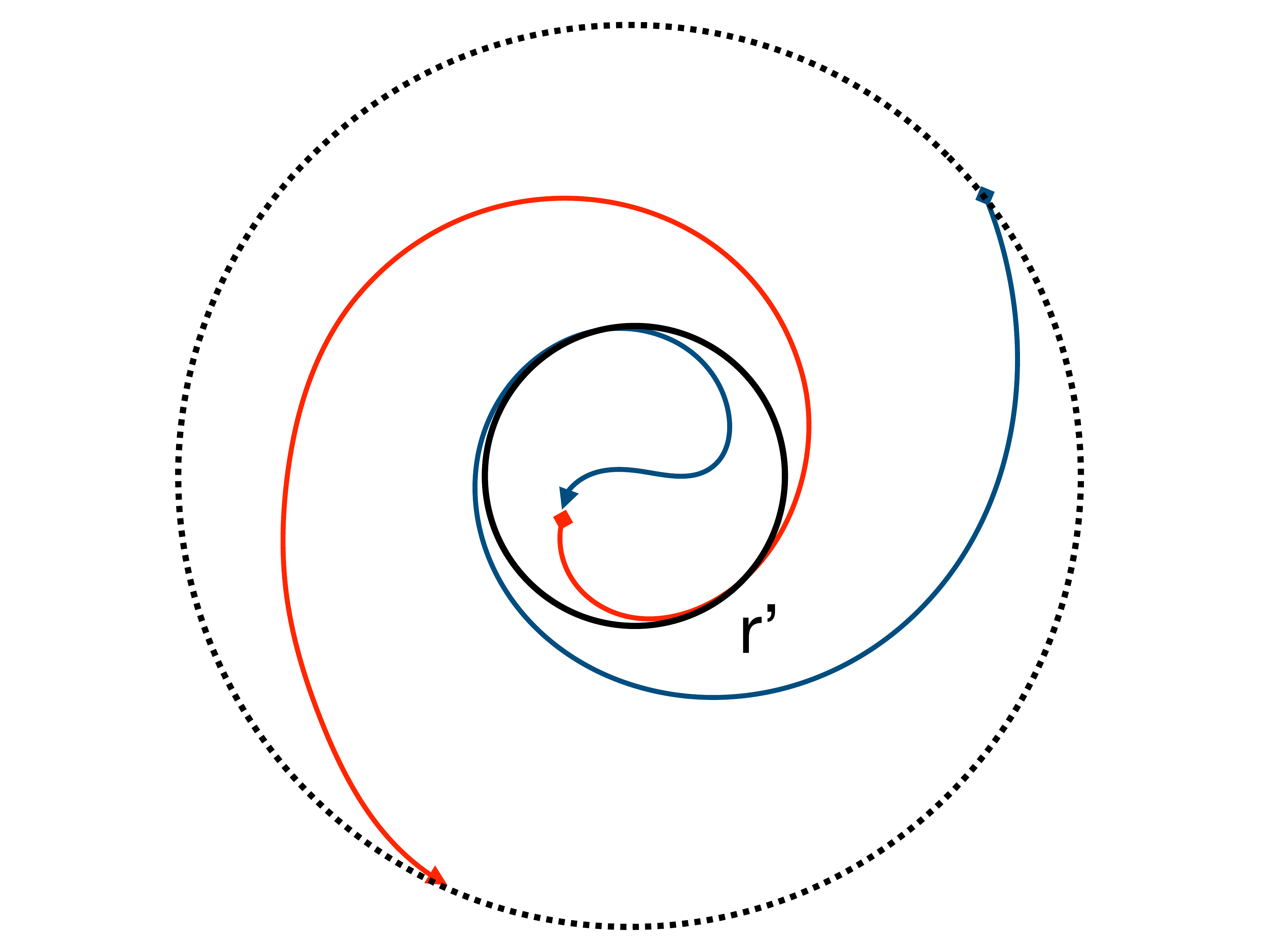}}
	\caption[]{Regularization of a bi-symmetric ($m=1$) spiral magnetic field that diverges for $r \to 0$.
Upper panel: Field as given by Equ.\ (\ref{BFieldUnmodified}), where $B_r \propto 1/r$, so that all field lines meet at $r= 0$.
Bottom panel: Modified field  inside radius $r'=R/10$ (Equ.\ (\ref{BFieldModified2})), so that field lines now avoid the center.}
	\label{SpiralField}
\end{figure}
\noindent Therefore, within a radius $[0,\,r']$ ($r'\ll R$), we introduce a modified field, which redistributes all incoming and outgoing field lines and therefore also changes the $r$ component. This correction is also already done for nonsolenoidal parts of the Galactic magnetic field model presented in \cite{Kleimann2019} and also in a similar way in \cite{Ferriere2017}. As discussed earlier, the divergence is supported by the factor $R/r$ in the $B_r$ component in Equ.\ (\ref{BFieldUnmodified}). We replace this factor by a differentiable function $p(r,r')$ at the boundary $r=r'$. We choose a second order polynomial including three coefficients which is fully determined due to the differentiability condition
\begin{equation}
\left.p(r,r')\right|_{r=r'}=\frac{\rho}{r'} \text{ and }\left.\frac{d p(r,r')}{dr}\right|_{r=r'}=-\frac{\rho}{r'^2}
\end{equation}
plus the request of equating $p(0,r')=0$.
In doing so, we obtain
\begin{equation}
p(r,r')= \frac{\rho}{r'}\left( 3\cdot\frac{r}{r'} -2\cdot\frac{r^2}{r'^2}\right)\, .
\end{equation}
With the help of this new function, we can now introduce the new modified field
\begin{equation}
\mathbf{B}_{\text{m}}=\frac{p(r,r')}{\frac{\rho}{r}}\cdot\mathbf{B}.
\end{equation}
In this representation, the modified field becomes non-solenoidal and violates the divergence-free condition.
Thus, we have to include an additional term into the $\phi$ component of the field that keeps the field divergence-free:
\begin{equation}
\begin{split}
&\vec{\nabla}\cdot(\mathbf{B}_{\text{m}}+B_{\phi,\text{new}}\cdot \vec{e}_{\phi})=\\ &\frac{1}{r}\frac{\partial (r\cdot B_{r,m})}{\partial r} +\frac{1}{r}\frac{\partial (B_{\phi,m}+B_{\phi,\text{new}})}{\partial \phi}=0\, .
\label{PhiNew-DivergenceFree}
\end{split}
\end{equation}
\begin{equation}
\begin{split}
B_{\phi,\text{new}}=& -\frac{\partial (r\cdot p(r,r'))}{\partial r} \cdot \int_{\phi_0}^{\phi} B_r(\psi(\phi '))\rm \, d\phi '\\
=&B_1 \cdot 12\cdot \frac{ \rho}{r'}\left( \frac{r}{r'}-\frac{r^2}{r'^2}\right) \cdot\exp\left(- \frac{z^2}{H_c^2} \right)\\ \cdot& \sin\left(  \pm \frac{m}{\eta}\cdot\ln\left( \frac{r+b}{\rho+b}  \right) + \frac{m}{2}\cdot{\phi }\right)\cos\left( \frac{m}{2}\cdot\phi\right) 
\end{split}
\end{equation}
Here, $\phi_0$ occurs as a free parameter which is set to zero in the last step.

Thus, for $r>r'$, the modified field in the MC regions $\vec{B}^{\pm}_{\text{MC}}$ is given by
\begin{equation}
\begin{split}
\vec{B}^{\pm}_{\text{MC}}=&B_1\cos\left(  \pm v(r) + m{\phi}\right)\\ \cdot&\exp\left(- \frac{z^2}{H_c^2} \right)
\begin{pmatrix}
\frac{\rho}{r} \\
\mp\eta^{-1} \cdot\frac{\rho}{r+b}  \\
0
\end{pmatrix}\\
\text{with } &v(r)=m\cdot\eta^{-1}\cdot\ln\left( \frac{r+b}{\rho+b}\right)
\end{split}
\label{BFieldMmodified1}
\end{equation}
For $r<r'$, it yields
\begin{equation}
\begin{split}
&\vec{B}^{\pm}_{\text{MC}}=B_1\exp\left(- \frac{z^2}{H_c^2} \right)\frac{\rho}{r'}\left( \frac{3r}{r'} -\frac{2r^2}{r'^2}\right)\cos\left(  \pm v(r)  + m{\phi}\right)\\
&\cdot 
\begin{pmatrix}
1\\
\mp \frac{r}{\eta(r+b)}\left( 1 +\frac{6(r-r')}{2\, r-3\, r'}\left(\frac{\sin(\pm v(r)+m\phi)-\sin(\pm v(r))}{\cos(\pm v(r)+m\phi)}\right)\right)  \\
0
\end{pmatrix}\, .
\end{split}
\label{BFieldModified2}
\end{equation}
This final expression of the horizontal magnetic field
has five free parameters: $H_c=H/\sqrt{\ln(2)}$, $\rho$, $\eta$, $m$ and $B_1$.
Here, the related Gaussian vertical scale height $H$, and the reference radius, $\rho$, are both
set to the MC radius, $R$, $\eta $ is given by a function of the MC gas density and intrinsic velocity. For ensuring the re-connection of the field lines, as illustrated in Fig. \ref{SpiralField} for bisymmetric fields, the wavenumber $m$ is set to 1. In doing so, we prevent the singularity at $r=0$. Further, from now on we assume that $r'$ corresponds to a fraction of the total radius of the MCs, here, $r'=R/10$.
The parameter $B_1$ is determined in the same way as done in Section \ref{PFM} and Equ.\ (\ref{B1}) where the average field values $\overline{B}_{\rm MC}$ are used to determine the individual $B_1$, that is, $B_1=\overline{B}_{\rm MC}/\alpha$.
$\overline{B}_{\rm MC}$ is listed in Table \ref{table1} and $\alpha$ depends on the parameter $\eta$, thus, $[\alpha_{\eta=0}, \alpha_{\eta=0.5}, \alpha_{\eta=0.77}, \alpha_{\eta=1.0}]=[59.40,1.29,0.91,0.75]$\\
\subsection*{Magnetic field at the event horizon of SgrA*}
Since the ambient condition around a supermassive black hole is completely different compared to a MC, we are not able to relate the $B_r$ and $B_{\phi}$ component. Thus, we merely consider the $\phi$ components of the magnetic field for this region and set $m=0$ in order to ensure the divergence-free property. 
For $ r>r'$
\begin{equation}
\vec{B}_{\rm SgrA^*}^{\pm}=\mp B_1\cdot\exp\left(- \frac{z^2}{H_c^2} \right)\cdot\frac{R}{r}\cdot
\begin{pmatrix}
0\\
1\\
0
\end{pmatrix} 
\end{equation}
and for $ r'>r$
\begin{equation}
\vec{B}_{\rm SgrA^*}^{\pm}=\mp B_1\cdot\exp\left(- \frac{z^2}{H_c^2} \right)\frac{R}{r'}\cdot \left( 3\frac{r}{r'} -2\frac{r^2}{r'^2}\right)\cdot
\begin{pmatrix}
0\\
1\\
0
\end{pmatrix}.
\end{equation}
In doing so, we ensure Equ.~(\ref{PhiNew-DivergenceFree}), and due to $B_r=0$ the total magnetic flux remains still zero. Additionally, we fulfill the $1/r$ dependency as suggested by \cite{Eatough2013} and \cite{Johnson2015}. In this construction the volume factor $\alpha_{\mathrm{SgrA*}}$ corresponds to 3.07. 
The magnetic field reaches a mG field at a radius of 0.024\,pc and a 10\,$\mu$G field at a radius of 2.4\,pc. Thus, the magnetic field induced by SgrA* is subdominant at latest after reaching the position of the MC CND, which has a field strength of 3\,mG.

\subsubsection*{Application of the horizontal field model to the inner 10\,pc}
\begin{figure}[H]
	\vspace*{-0cm}
	\hspace*{-0cm}
	\centering
	\subfigure{\includegraphics[width=0.9\linewidth]{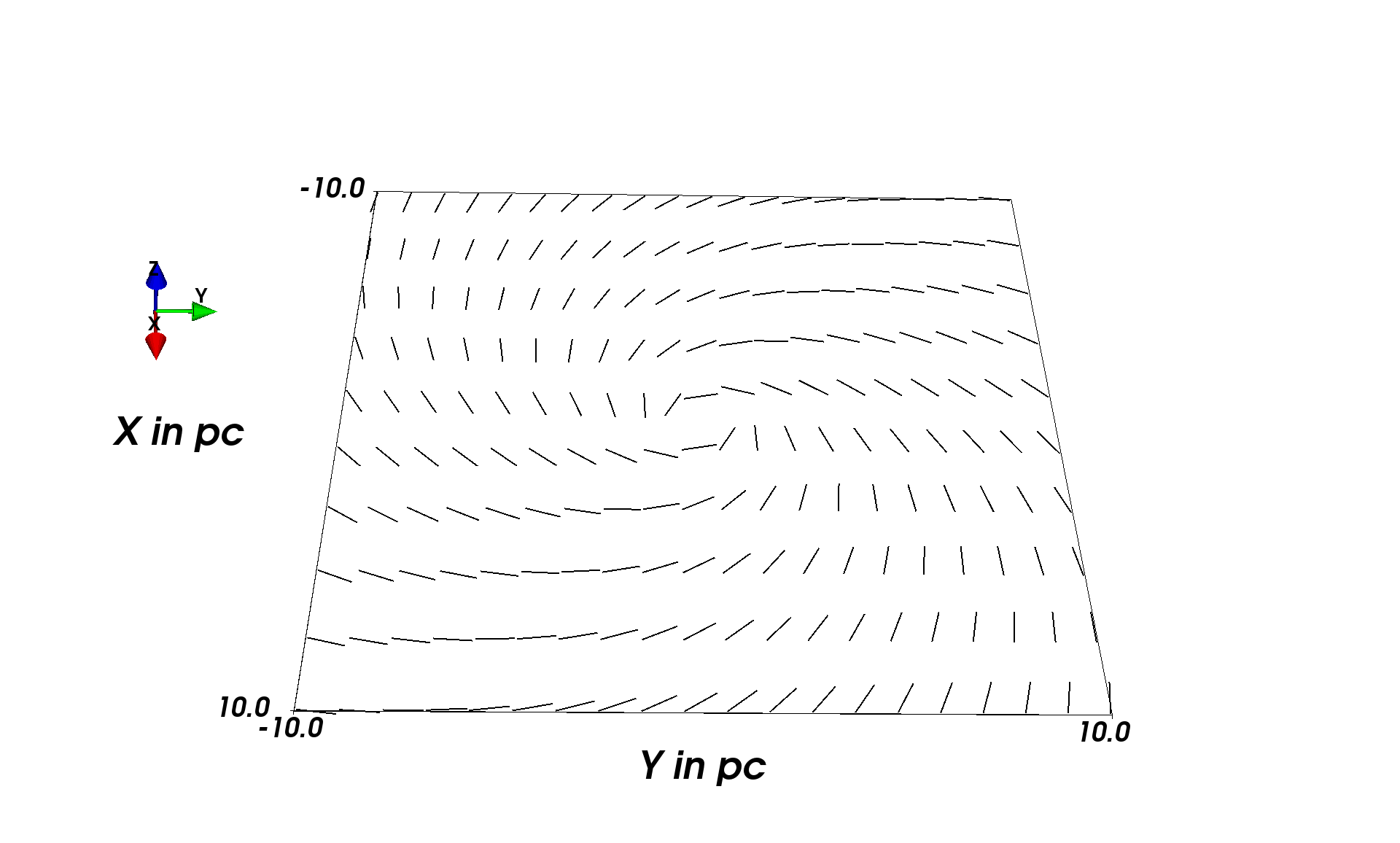}}
	\subfigure{\includegraphics[width=0.75\linewidth]{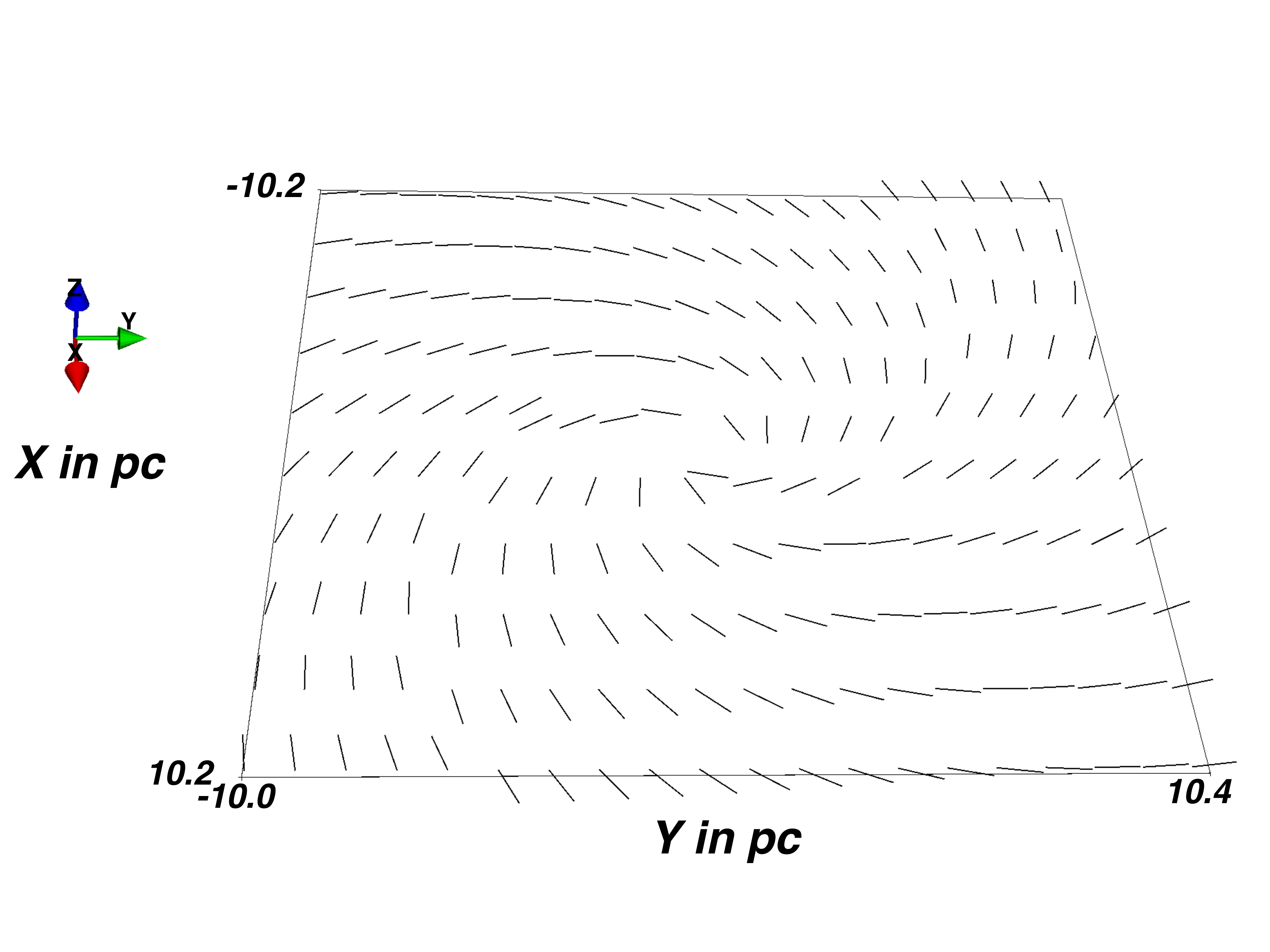}}
	\subfigure{\includegraphics[width=0.8\linewidth]{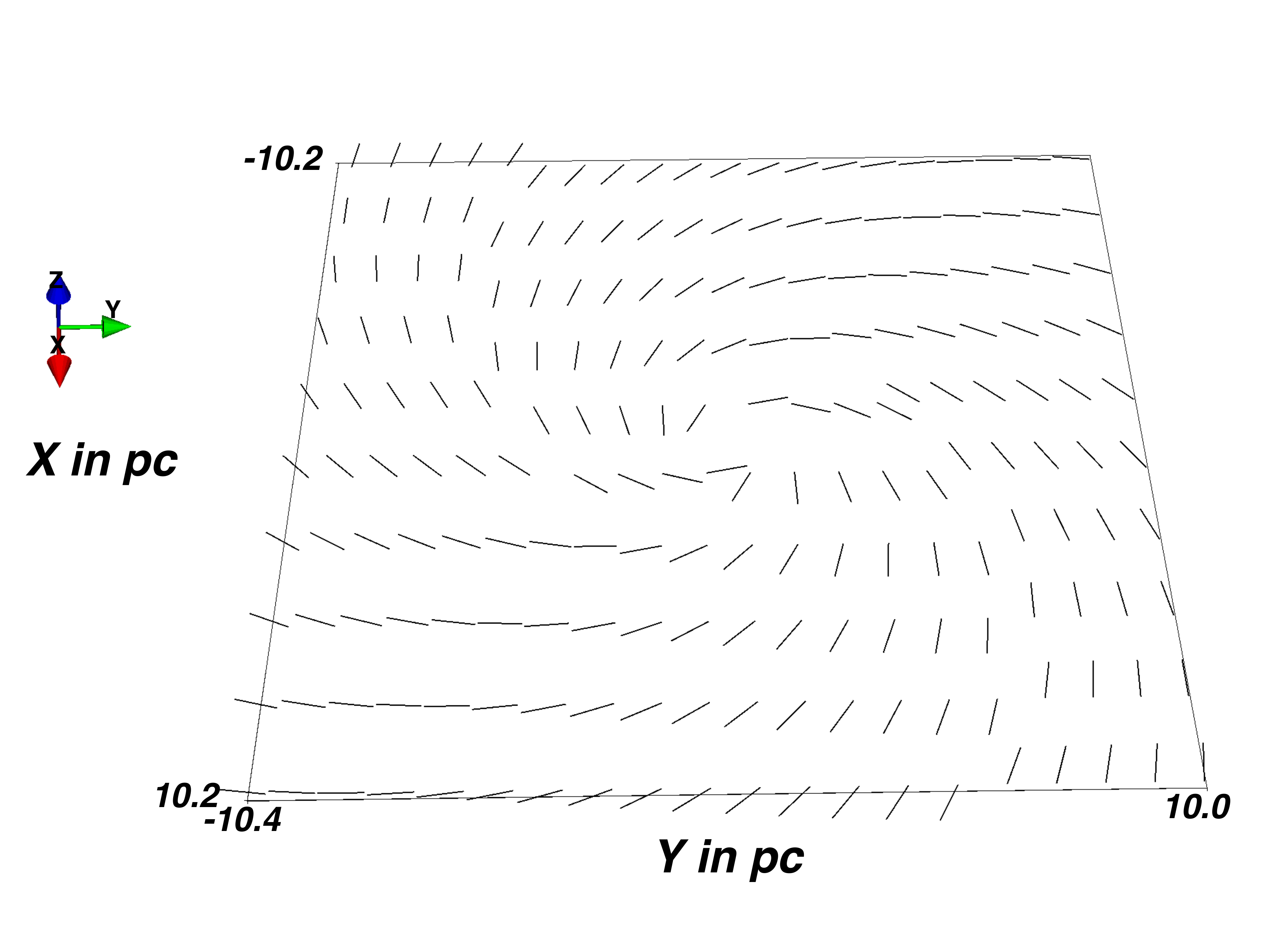}}
	\caption{Vizalization of the MC field - Upper panel: model of \cite{CNRMagneticField}; middle panel: visualization of $\vec{B}^-$ (left) and bottom panel: visualization of  $\vec{B}^+$ (right).}
	\label{FieldCNDModel}
\end{figure}

As discussed above, two solutions $\vec{B}^{-}_{\text{mod}}$ and $\vec{B}^{+}_{\text{mod}}$ can mathematically describe the horizontal field. The difference between these two is based on the rotational direction, and the observations should constrain the solutions. We therefore apply our model to the central 10\,pc around SgrA* and compare the result to the model of \cite{CNRMagneticField} who fit their model to the polarization measurements of \cite{Hildebrand1990}.\\
\noindent Figure \ref{FieldCNDModel} sketches the resulted two configurations of our model using $R=5$\,pc and $\eta=0.77$ which are the parameters around the CND. The comparison of these configurations reveals the model $\vec{B}^+$ to fit most suitable to the data of \cite{Hildebrand1990}. By assuming that all MCs in CMZ are following the same rotational direction, we adopt $\vec{B}^+$ for all MCs.
\subsection{Total field and comparison with data}
\label{Section:DataComparison}
\begin{figure}[H]
	\vspace*{+0.0cm}
	\centering
	\hspace*{-1.0cm}
	\subfigure{\includegraphics[width=1.2\linewidth]{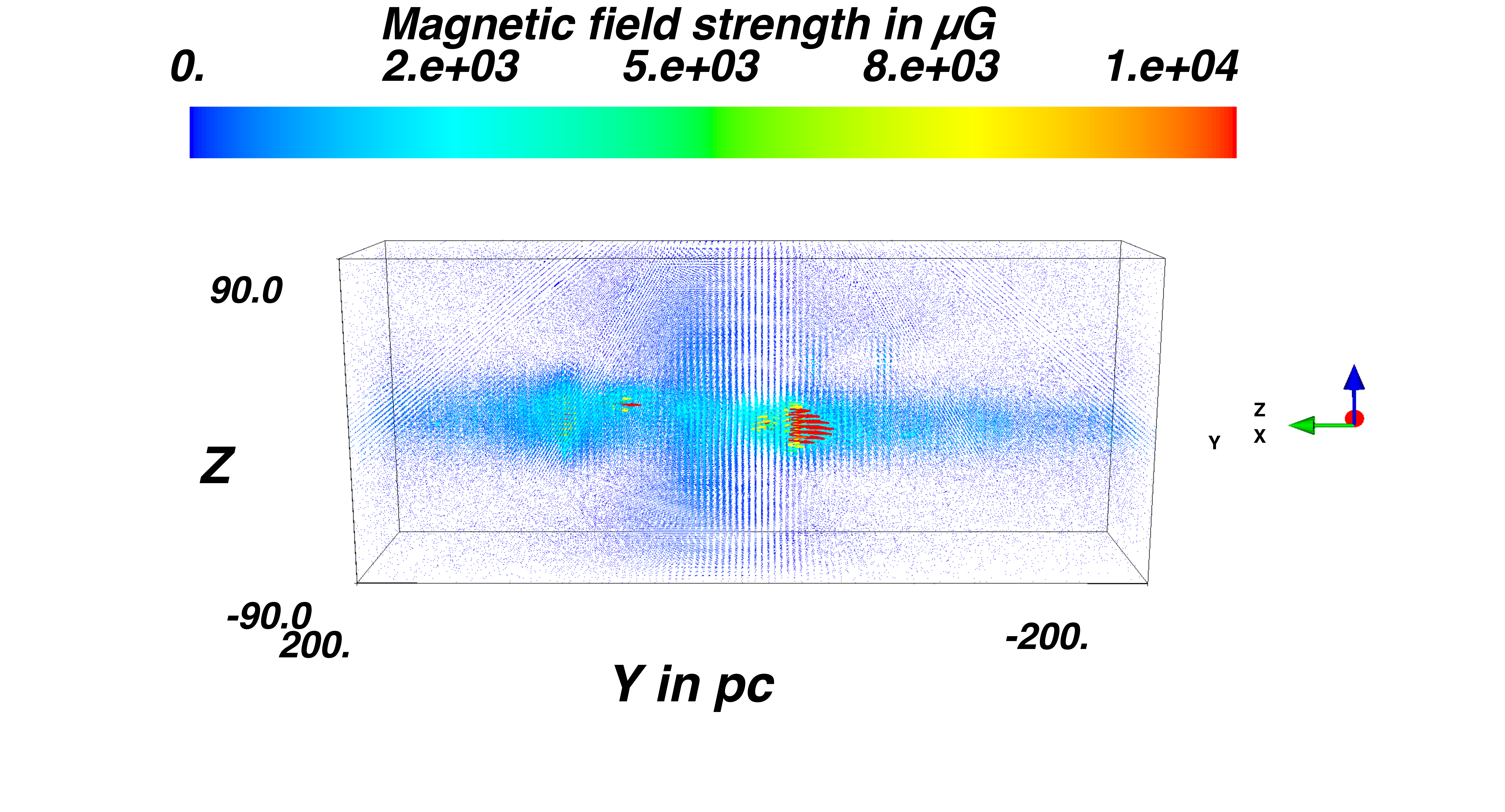}}
	\subfigure{\includegraphics[width=1.\linewidth]{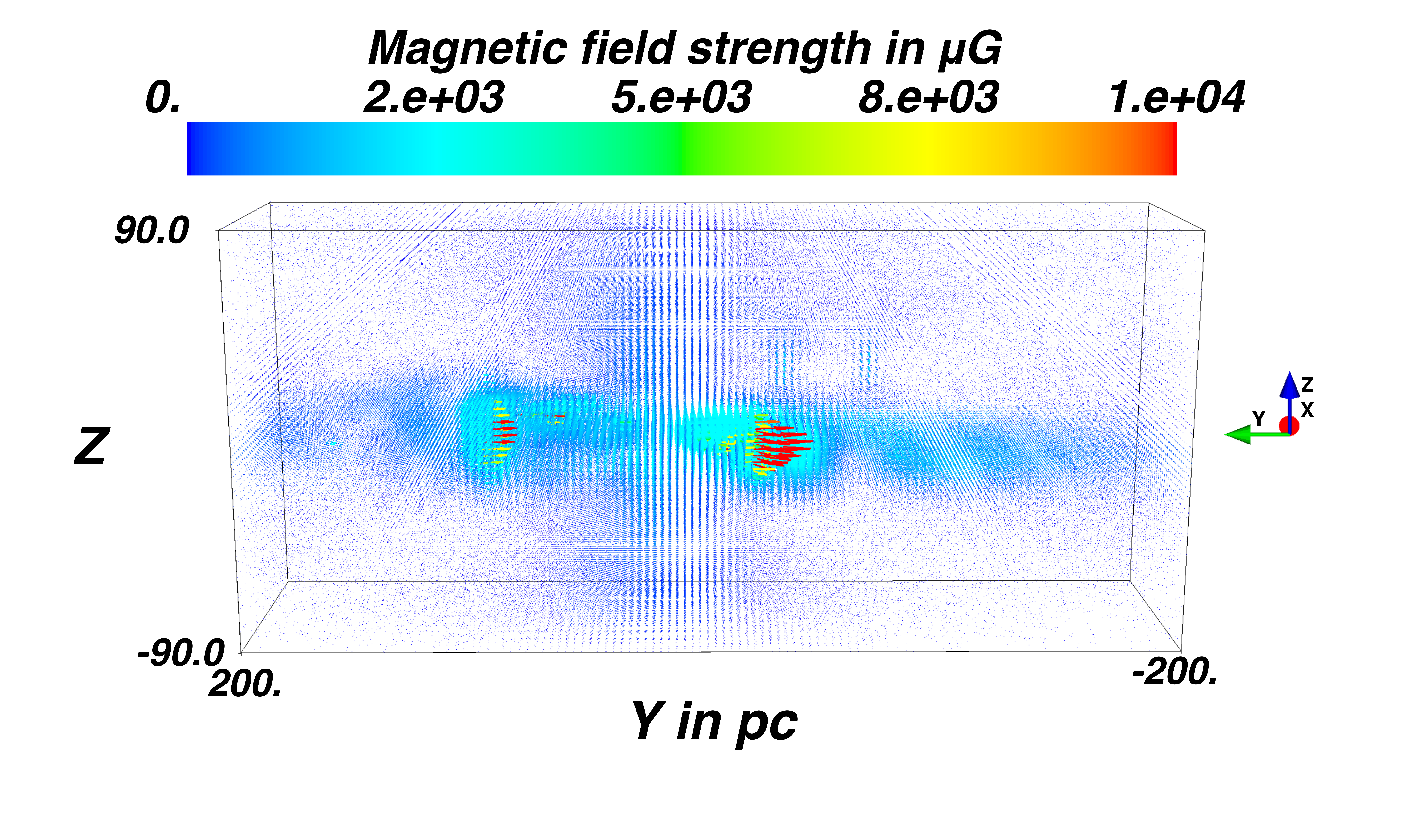}}
	\subfigure{\includegraphics[width=1.\linewidth]{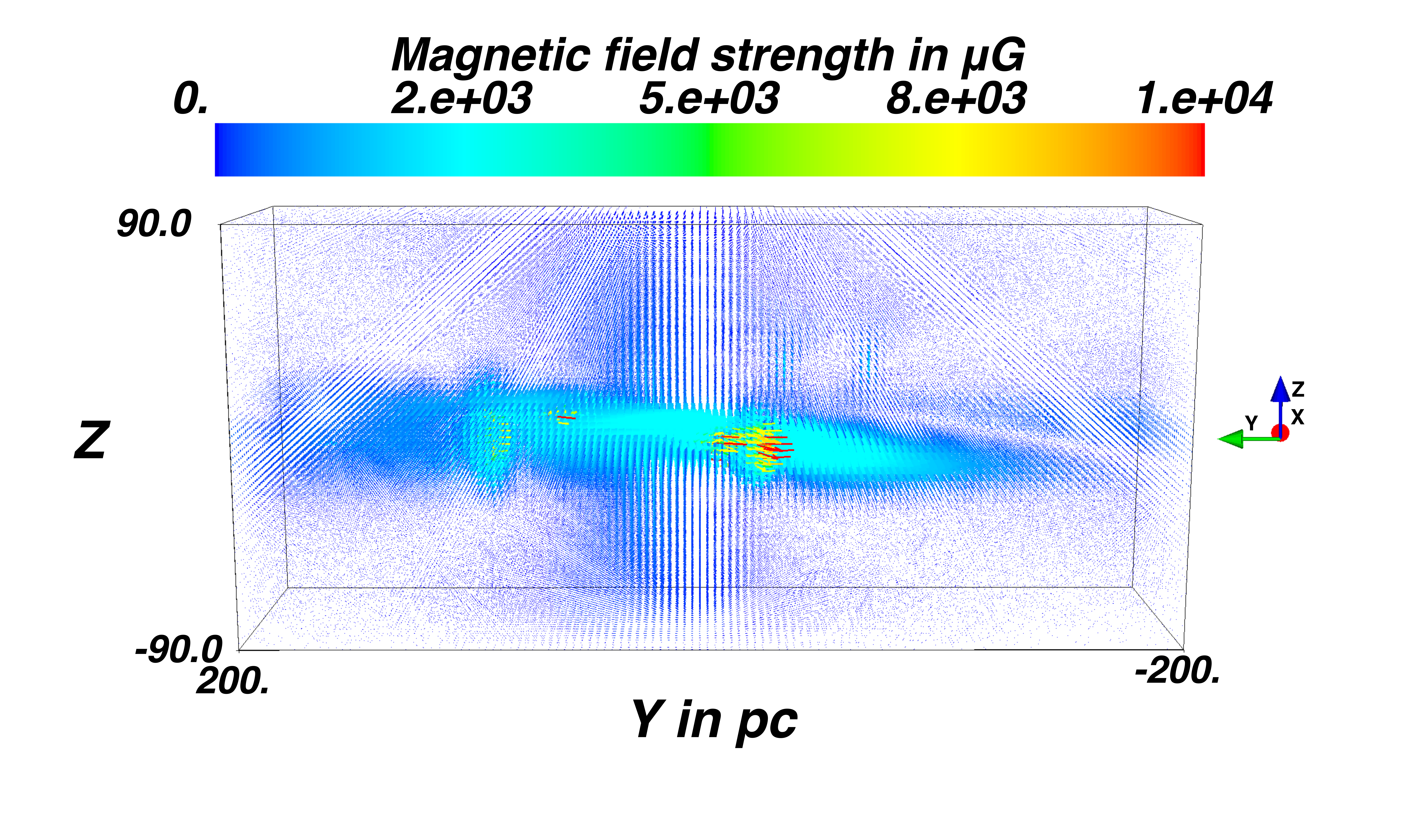}}
	\caption[]{Total magnetic field strength in the CMZ for $\eta=0$ (upper panel), $\eta=0.5$ (middle panel) and $\eta=1.0$ (lower panel). Relative length of the arrows/pixels represents the relative magnetic field strength.}	\label{FieldTotal}
\end{figure}

The total field in the CMZ is obtained by a superposition of magnetic field components as derived in Sections \ref{PFM} and \ref{MC}.
\begin{equation}
\mathbf{B}_{\text{tot}}=\mathbf{B}_{\text{IC}}^C+\sum_{i=1}^{8}\mathbf{B}_{\text{NTF,i}}^C+\sum_{i=1}^{12}\mathbf{B}^{+}_{\text{MC,i}}+\mathbf{B}_{\rm SgrA^*}^+
\end{equation}
The following figures 
show the total magnetic field strength and the related configuration including the ICM, NTF and MC components for different values of $\eta$. Furthermore, the dense MCs and the NTFs are added in Fig. \ref{FieldTotal}.

\begin{figure}[H]
	\hspace*{-0.8cm}
	\centering
	\subfigure{\includegraphics[width=1.2\linewidth]{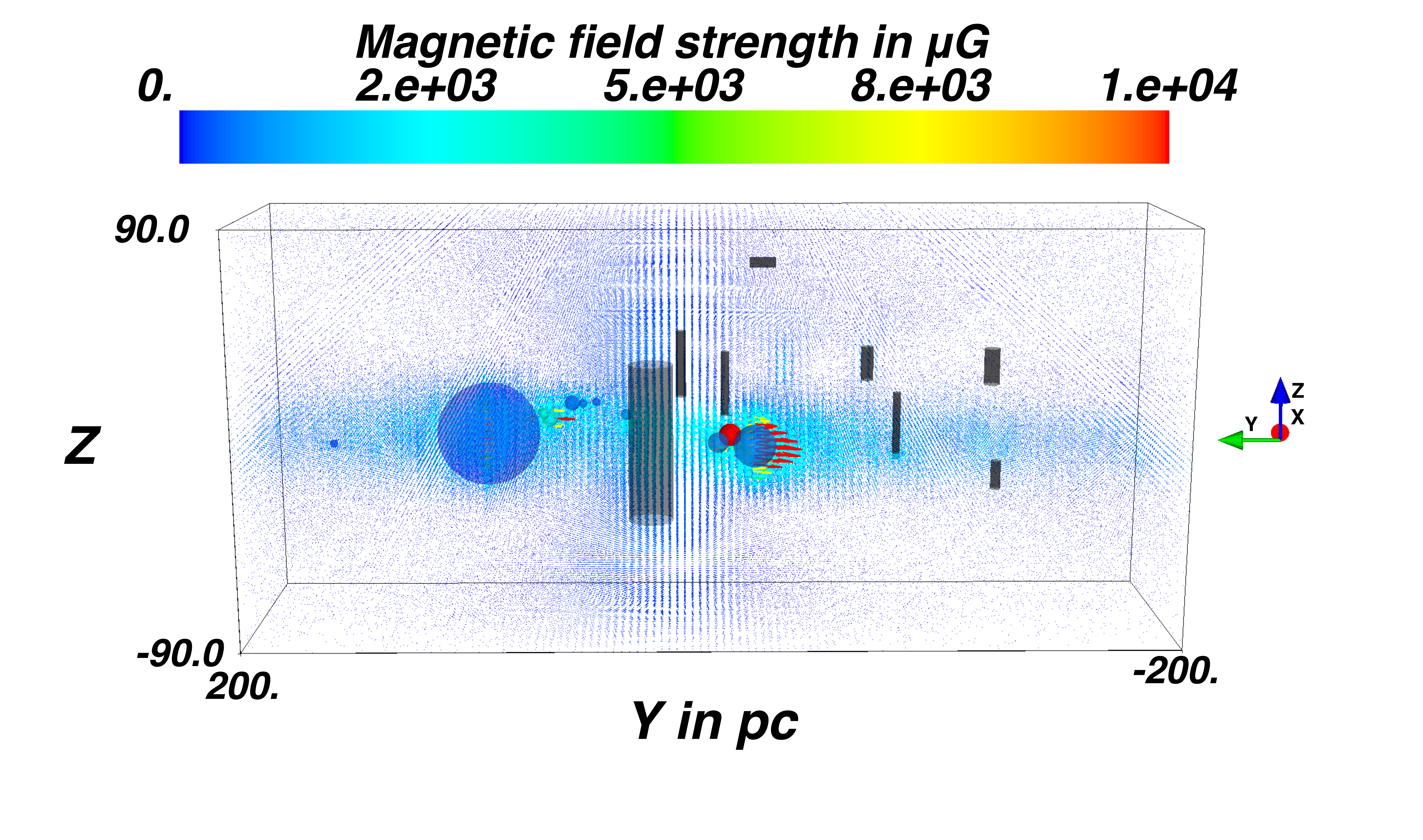}}
	\hspace*{-0.8cm}
	\subfigure{\includegraphics[width=1.2\linewidth]{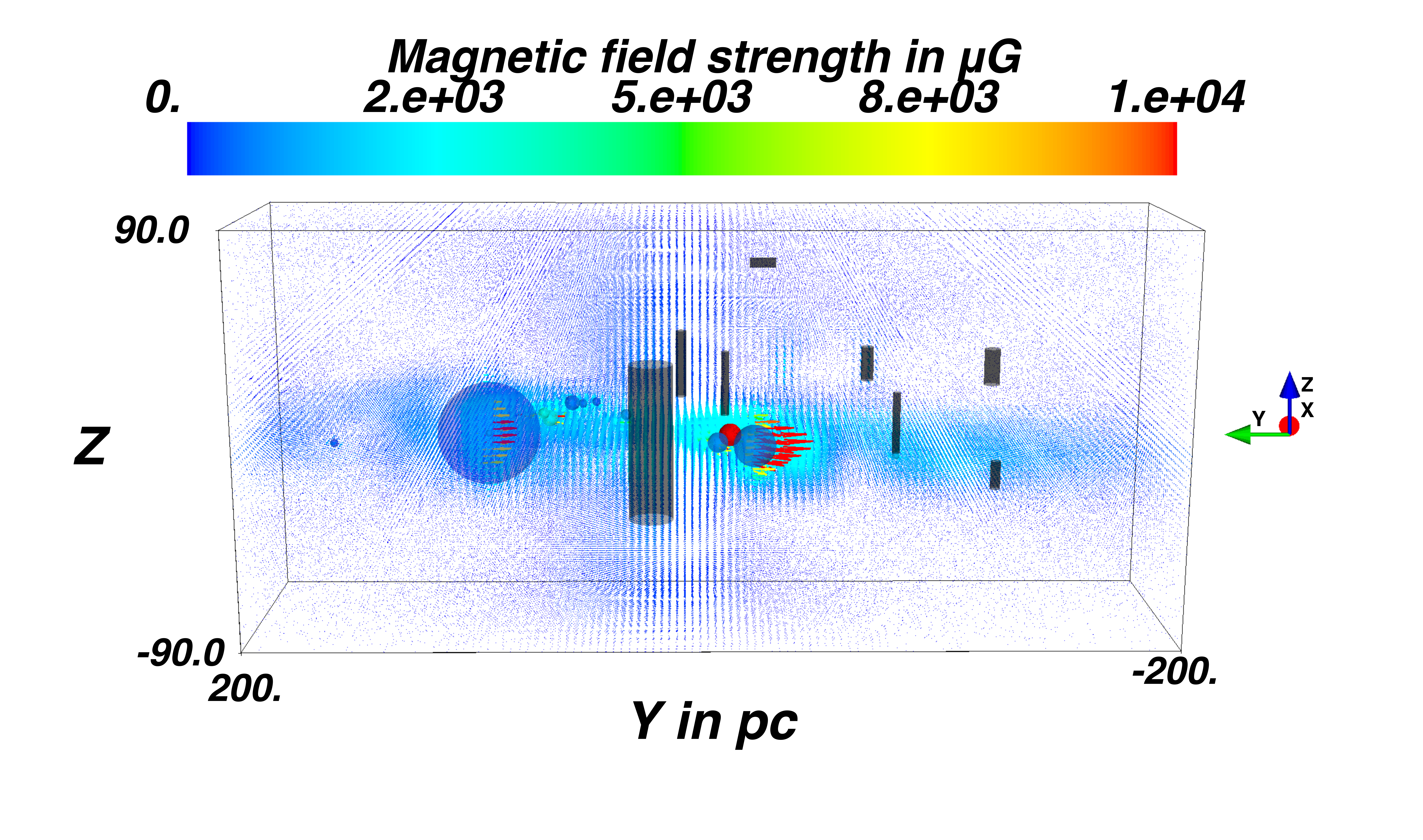}}
	\hspace*{-0.8cm}
	\subfigure{\includegraphics[width=1.2\linewidth]{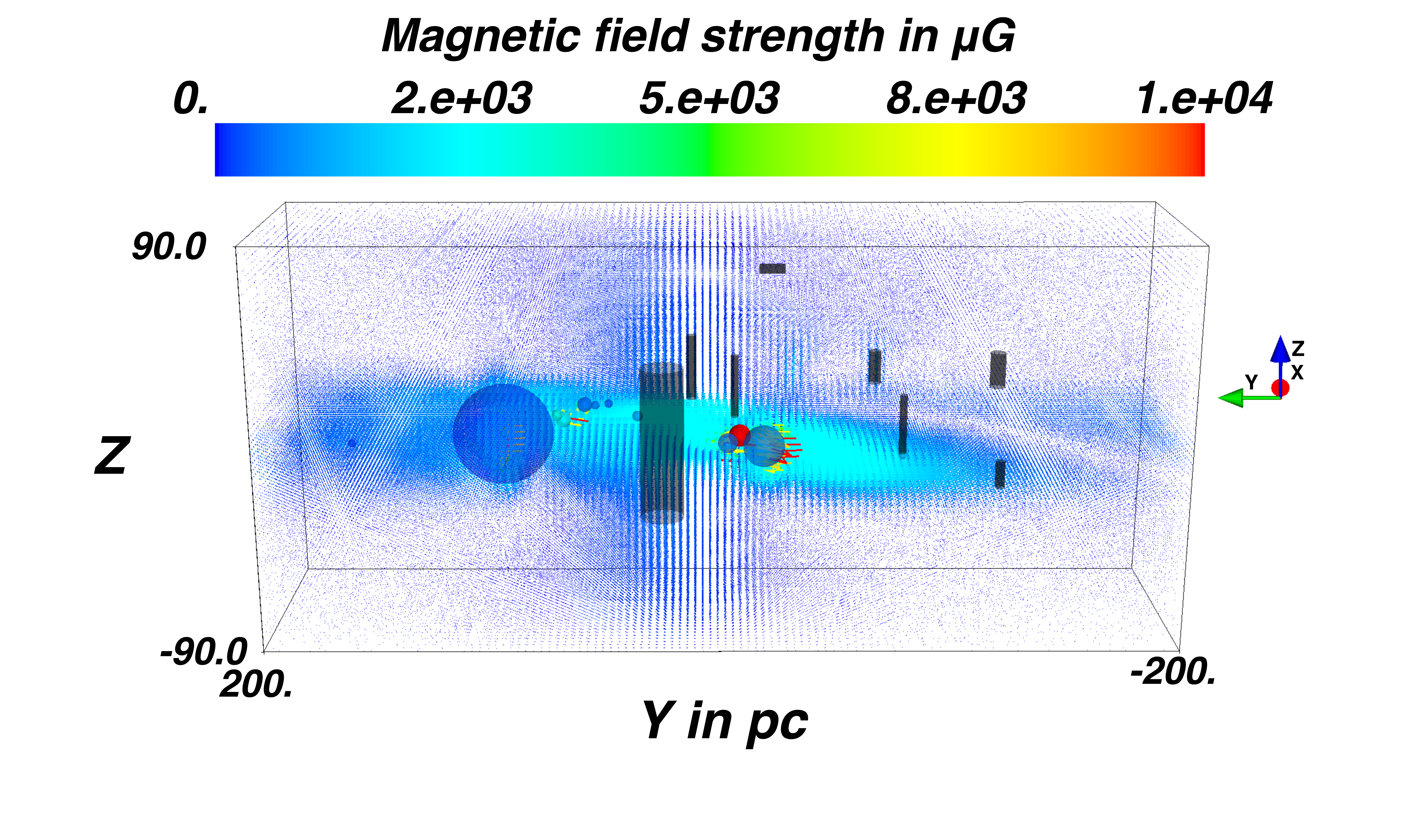}}
	\caption[]{Same as Fig. \ref{FieldTotal}. Additionally, the figures show the particle densities for the MCs in units of cm$^{-3}$ and the NTFs are shown schematically.}
	\label{FieldTotal_with_masses}
\end{figure}
\begin{figure}[H]
	\hspace*{-0.4cm}
	\subfigure{\includegraphics[width=1.1\linewidth]{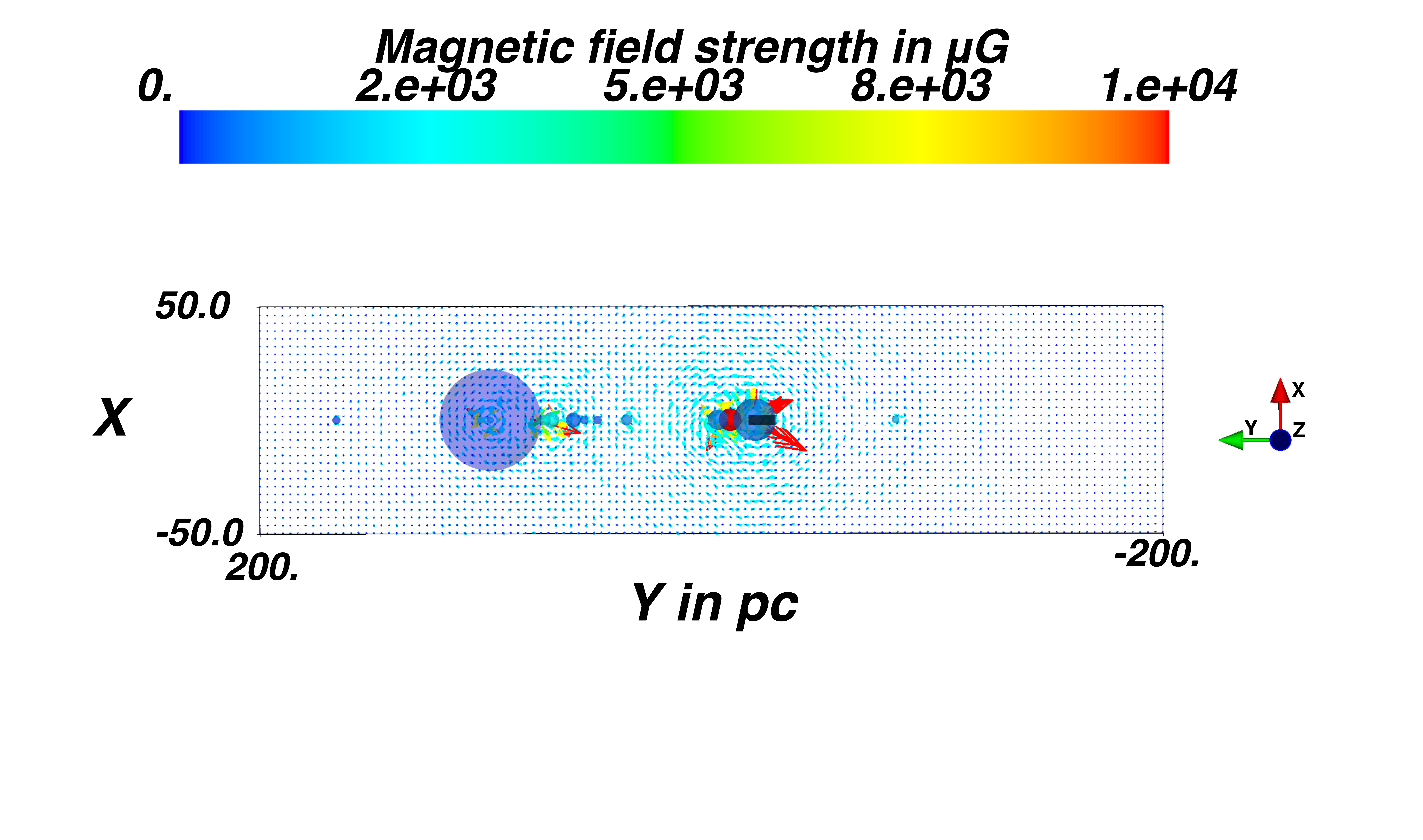}}
	\subfigure{\includegraphics[width=1.1\linewidth]{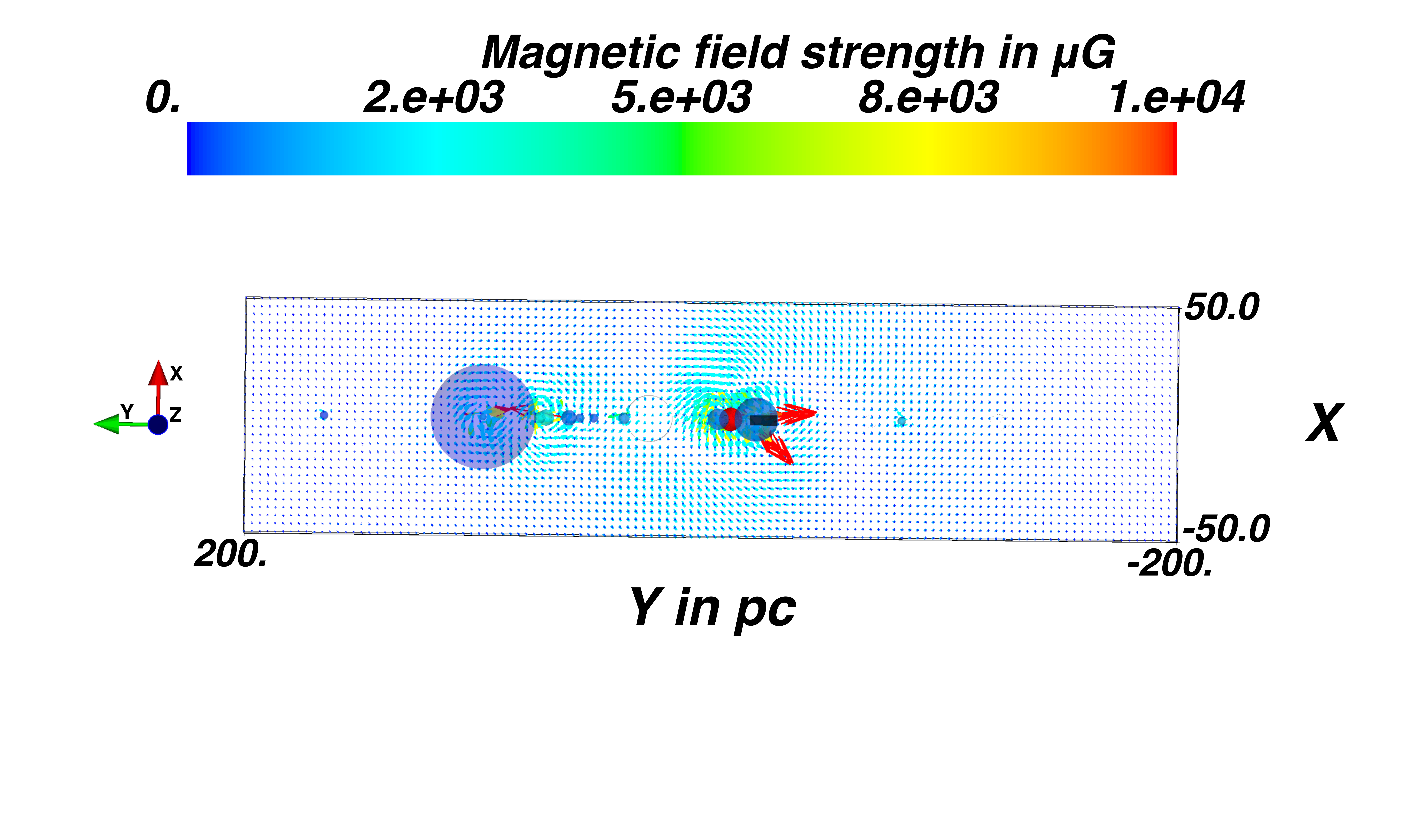}}
	\subfigure{\includegraphics[width=1.1\linewidth]{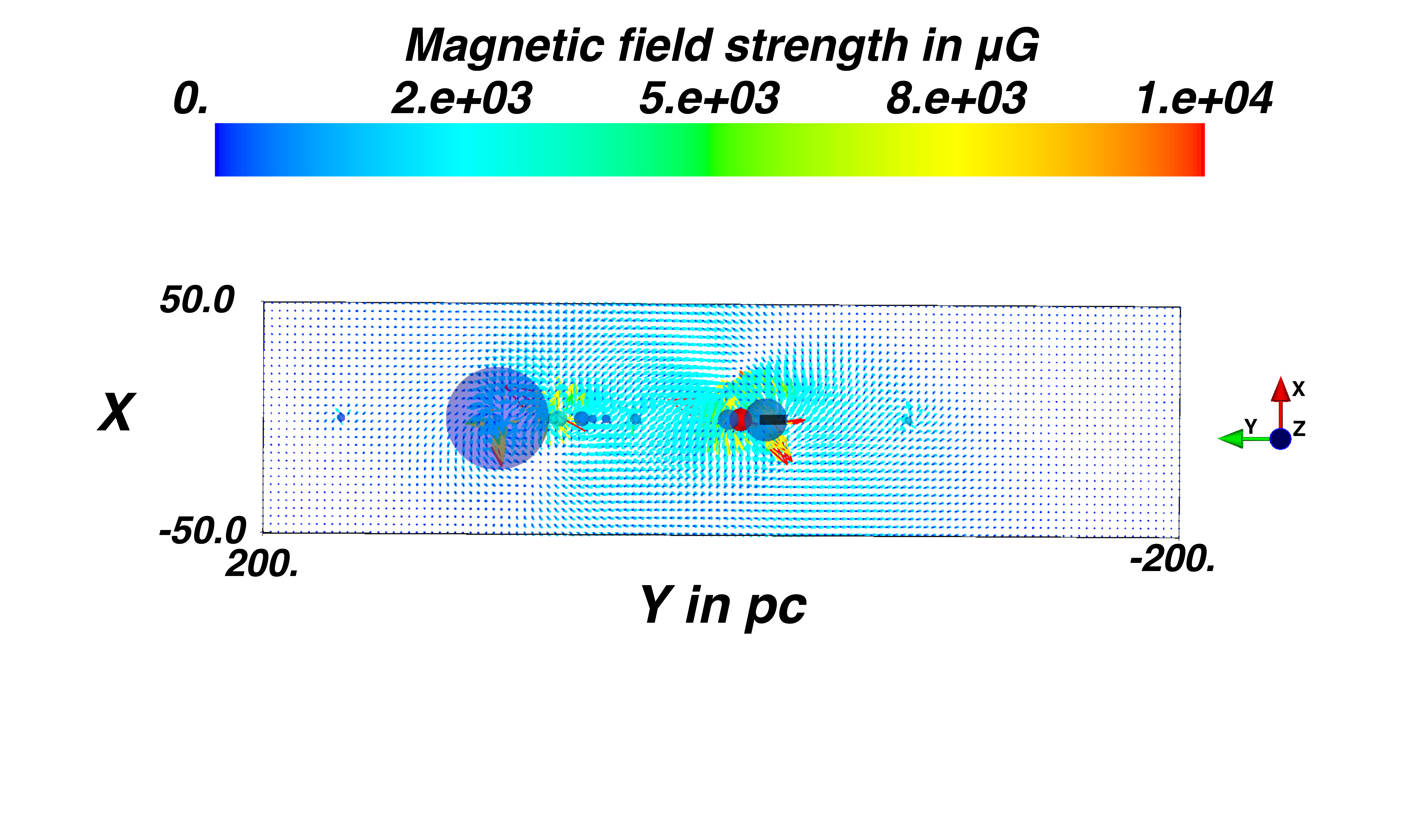}}
	\caption[]{Same as Fig. \ \ref{FieldTotal}, but with a view along the $z$-axis.}
	\label{FieldTotalZoom}
\end{figure}
\begin{figure}[H]
	\centering
	\hspace*{-0.3cm}
	\subfigure{\includegraphics[width=1.05\linewidth]{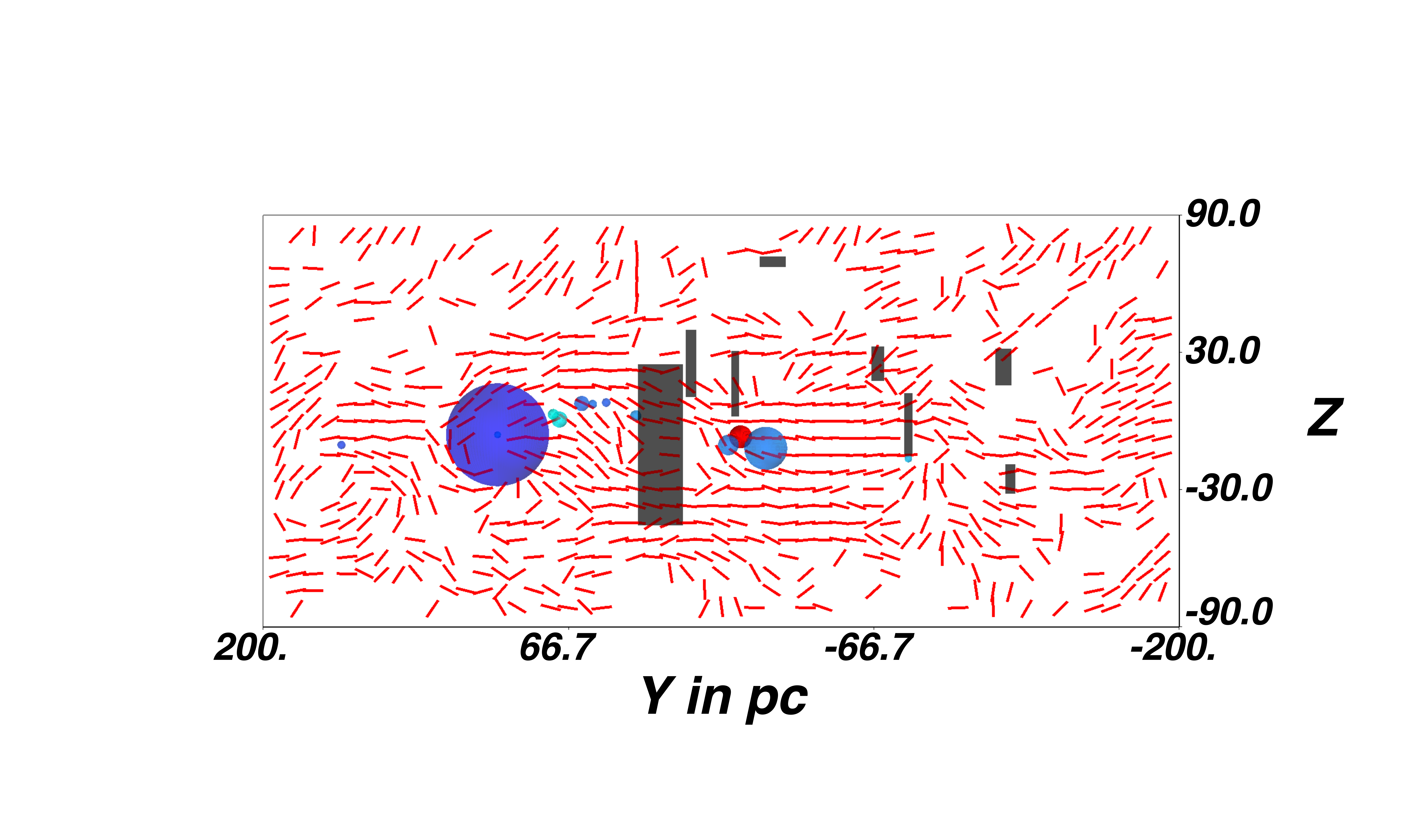}}
	\caption[]{Visualization of polarization measurement from \cite{Nishiyama2} together with the MCs and NTFs.}
	\label{Polarization}
\end{figure}

\newpage
\noindent In order to better recognize the influence of local objects such as the MCs on the field structure, Fig.\ \ref{FieldTotalZoom} provides the view Fig. \ref{FieldTotal} along the $z$-axis.\\
The polarization map of \cite{Nishiyama2} (see Fig.\ \ref{Polarization}) can be used to cross-check our results with measurements. While these specific structures in dense regions were observed in previous studies \citep{Chuss2003,Polarization2}, they are not visible in the recent measurements presented by \cite{Mangilli2019}, who reported a much smoother field configuration in dense regions in the CMZ.  Moreover, \cite{Mangilli2019} did not report any significant change in field orientation at the position of the prominent Radio Arc NTF, where a vertical orientation of the field dominates. These discrepancies further emphasize that this region is not fully understood. In this work, we present a first magnetic field configuration that can be tested against other data in the future.\par
\begin{figure}[H]
	\subfigure{\includegraphics[width=1.1\linewidth]{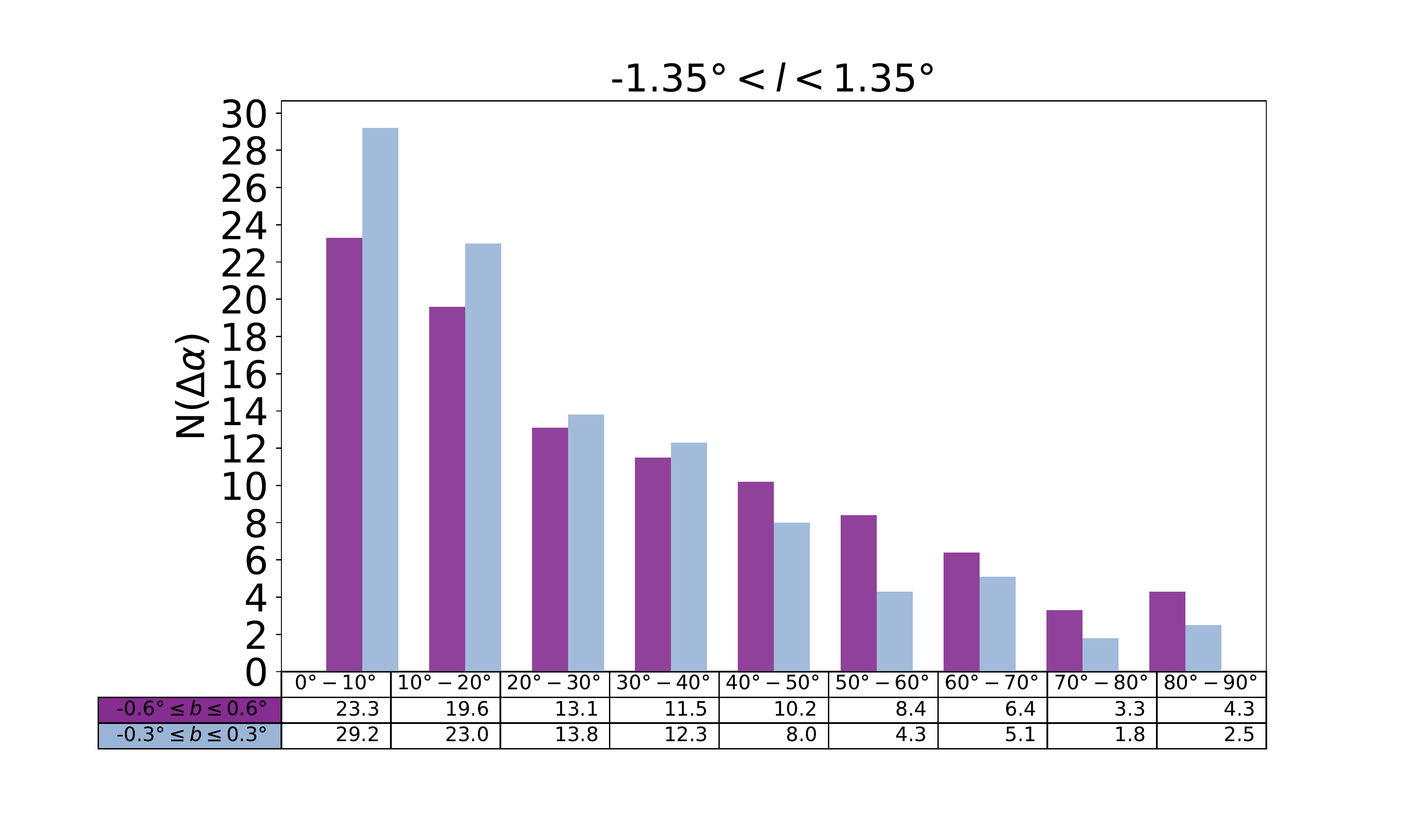}}
	\subfigure{\includegraphics[width=1.1\linewidth]{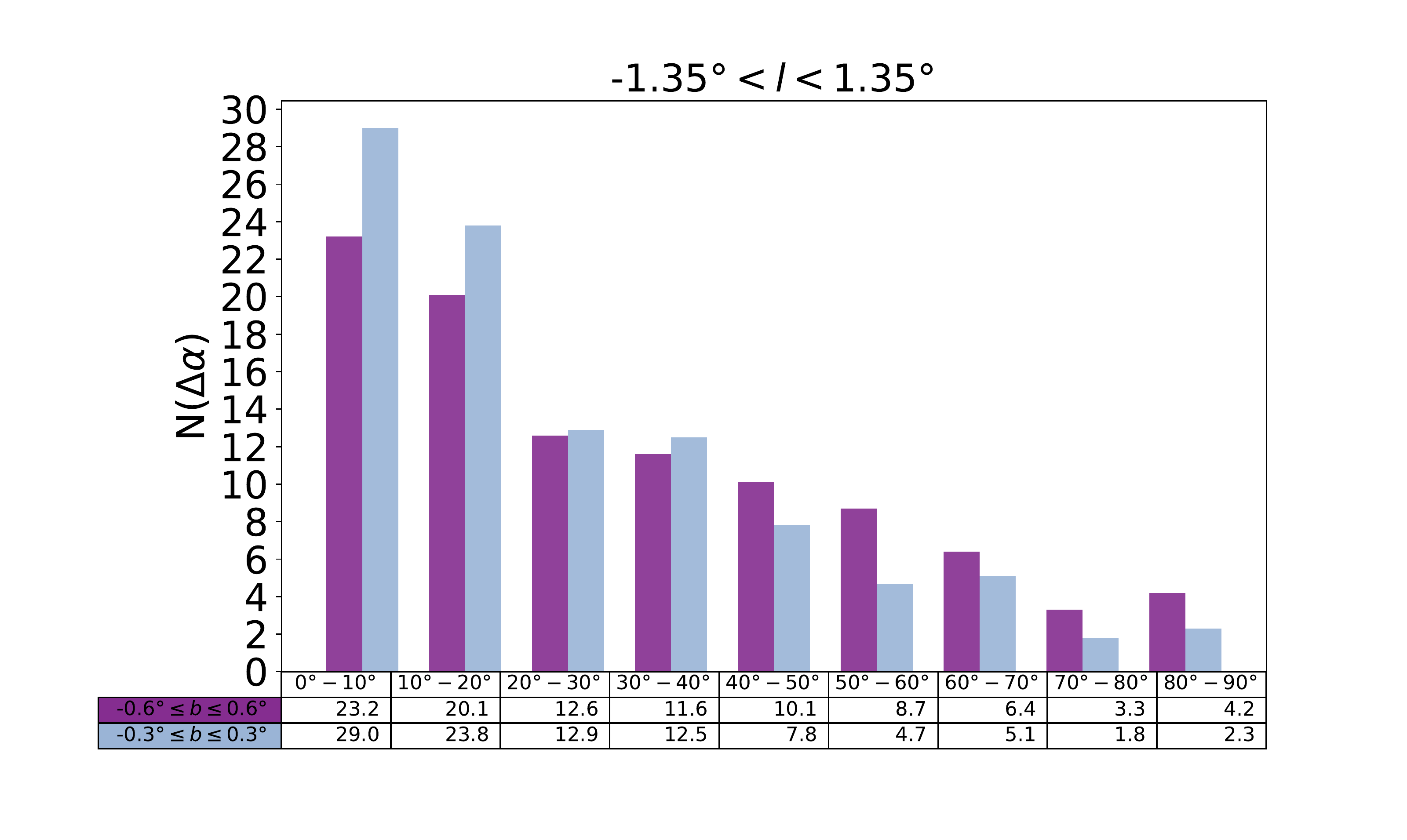}}
	\subfigure{\includegraphics[width=1.1\linewidth]{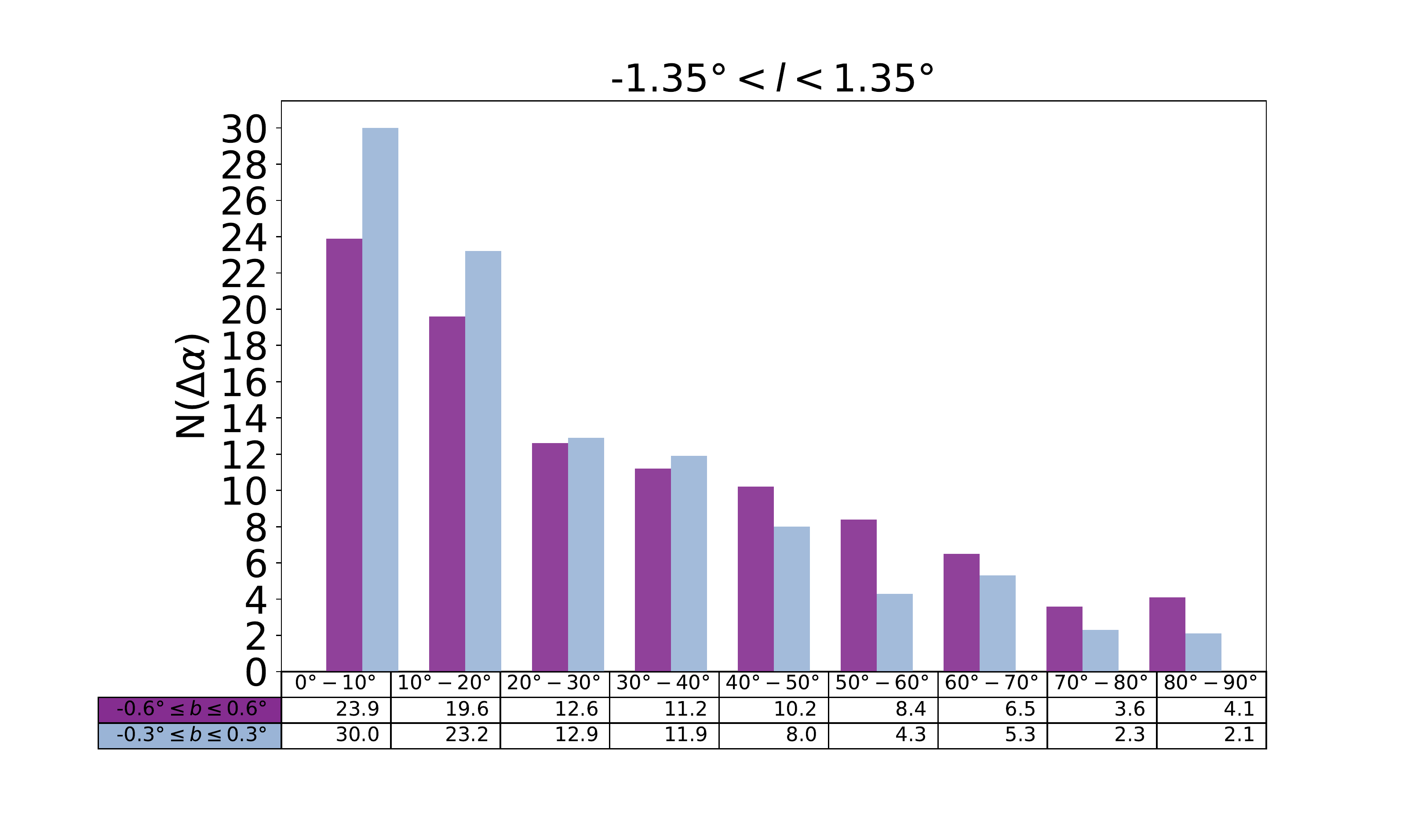}}
	\caption[]{
	Deviations of polarization vectors from the {GBFD20} field model, separated in $10 \degree$ steps extending from $0\degree$ to $90\degree$. The deviation is calculated for a regions extending from $-1.35\degree<l<1.35\degree$ in the longitude and for two different latitudes: $-0.6\degree \leq b\leq0.6\degree$ and  $-0.3\degree \leq b\leq0.3\degree$.
	Upper panel: $\eta=0$, middle panel: $\eta=0.5$ and lower panel: $\eta=1.0$}
	\label{Comparison}
\end{figure}

A qualitative comparison is reached by calculating the deviation of the measured polarization angle $\alpha_{\rm Nishiyama}$ from the expected polarization angle $\alpha_{\rm CMZF}$ from our model following $\Delta \alpha=|\alpha_{\rm Nishiyama}-\alpha_{\rm CMZF}|$. In Fig.\  \ref{Comparison}, $N(\Delta\alpha)$ represents the relative amount of the deviation $\Delta \alpha$ which occurs within a specific range ($0\degree -10\degree$, $10\degree -20\degree$,... , $80\degree -90\degree$) with respect to the total number of polarization measurements. It should be noted that the field orientation from our model is deduced considering the line of sight with a depth of 400\,pc. 
Thus, we do not present the expected polarization from our model, but only the expected field orientation, which makes this study sensitive to the polarization angle only. This comparison should therefore only be considered as a first, qualitative comparison to the data. 

\begin{figure}[H]
	\subfigure{\includegraphics[width=1.1\linewidth]{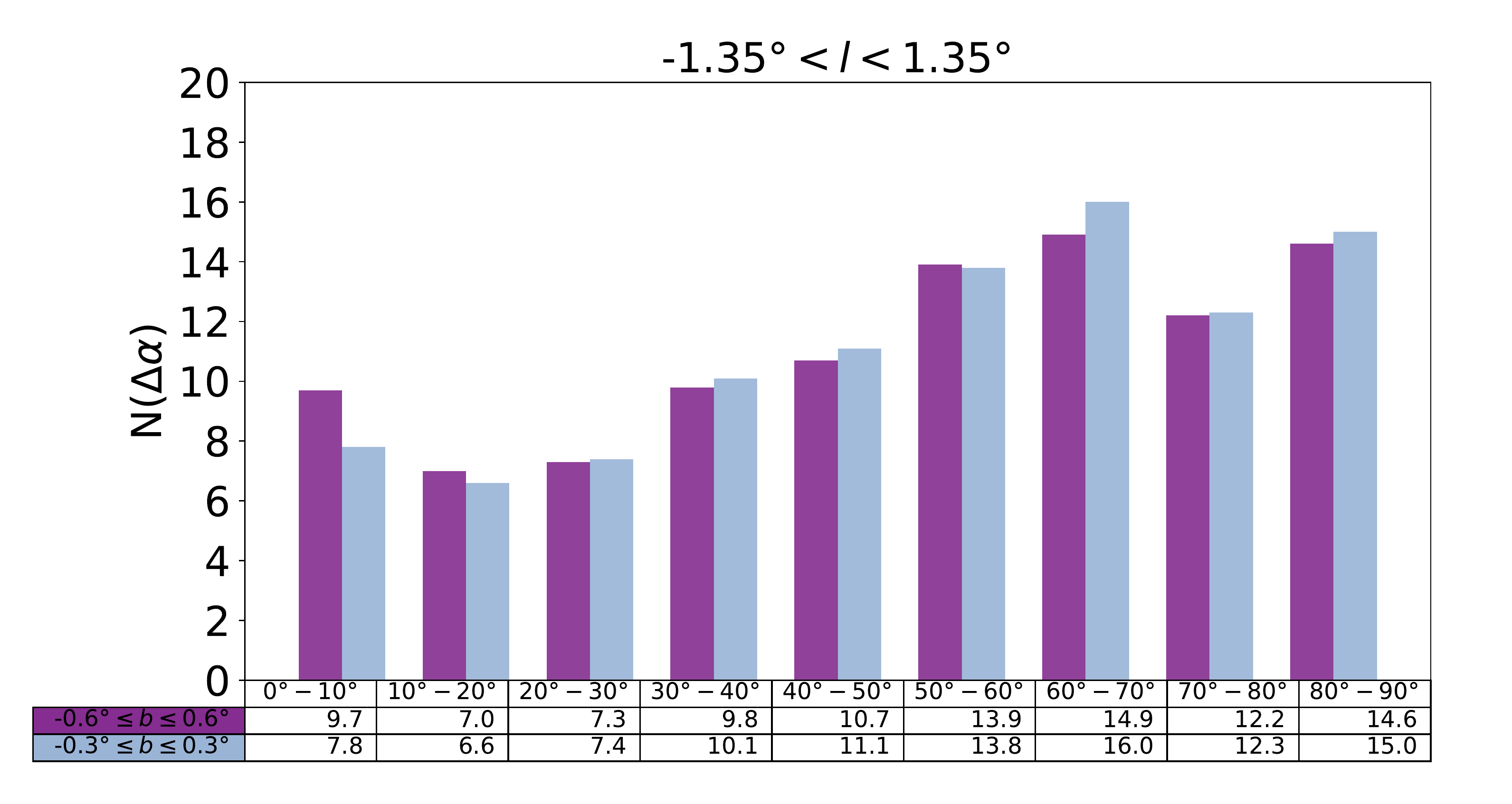}}
	\caption[]{
	Deviation of the \textit{JF12} polarization vectors from the polarization map, separated in $10 \degree$ steps extending from $0\degree$ to $90\degree$. The deviation is calculated for a regions extending from $-1.35\degree<l<1.35\degree$ in the longitude and for two different latitudes: $-0.6\degree \leq b\leq0.6\degree$ and  $-0.3\degree \leq b\leq0.3\degree$.}
	\label{ComparisonJF}
\end{figure}
\begin{figure}[H]
	\vspace*{-0cm}
	\hspace*{-0.5cm}
	\subfigure{\includegraphics[width=1.05\linewidth]{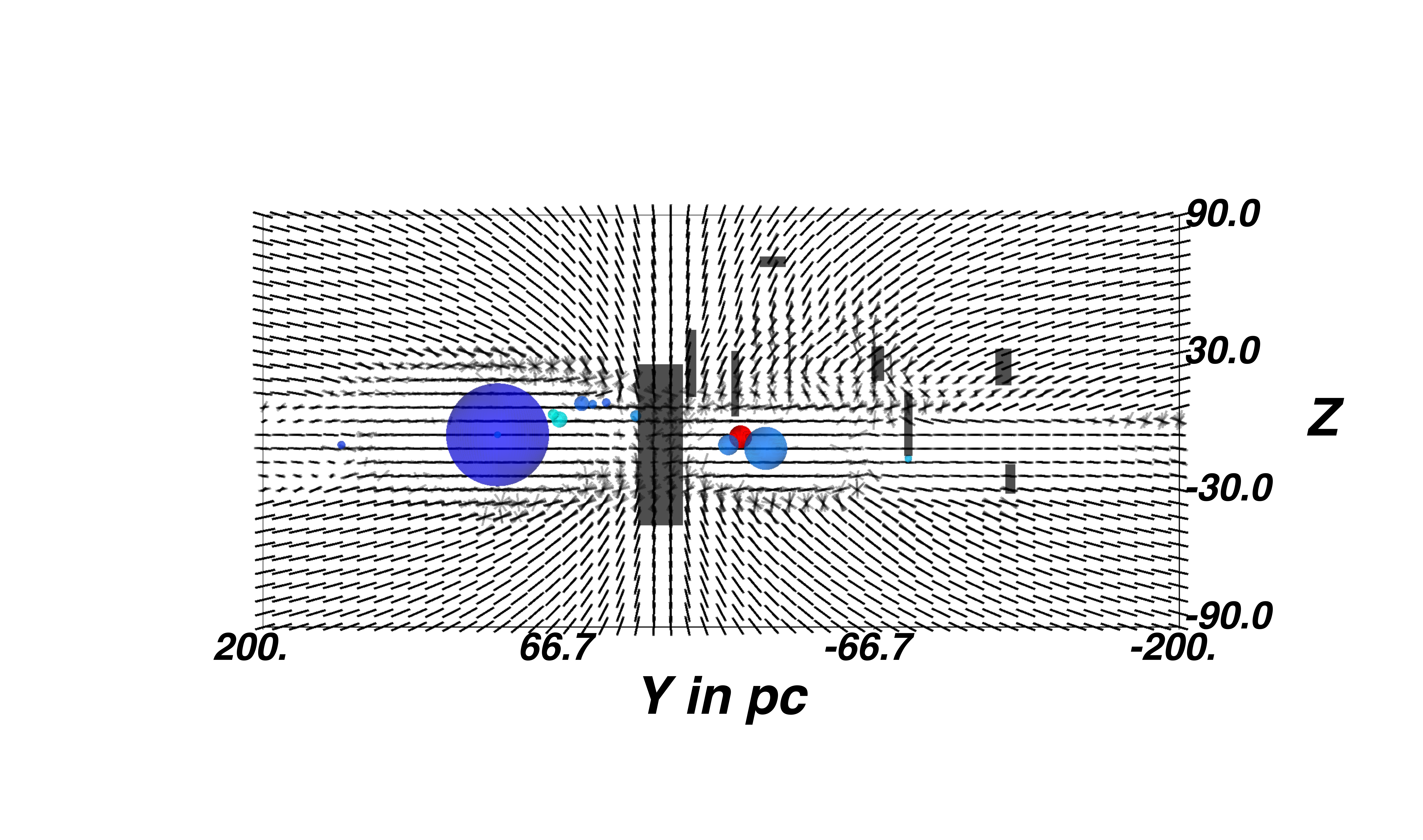}}
	\caption[]{Magnetic field configuration as derived in this work visualized together with the locations of the MCs and NTFs.}
	\label{Polarization_Model}
\end{figure}

The same procedure is also done for the {JF12} extrapolation in the GC region and the result are visualized by Fig. \ref{ComparisonJF}.
\begin{figure*}[htbp]
	\hspace*{0.5cm}
	\includegraphics[width=1.0\textwidth]{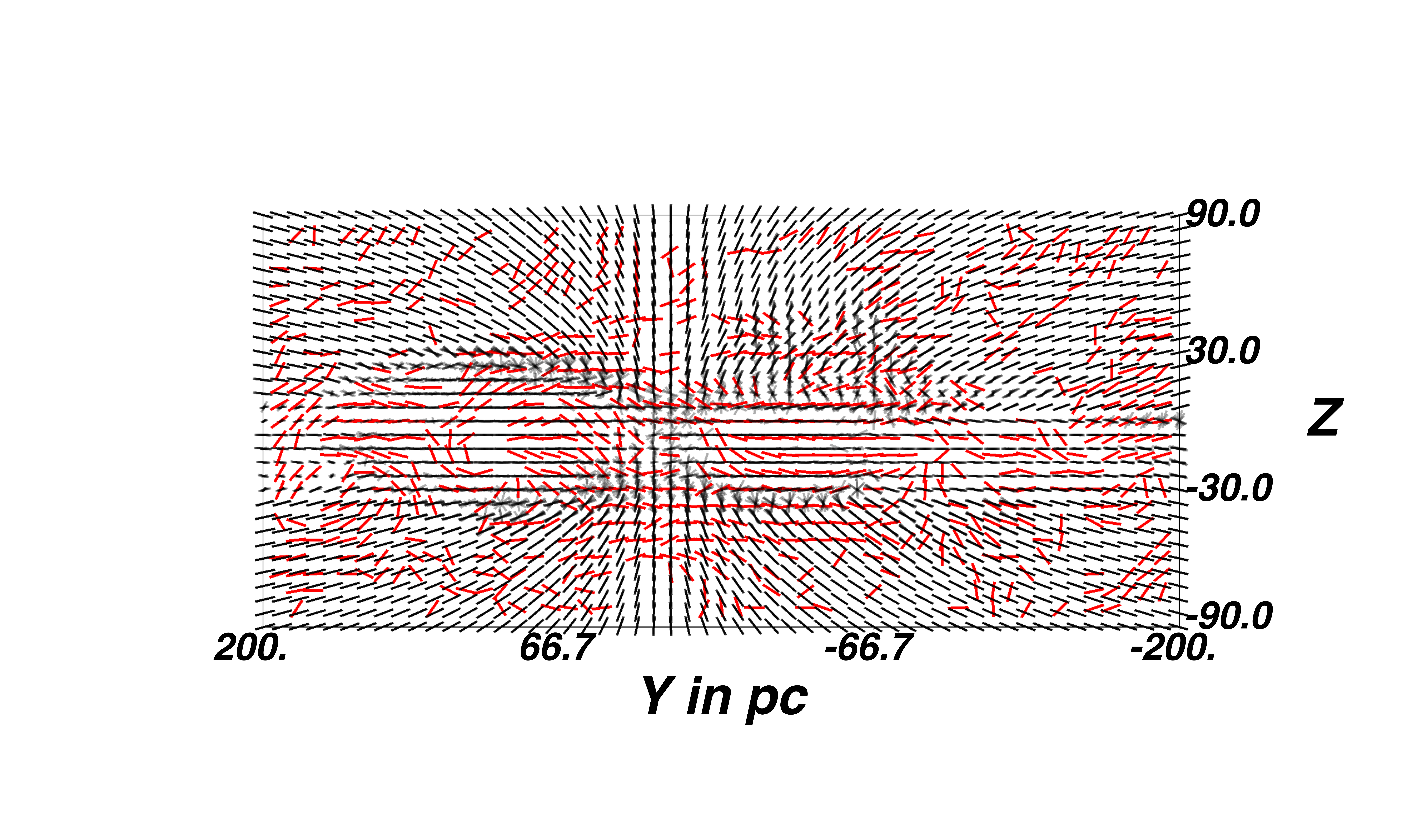}
	\caption[]{
	Magnetic field configuration as derived in this work is visualized by black dashes lines. The polarization as measured by \cite{Nishiyama2} is shown as the red colored dashed lines.}
	\label{PolarizationMass}
\end{figure*}
What can be learned from this quantitative comparison is that the model developed here follows the polarization map to a better degree (deviations from the measured polarization direction are generally smaller here than for the JF12 field). Also, the choice of $\eta$ has no significant influence on the results as can be seen from Fig.\ \ref{Comparison}. This indicates that the field model presented here can provide a better representation of the large-scale field than typical global magnetic field models.
Figure \ref{Polarization_Model} illustrates the field configuration from this work. On this map, all configurations along the line of sight with a depth of 400\,pc were added together. Therefore, darker lines represent more frequently appearing configurations along the line of sight.

At first glance, the field configuration measured by \cite{Nishiyama2} seems to be chaotically distributed and some magnetic structures are not clearly attributable to their physical origin.
For a better identification, Fig. \ref{Polarization} visualizes the MCs and NTFs additionally.
Figure \ref{PolarizationMass} simplifies the comparison for the reader by an overlay of Fig.s \ref{Polarization_Model} and \ref{Polarization}.
\noindent The potential of the model investigated in this work can be seen in the detailed reproduction of the configuration. Many of the field lines seem to be chaotically oriented on first glance, but in fact, even these are caused by some specific components in the CMZ or a superposition of them. 
A good example is the measured field line configuration at $l=0.6-0.7 \deg$ and $b=-0.1 \deg$ which is eye-catching due to the circulated configuration. This feature, in reality, is caused by a superposition of Sgr B2's and ICM's magnetic field. Many other feature can be explained in the same way.\\
The development of some discrepancies at higher longitudes, for example, in the corners, can have different reasons:
\begin{itemize}
	\item GBFD20 is mainly based on the mass distribution which decreases at higher longitude and is therefore in these regions subdominant as the CMZ region ends there
	\item 
	the dynamical effect of the MCs varies due to their orbit within the CMZ \citep{CMZMC} and  can lead to discrepancies
\end{itemize}
However, the most relevant structures in the region $|l|<1.35\degree$ and $|b|<0.6\degree$ are well described by the model derived in this work and thus indicate that the GBFD20 model is a good representation for the field in the GC.

\section{Impact on cosmic-ray propagation - a first comparison}
\label{Application}

In this section, the impact of the magnetic field configuration on cosmic-ray propagation is tested by comparing the magnetic field model of the CMZ as derived in this paper with the standard field of {JF12} and the  cosmic-ray propagation presented therein. The {JF12} model has been developed in order to have a good description of the global Galactic field model. Updates of this original field that improve the divergence-freeness as well as the continuity of the field have been presented recently by \cite{Unger2019} and \cite{Kleimann2019}. It is expressly mentioned by the authors that the {JF12} model is not optimized for propagation in the region of the GC. Nevertheless, we want to present our results with the {JF12} field to quantify the influence of the field configurations. Secondly, the {JF12} field is often used for global cosmic-ray propagation and that includes the central region in which the source density is high. Some influence of the local field configuration on the large-scale cosmic-ray picture can therefore still be present.

Here we simulate the propagation of particles with the open-source tool CRPropa \citep{CRPropa2007,CRpropa2016,CRpropa2017}. CRPropa was written originally for the propagation of extragalactic particles in intergalactic magnetic fields \citep{CRPropa2007} and updated with a modern, modular structure in \cite{CRpropa2016}. In its original form, propagation is done in discrete steps by solving the equation of motion.  For low-energy particles in the Galacitc magnetic field, that is for example, below energies of $10^{16}-10^{17}$\,eV, this method is inapplicable due to the highly diffusive nature and connected to that computationally long propagation times of the particles. Therefore,  the tool has been extended by adding an alternative propagation scheme, in which the transport equation is solved via the method of stochastic differential equations (SDEs) \citep{CRpropa2017}. The latter method works with pseudo-particles and is therefore well compatible with the original CRPropa framework. This software is well suitable for our tests as the propagation environment can be adjusted due to the modular structure of the code.
 
In each of the simulations presented here, we use the multiparticle picture within the module {Diffusion SDE}, the solution of the transport equation:
\begin{equation}
    \frac{\partial n}{\partial t}=\nabla (\hat{D}\nabla n)+S(\vec{r},\,t)
\end{equation}
with $S(\vec{r},\,t)$ as the source term. In general, the code works with the nonstationary solution of the equation  $\partial n/\partial t$. In this test scenario, we inject particles at the position of SgrA* and propagate them until they leave the Galactic Center region, defined by a box of $(\Delta X\times\Delta Y\times\Delta Z)=(234\times234\times234)$\,pc$^3$. We then produce the figures shown below by adding up the particle density for different time steps with weights as defined in \cite{CRpropa2017} in order to determine the stationary solution.
The code is able to handle arbitrary diffusion tensors $\hat{D}$. Here, the standard diffusion coefficient used in this formalism is 
\begin{equation}
D_{\parallel}\simeq6.1\cdot10^{28} \left(\frac{R}{4\,\mathrm{GV}}\right)^{1/3} \ \mathrm{cm^{2} \ s^{-1}}\, ,
\label{eq:DiffusionCoefficient}
\end{equation} 
where $R$ denotes the rigidity. The perpendicular component of the diffusion coefficient is then determined by assuming a fixed ratio 
$\epsilon=D_{\perp}/D_{\parallel}=0.1$. The transport equation is solved in the orthonormal bases corresponding to the background magnetic field such that the parallel diffusion path follows the magnetic field lines. 

\begin{figure}[H]
	\centering
	\hspace*{-0.cm}
	\subfigure{\includegraphics[width=0.9\linewidth]{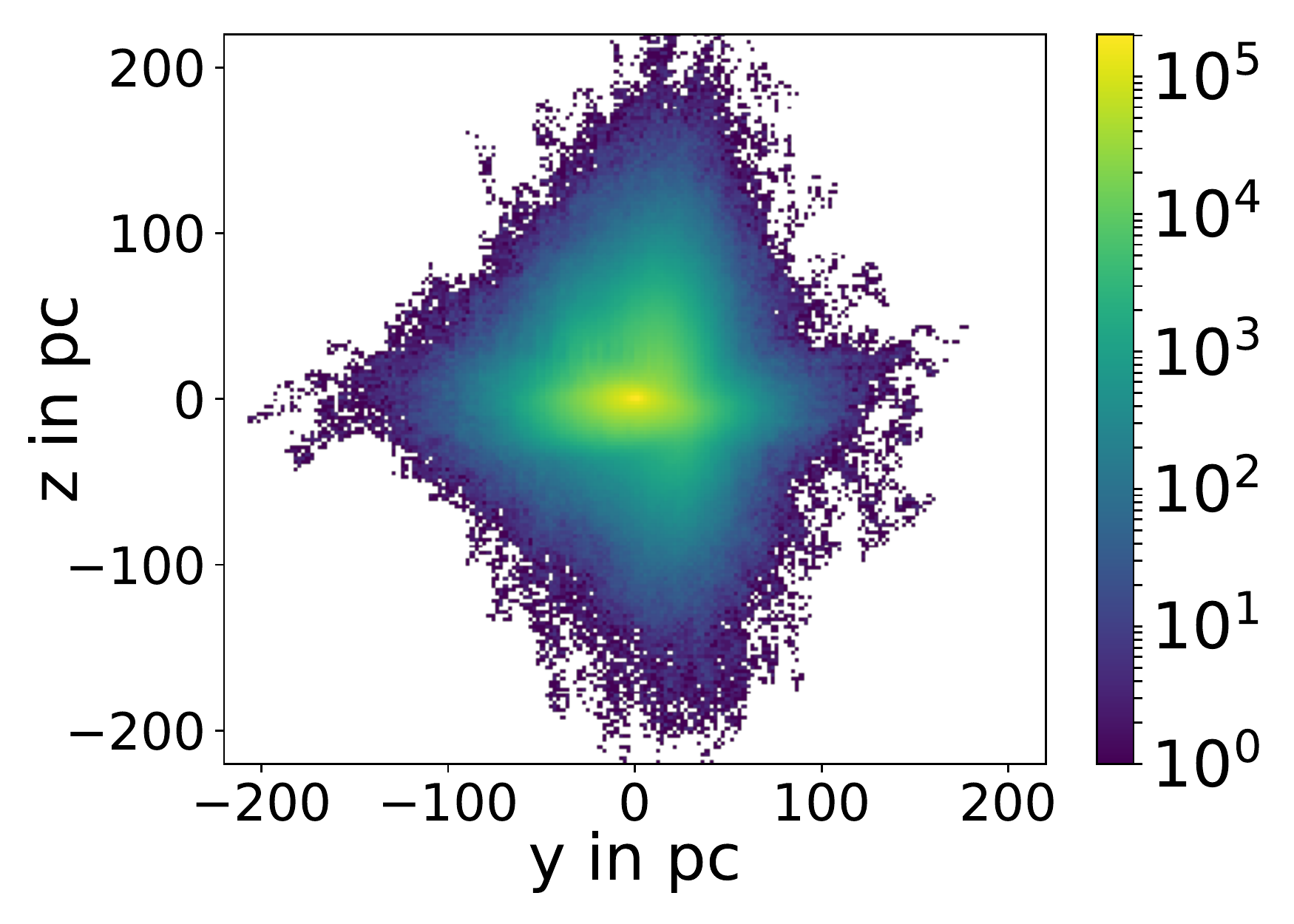}}
	\subfigure{\includegraphics[width=0.9\linewidth]{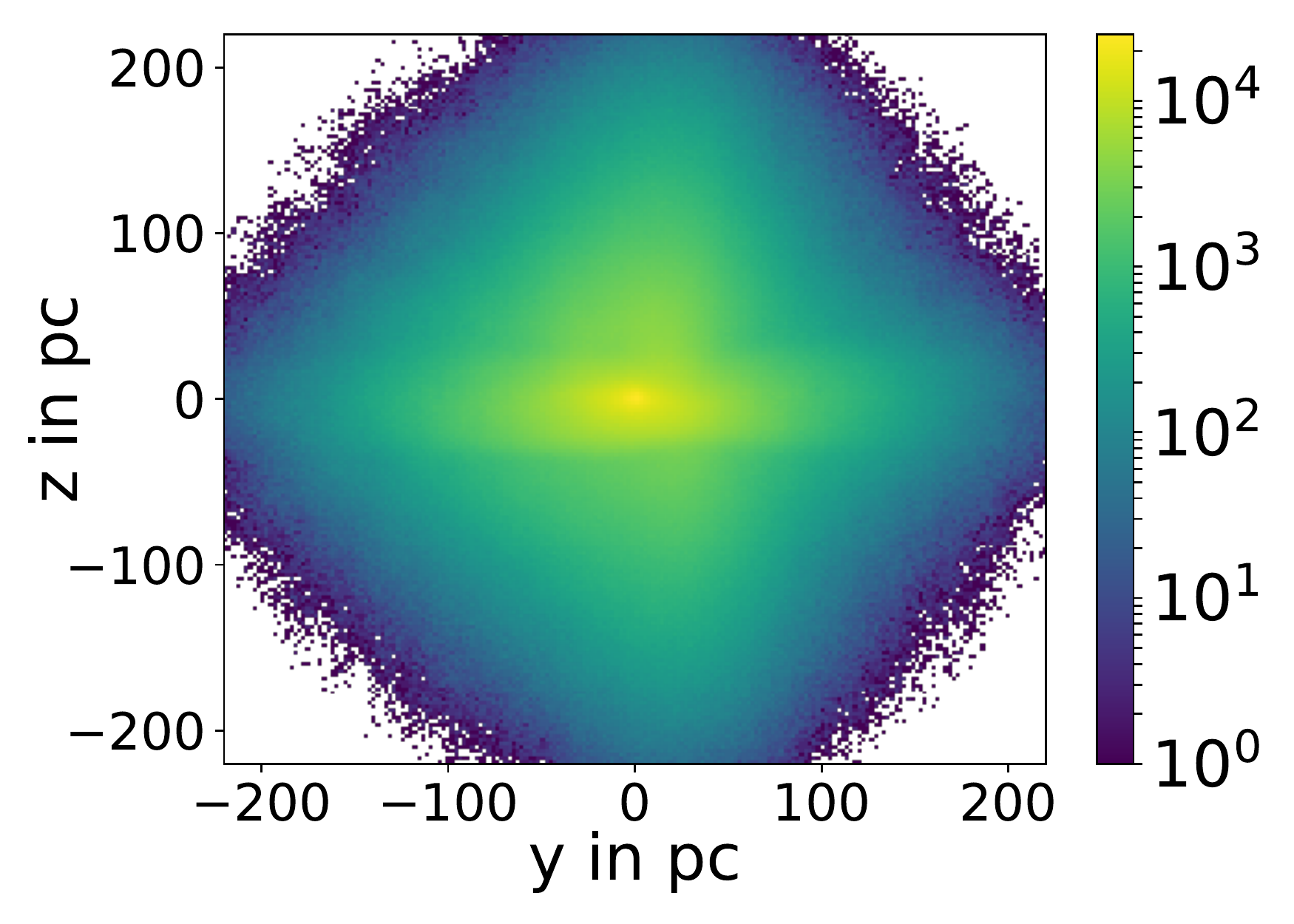}}
	\caption[]{Density of protons after propagation in the {GBFD20} field. Upper panel: $\eta=0$ with proton energies between $1-1000$\,TeV; Lower panel: $\eta=0$ with proton energies between $1-100$\,PeV.}
	\label{Traj}
\end{figure}

\begin{figure}[H]
	\centering
	\hspace*{-0.0cm}
	\subfigure{\includegraphics[width=0.9\linewidth]{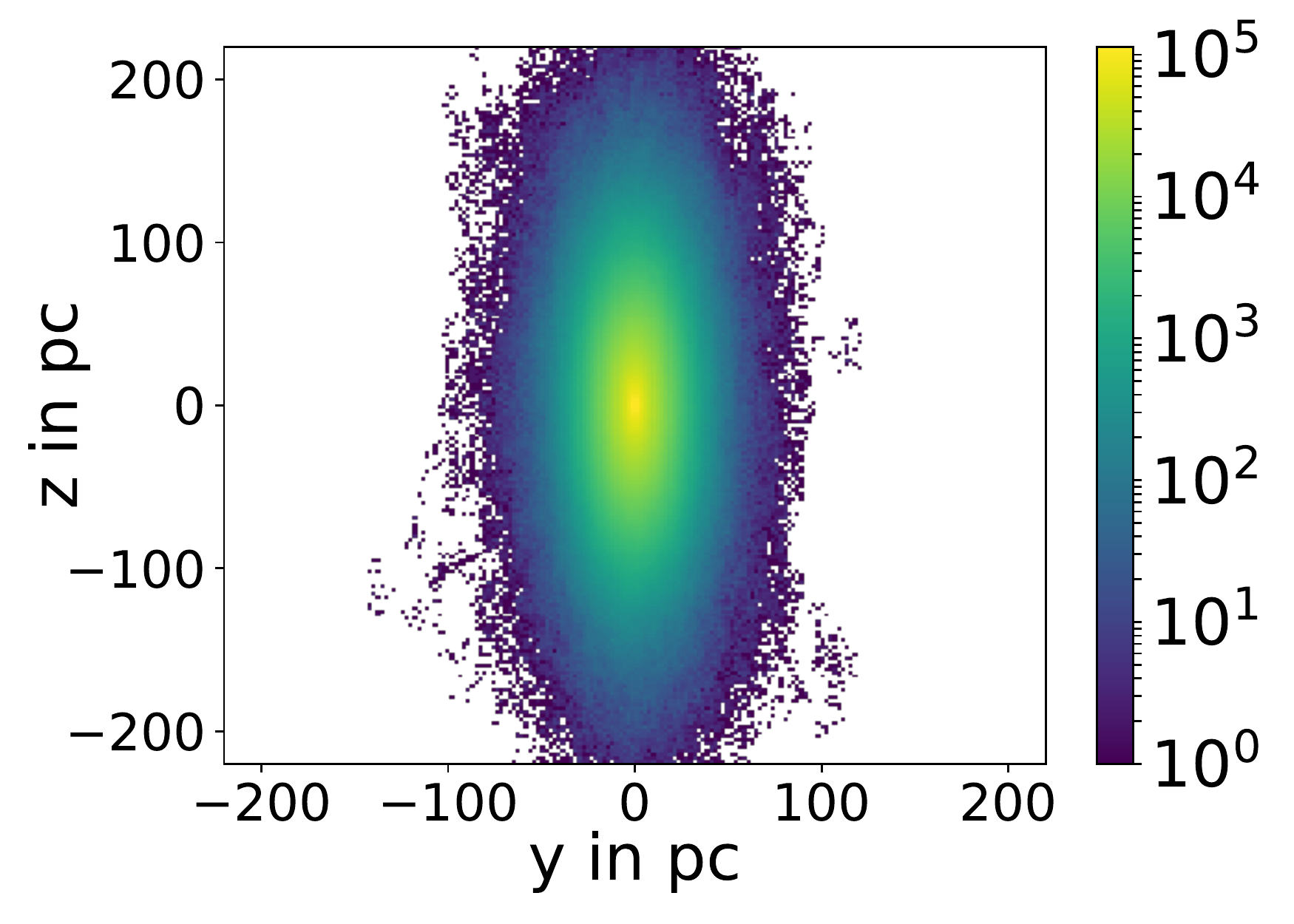}}
	\subfigure{\includegraphics[width=0.9\linewidth]{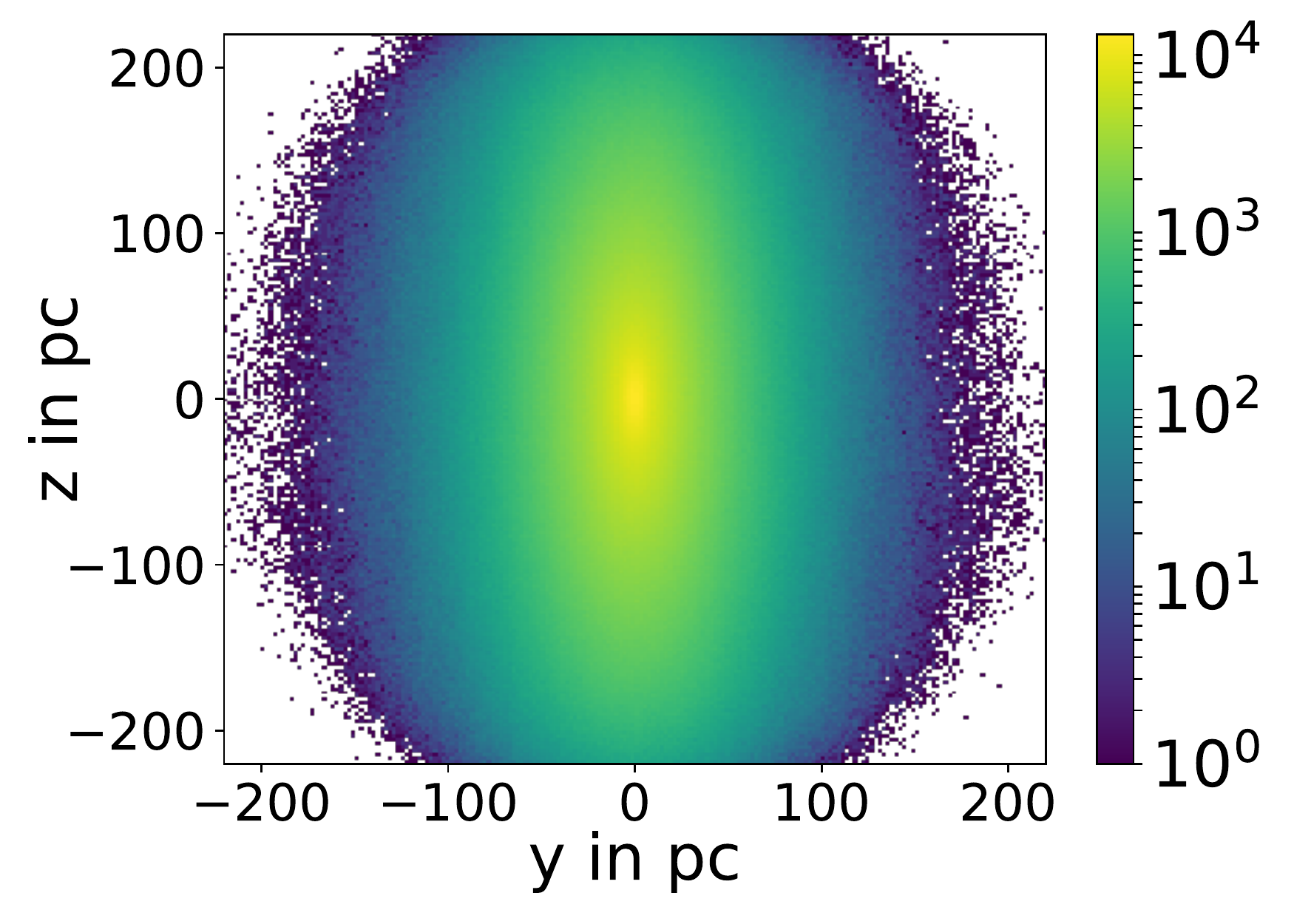}}
	\caption[]{Same as Fig.\ \ref{Traj}, but using the \textit{JF12} field in the  energy range $1-1000$\,TeV (upper panel) and $1-100$\,PeV (lower panel) }
	\label{TrajJF}
\end{figure}

\begin{figure}[H]
	\centering
	\subfigure{\includegraphics[width=1.1\linewidth]{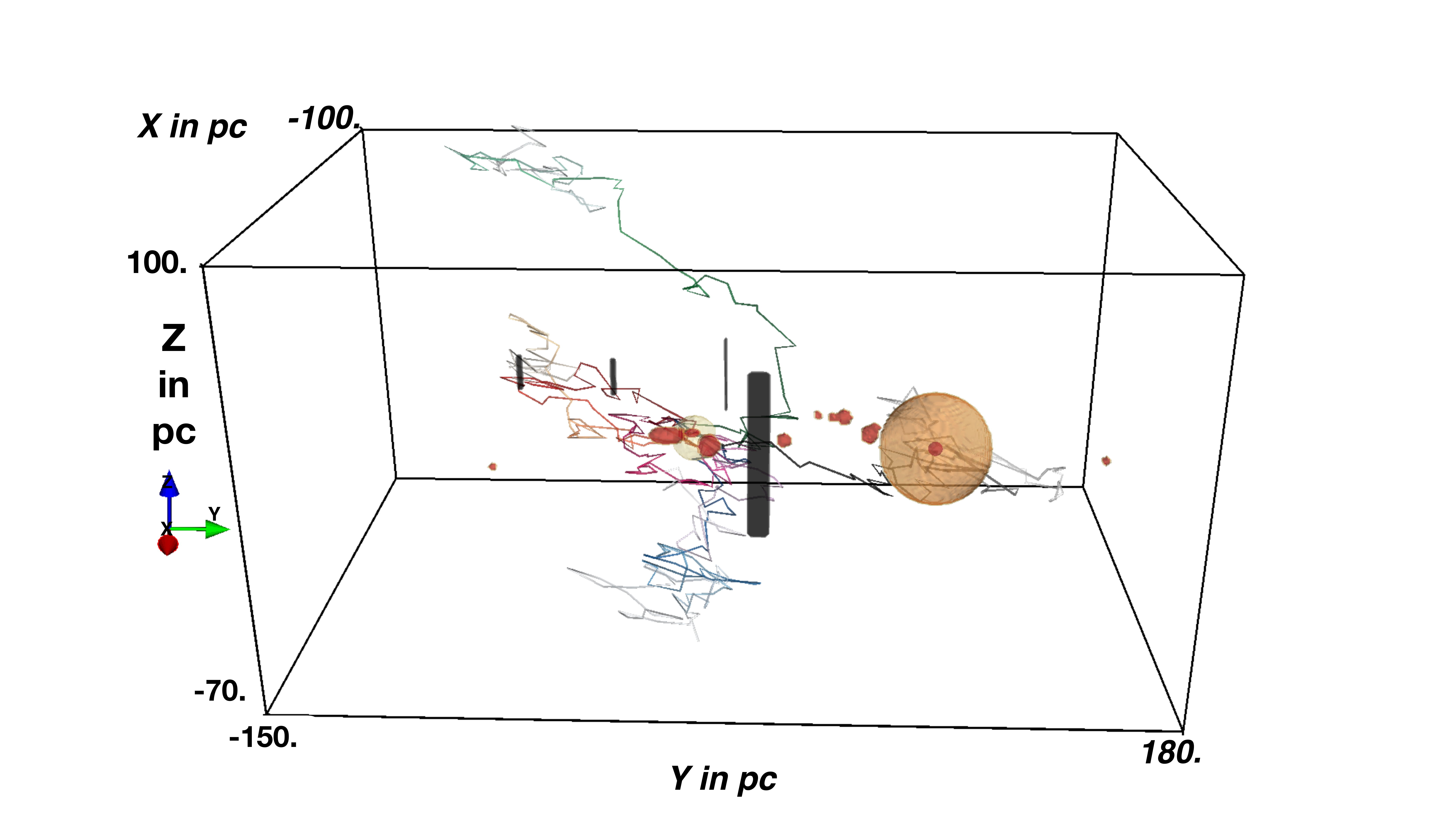}}
	\subfigure{\includegraphics[width=1.1\linewidth]{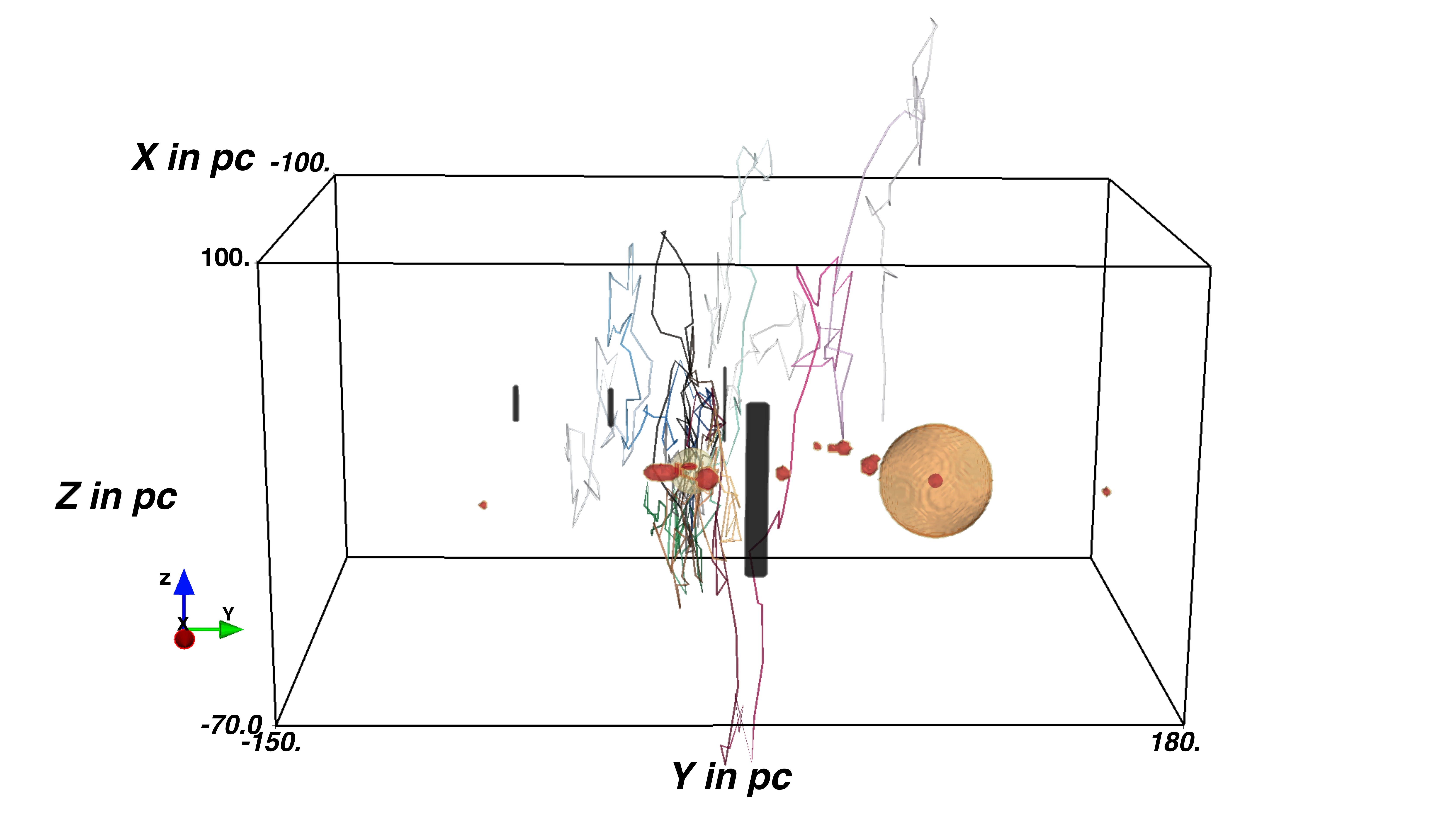}}
	\hspace*{-1.4cm}
	\subfigure{\includegraphics[width=1.4\linewidth]{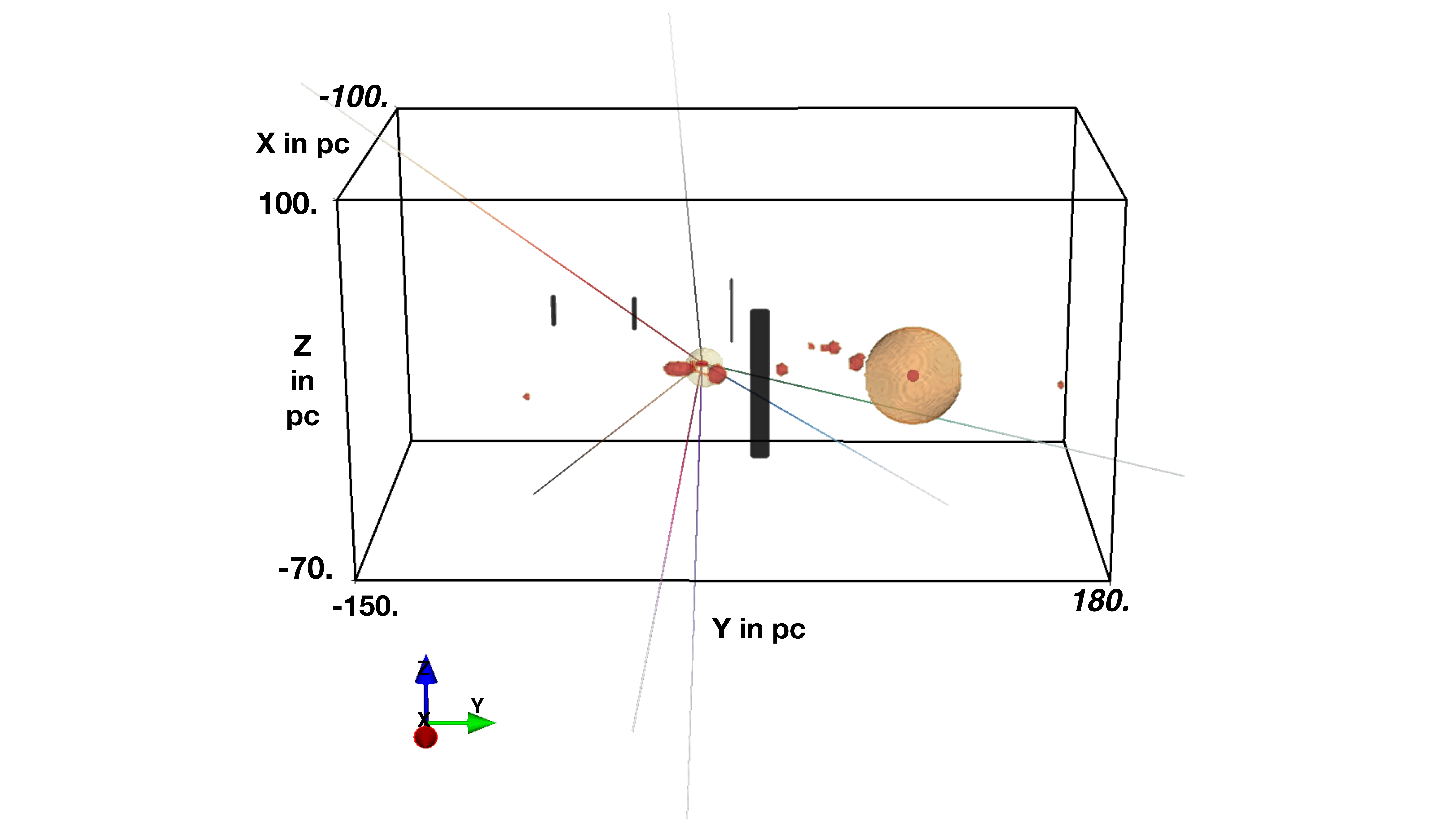}}
	\caption[]{Examples of seven proton trajectories at energies from 1\,TeV to 10\,PeV with a step length of 0.5\,pc: upper- {GBFD20} with $\eta=0.5$ (this work), middle- {JF12}, bottom (No field). The passed time is proportional to the color batch.}
	\label{Traj2}
\end{figure}

We perform test simulations for two different magnetic field models: (1) {GBFD20} [this work] and (2) {JF12}. Into this environment, we inject $10^5$ particles in the center of the simulation box, $(X,\,Y,\,Z)=(0,\,0,\,0)$. These particles are propagated in a discrete time step of $\Delta t_i=1$\,pc/c for 234 times. This way, 234 time steps exist and in each time step, the particles start at the position that they had reached the time step before, $\Delta t_{i-1}$. The maximum distance from the center that can be reached this way by a single particle corresponds to $234$\,pc, in case of a diffusive influence by the magnetic field, this distance can be significantly shorter.  The results are shown in Fig.s \ref{Traj} ({GBFD20}) and \ref{TrajJF} ({JF12}). In these figures, the images for each time step ($i=1,\ldots 234$) are stacked.\\
\noindent The figures reveal that in the {JF12} field particles diffuse in an approximately cylindrically symmetric structure, while this work ({GBFD20}) leads to an approximately dagger-shaped propagation pattern.
As an example, at a distance  $Y=\pm100$\,pc and $Z=0$\,pc and energies between $1-1000$\,TeV, the {JF12} configuration results in a flux level close to zero particles. The {GBFD20} with $\eta=0.5$ field, on the other hand, reaches a level of $10^4$ protons at the same energy and distance. This show that diffusion in the $Y-$direction is therefore much stronger in the field presented in this work. In the {JF12} model, the particles rather propagate along the $Z$-axis. 
This picture is further emphasized by Fig. \ref{Traj2}, in which random example trajectories are shown for the case of the {GBFD20} field (upper panel), the {JF12} field (middle panel) and no field at all (lowest panel). All particles start at the ($X,\,Y,\,Z)=(0,\,0,\,0)$. The colors of the trajectories go from dark (early) to light (late times).
\begin{figure}[H]
	\centering
	\hspace*{-0.cm}
	\subfigure{\includegraphics[width=0.9\linewidth]{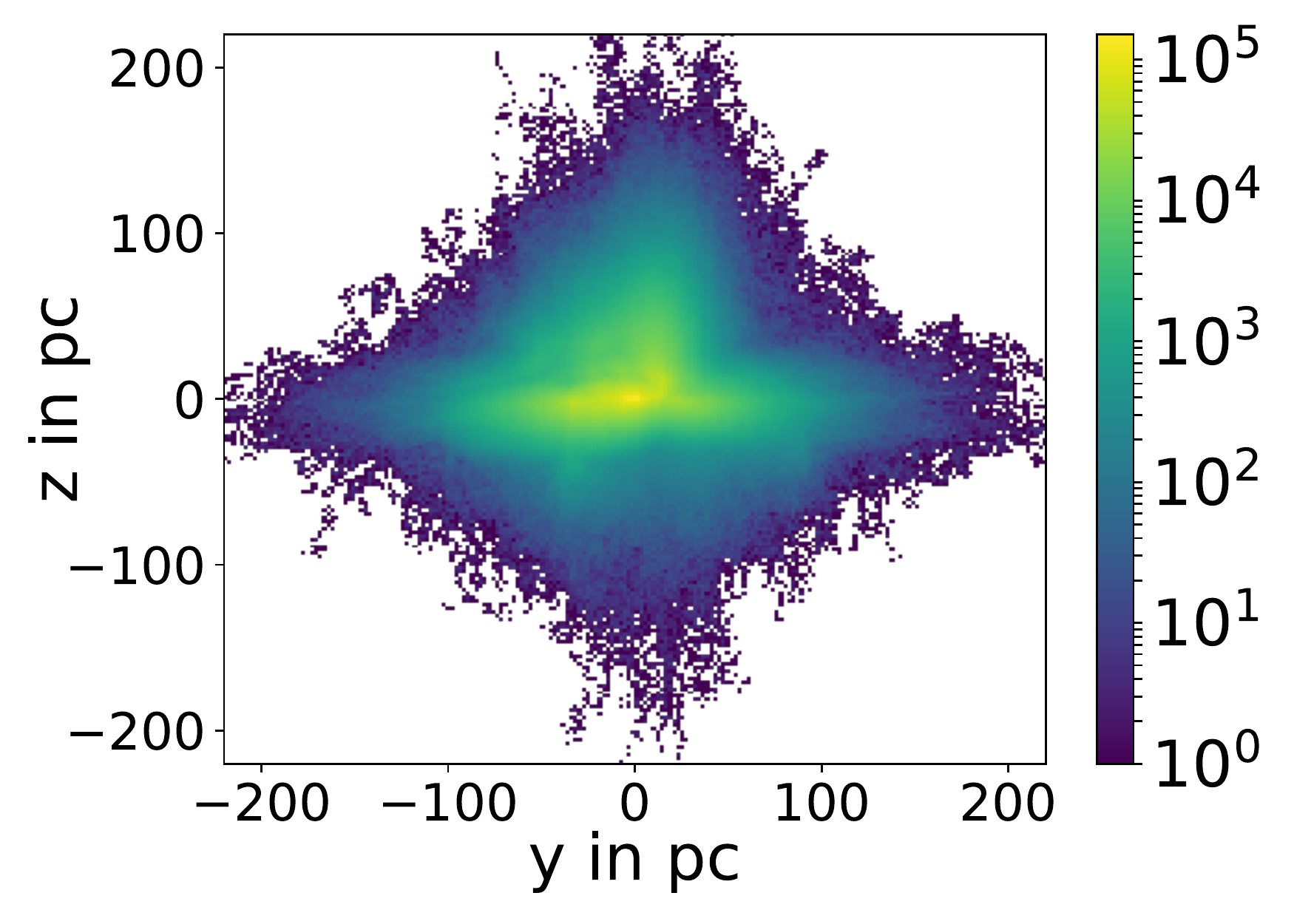}}
	\subfigure{\includegraphics[width=0.9\linewidth]{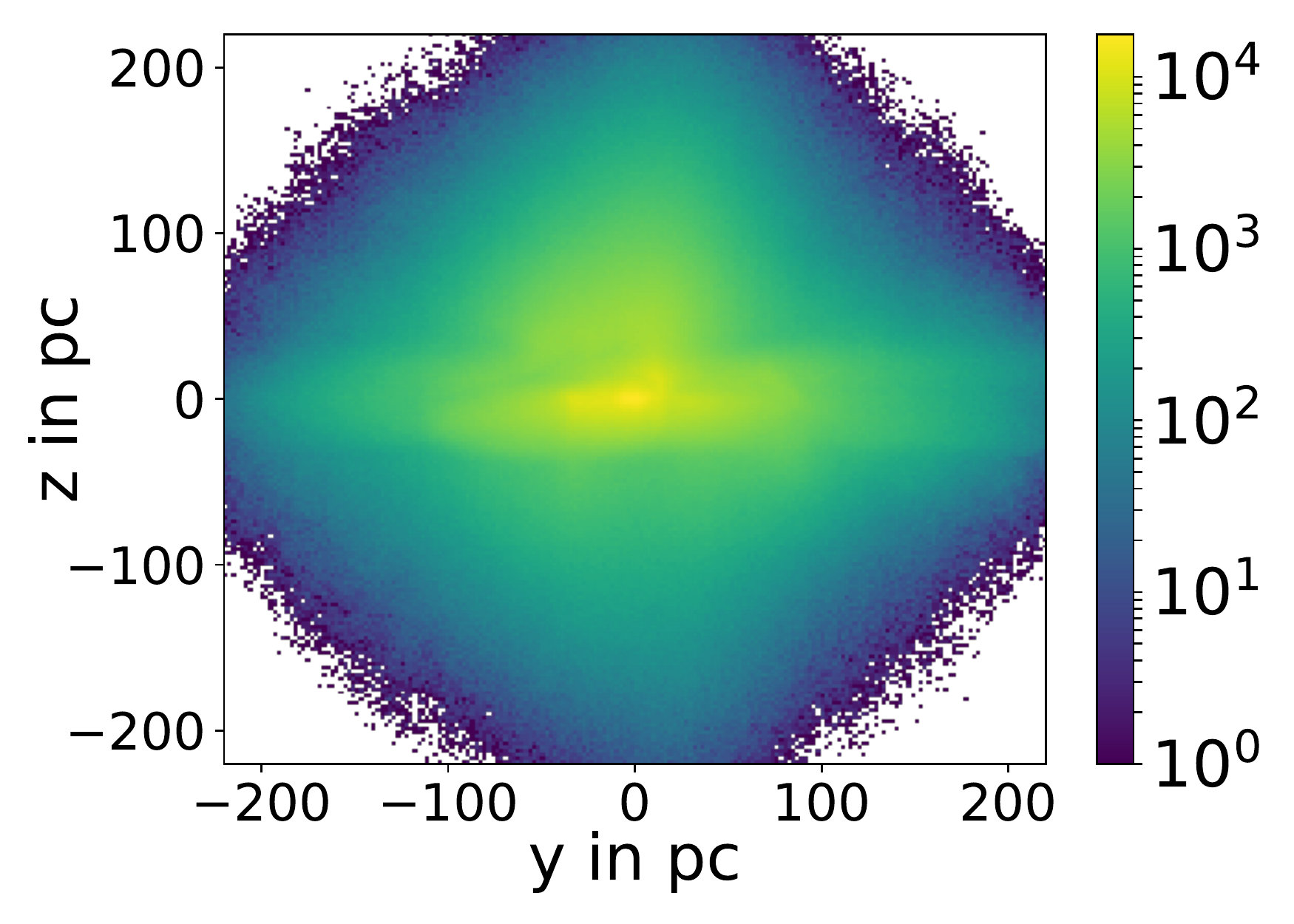}}
	\caption[]{Same as Fig.\ \ref{Traj}, but with $\eta=0.5$ (upper panel: energies between $1-1000$\,TeV and lower panel: energies between $1-100$\,PeV).}
	\label{Traj_2}
\end{figure}

\begin{figure}[H]
	\centering
	\hspace*{-0.cm}
	\subfigure{\includegraphics[width=0.9\linewidth]{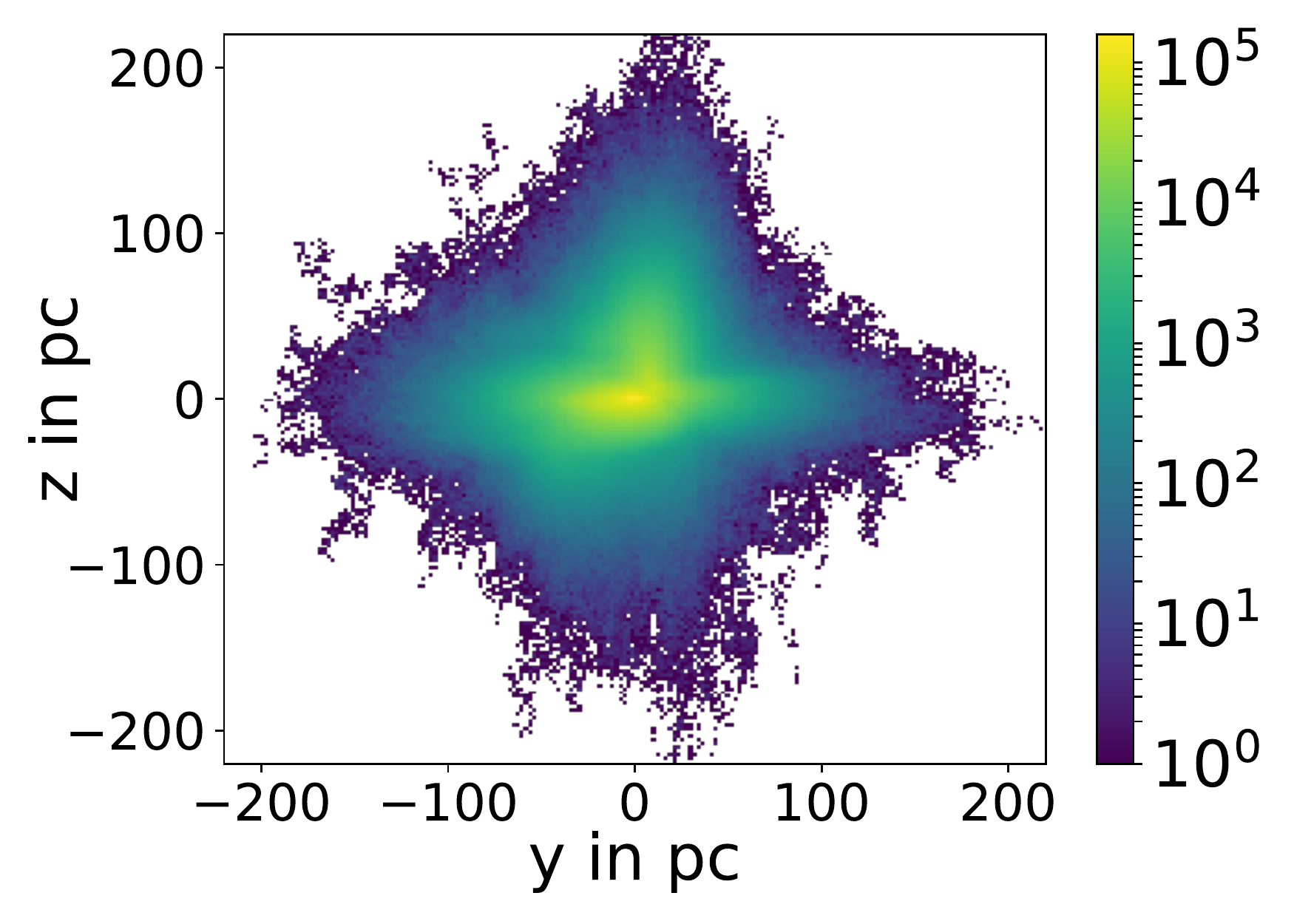}}
	\subfigure{\includegraphics[width=0.9\linewidth]{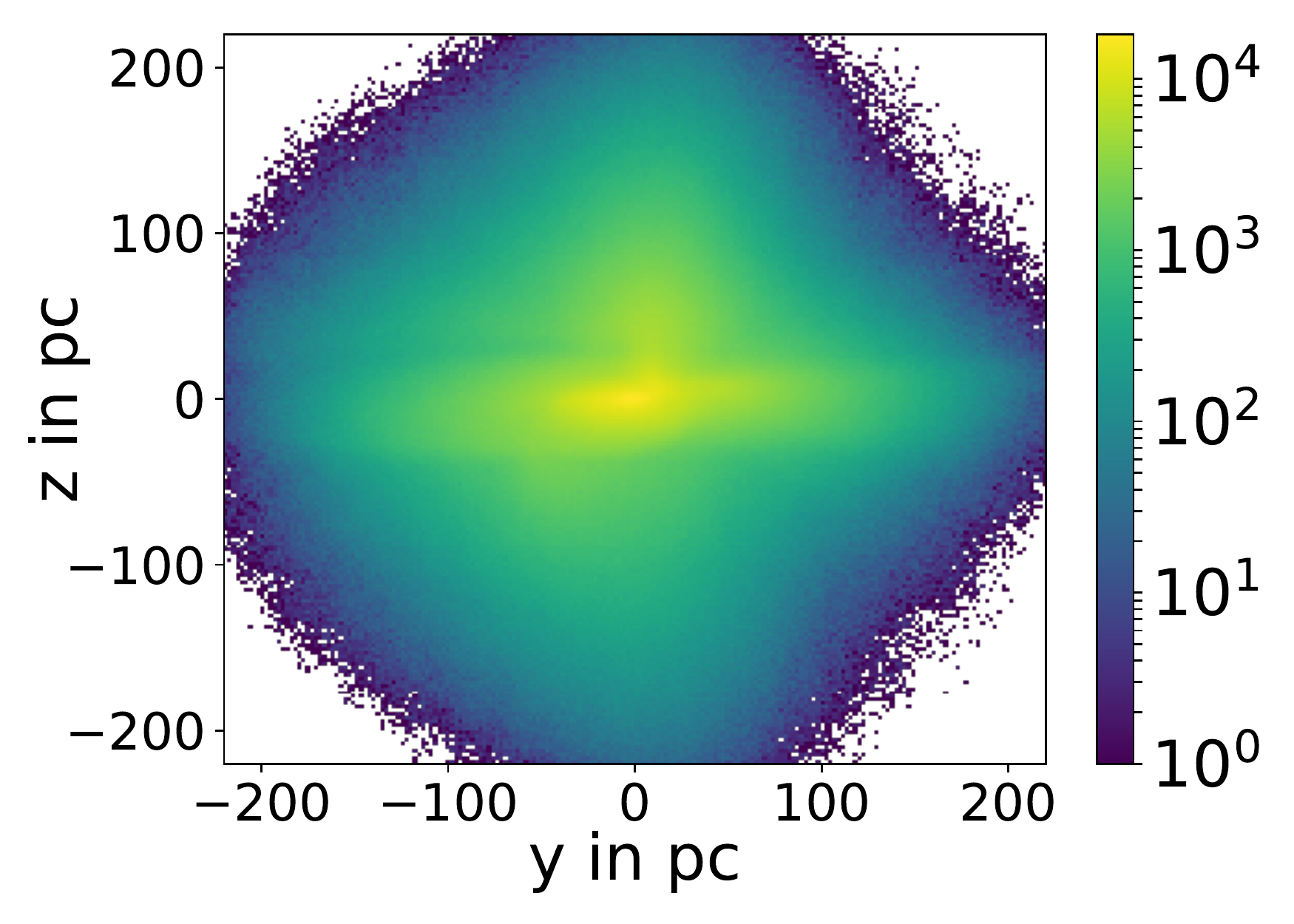}}
	\caption[]{Same as Fig.\ \ref{Traj}, but with $\eta=1.0$ (upper panel: energies between $1-1000$\,TeV; lower panel: energies between $1-100$\,PeV).}
	\label{Traj3}
\end{figure}

It is interesting to note that the diffuse $\gamma$-ray detection by H.E.S.S. shows a latitudinal extension up to 220\,pc \citep{HESS2018GC}, that is, approximately the same distance which CRs can reach within the {GBFD20} model with $\eta=0.5$ model. Further studies are required to show if this broad distribution could be due to the enhanced latitudinal propagation of cosmic rays.
\section{Discussion on the model parameters}
Our model is constructed with a certain parametric freedom that can be easily adapted in the future to new observational data. 
Here, we summarize all parameters related to the different magnetic field components together in modeling the ICM, NTFs and MCs. We also discuss the uncertainties in our results.

In general, the {poloidal magnetic field} component has four free parameters:
\begin{enumerate}
	\item $B_1$ is the normalization factor, which is determined by assuming the observed and accordingly, the expected field strength is an average value in the region of interest.
	\item $L$ the radial exponential scale length is determined by assuming the half-max value of $|B|$ at the maximum longitudinal extent of the region of interest.
	\item The parameter $a$ is governing the opening of field lines away from the z-axis.
	\item $m$ is the wavenumber which is set to zero for an axisymmetric description.
\end{enumerate}
The poloidal field is applied to the inter-cloud and non-thermal filament regions:
\begin{itemize}
    \item  For the {inter-cloud region}, the parameter $B_1$ is determined by the expected average field strength of $10$\,mG based on minimum-energy analysis of the non-thermal sources with an approximate uncertainty of a factor 10 \citep{LaRosa330MHz,FerriereMagneticField2009}. In the same context, the parameter $L$ is determined by assuming the half-max value of $B$ at the maximum radius $R_{IC}=L\cdot \ln{2}=158$\,pc of the CMZ following the CMZ mass distribution from \cite{MassGalacticCenter2}. Here, $R_{IC}$ is in the range of $100-250$\,pc \citep{MassGalacticCenter2}. The parameter $a$ is determined by a best-fit to the polarization data \citep{Nishiyama2}. Here, the error is quantified to less than 1\% \citep{Nishiyama}.
\item For the {NTF region}, $B_1$ is determined by the expected field
 strengths derived from equipartition. The error
 in the measurement of the flux is approx. 15\% \citep{LaRosaFilaments2001} which gives us an error of $\sim 5\%$ for the equipartition magnetic field.  Here, the systematic uncertainty might be more significant, in particular connected to the assumption of equipartition. The parameter $L$ is determined by assuming the half-max value of $B$ at the
 maximum radius $R_{NTF}=L\cdot \ln{2}$ of the NTFs following the geometry
 distribution observed by several groups. The error for $R_{NTF}$ is given by the resolution of $24''$, corresponding to approximately $0.5$\,pc \citep{LaRosaFilaments}.  The parameter $a$ is determined by a vertical Gaussian scale assuming the half-max value of $B$ at the vertical edges of each NTF. The vertical
 sizes are taken as $A = \sqrt{2}/V$, where V is the vertical size which is taken from \cite{LaRosaFilaments2001} with an error of $1.8$\,pc (a resolution of $43''$).
\end{itemize}

The {horizontal magnetic field} component has five free parameters:
\begin{enumerate}
	\item $B_1$ is the normalization factor which is determined by assuming the observed and accordingly, the expected field strength is the average value in the system under consideration.
	\item $\rho$ the reference radius at which the field line is crossing the reference angle $\psi$. In our model for the MCs, this parameter is set to the radius of the cloud.
	\item $H_c$ vertical scale length which is determined by assuming the half-max value of $|B|$ at the edges of the MC.
	\item $m$ the wavenumber which is set to one for ensuring the solenoidal property.
	\item $\eta$ is the ratio $|B_r|/|B_{\phi}|$. In our MC model, it is a free parameter expect for the MC CND.
\end{enumerate}
The horizontal model has been developed in this paper specifically to describe the magnetic field in the MCs. The parameter values have to be set individually and so are the errors. Here is a short discussion of the general range of uncertainties 
\begin{itemize}
\item $B_1$ - since the normalization factor is proportional to the expected field strength $\overline{B}_{\mathrm{MC}}$, the associated uncertainty depends on the $\delta \overline{B}_{\mathrm{MC}}$ which is listed in Table \ref{table1}.
\item the MC radius $R$ - the related uncertainty depends on the detector resolution and is listen in Table \ref{table1}.
\item The error in the vertical Gaussian scale length is fully determined by the error in the radius, that is,  $H_c=R/\sqrt{\ln{2}}$ 
\item $\eta$ depends on the azimuthal shearing of the field lines and is in the range between zero and one.
\item  
The azimuthal wave number $m$ 
is generally not known, but set to one to reach a net magnetic flux of zero which is important for ensuring the solenoidal property.
\end{itemize}

\noindent 
It should be noted, that for all components an additional degree of freedom comes from the positions of MCs and NTFs.

All these parameters can be adjusted in order to fit specific data. For those who wish a single global model, they can consult the x-shape models of \cite{X-ShapeModel} including the azimuthal component. However, such a model is not appropriate for the GC since they cannot reproduce local configurations which are significant, for example, in NTF regions.
All relevant parameters are summarized in Table \ref{table1}, \ref{table2} and \ref{table3}. 
\section{Summary \& conclusion}
At first, we summarize molecular cloud surveys from different works and combine them with the proper gas structure in the central 10\,pc by \cite{Inner10pc}. Further, we add an adjusted diffuse gas model by \cite{MassGalacticCenter2} and identify this model as the ICM. Putting them all together delivers an accurate 3d gas distribution in the CMZ.\\
We build a superposition of all components and reconstruct a polarization map considering $\eta=0.5$, which leads to a very good agreement with \cite{Nishiyama2}. The application of the first magnetic field model in the CMZ is carried by the CR propagation tool CRPropa, which delivers the CR trajectories.
The results are compared with the extrapolated model in JF12. 
Throughout this work, the composed model exhibits its power and ability by very good agreement with the measured data.\\
Future magnetic field models of our Galaxy demand a continuous description without any blank areas. For this purpose, this work can be connected with any previous models such as {JF12}. For doing so, $\mathbf{B}_C^{\rm IC}$ has to be modified and a further azimuthal component has to be included which ensures the continuous differentiability and the divergence-free property at the break point of the former field. At the same time, the average value of 10\,$\mu$G within the CMZ still has to be maintained.
\begin{acknowledgements}
	We would like to thank Anvar Shukurov and Jens Kleimann for very helpful discussions. We want to thank Lukas Merten for helpful comments on the application of CRPropa. We further acknowledge the support from Rosa-Luxemburg Stiftung.
\end{acknowledgements}

	\bibliographystyle{aa}
	\bibliography{literaturNew}

\begin{thebibliography}{80}
\expandafter\ifx\csname natexlab\endcsname\relax\def\natexlab#1{#1}\fi

\bibitem[{{Abdalla} {et~al.}(2018){Abdalla}, {Abramowski}, {Aharonian}, {Ait
  Benkhali}, {Akhperjanian}, {Andersson}, {Ang{\"u}ner}, {Arakawa}, {Arrieta},
  \& et~al.}]{HESS2018GC}
{Abdalla}, H., {Abramowski}, A., {Aharonian}, F., {et~al.} 2018, Astron. \&
  Astroph., 612, A9

\bibitem[{Abdalla {et~al.}(2018)}]{HessGC2017}
Abdalla, H. {et~al.} 2018, Astron. Astrophys., 612, A9

\bibitem[{Abramowski {et~al.}(2016)}]{AbramowskiNature}
Abramowski, A. {et~al.} 2016, Nature, 531, 476

\bibitem[{Ackermann {et~al.}(2017)}]{FermiGCGammaRayExcess2017}
Ackermann, M. {et~al.} 2017, Astrophys. J., 840, 43

\bibitem[{Alves~Batista {et~al.}(2016)Alves~Batista, Dundovic, Erdmann,
  Kampert, K{\"u}mpel, M{\"u}ller, Sigl, van Vliet, Walz, \&
  Winchen}]{CRpropa2016}
Alves~Batista, R., Dundovic, A., Erdmann, M., {et~al.} 2016, JCAP, 1605, 038

\bibitem[{{Anantharamaiah} {et~al.}(1999){Anantharamaiah}, {Lang}, {Kassim},
  {Lazio}, \& {Goss}}]{AnantharamaiahPelican1999}
{Anantharamaiah}, K.~R., {Lang}, C.~C., {Kassim}, N.~E., {Lazio}, T.~J.~W., \&
  {Goss}, W.~M. 1999, in Astronomical Society of the Pacific Conference Series,
  Vol. 186, The Central Parsecs of the Galaxy, ed. H.~{Falcke}, A.~{Cotera},
  W.~J. {Duschl}, F.~{Melia}, \& M.~J. {Rieke}, 507

\bibitem[{{Anantharamaiah} {et~al.}(1991){Anantharamaiah}, {Pedlar}, {Ekers},
  \& {Goss}}]{Anantharamaiah1991}
{Anantharamaiah}, K.~R., {Pedlar}, A., {Ekers}, R.~D., \& {Goss}, W.~M. 1991,
  Mon.~Not.~Roy.~Astron.~Soc., 249, 262

\bibitem[{{Armengaud} {et~al.}(2007){Armengaud}, {Sigl}, {Beau}, \&
  {Miniati}}]{CRPropa2007}
{Armengaud}, E., {Sigl}, G., {Beau}, T., \& {Miniati}, F. 2007, Astroparticle
  Physics, 28, 463

\bibitem[{{Benford}(1988)}]{Benford1987}
{Benford}, G. 1988, Astroph.~J., 333, 735

\bibitem[{{Bland-Hawthorn} \& {Reynolds}(2000)}]{bland_hawthorn2000}
{Bland-Hawthorn}, J. \& {Reynolds}, R. 2000, {Gas in Galaxies} (Bristol:
  Institute of Physics Publishing), 2636

\bibitem[{{Boehle} {et~al.}(2016){Boehle}, {Ghez}, {Sch{\"o}del}, {Meyer},
  {Yelda}, {Albers}, {Martinez}, {Becklin}, {Do}, {Lu}, {Matthews}, {Morris},
  {Sitarski}, \& {Witzel}}]{SgrA*Mass}
{Boehle}, A., {Ghez}, A.~M., {Sch{\"o}del}, R., {et~al.} 2016, Astroph.~J.,
  830, 17

\bibitem[{{Chuss} {et~al.}(2003){Chuss}, {Davidson}, {Dotson}, {Dowell},
  {Hildebrand}, {Novak}, \& {Vaillancourt}}]{Chuss2003}
{Chuss}, D.~T., {Davidson}, J.~A., {Dotson}, J.~L., {et~al.} 2003, The
  Astrophysical Journal, 599, 1116

\bibitem[{Crutcher {et~al.}(1996)Crutcher, Roberts, Mehringer, \&
  Troland}]{Crutcher1996}
Crutcher, R.~M., Roberts, D.~A., Mehringer, D.~M., \& Troland, T.~H. 1996, The
  Astrophysical Journal Letters, 462, L79

\bibitem[{{Dexter} {et~al.}(2010){Dexter}, {Agol}, {Fragile}, \&
  {McKinney}}]{Dexter2010}
{Dexter}, J., {Agol}, E., {Fragile}, P.~C., \& {McKinney}, J.~C. 2010, The
  Astrophysical Journal, 717, 1092

\bibitem[{{Dobler}(2012)}]{WmapHazeDobler}
{Dobler}, G. 2012, Astrophys. J., 750, 17

\bibitem[{{Dobler} {et~al.}(2011){Dobler}, {Cholis}, \&
  {Weiner}}]{DMAnnihilationGC}
{Dobler}, G., {Cholis}, I., \& {Weiner}, N. 2011, Astrophys. J., 741, 25

\bibitem[{{Eatough} {et~al.}(2013){Eatough}, {Falcke}, {Karuppusamy}, {Lee},
  {Champion}, {Keane}, {Desvignes}, {Schnitzeler}, {Spitler}, {Kramer},
  {Klein}, {Bassa}, {Bower}, {Brunthaler}, {Cognard}, {Deller}, {Demorest},
  {Freire}, {Kraus}, {Lyne}, {Noutsos}, {Stappers}, \& {Wex}}]{Eatough2013}
{Eatough}, R.~P., {Falcke}, H., {Karuppusamy}, R., {et~al.} 2013, Nature, 501,
  391

\bibitem[{Federrath {et~al.}(2016)Federrath, Rathborne, Longmore, Kruijssen,
  Bally, Contreras, Crocker, Garay, Jackson, Testi, \& Walsh}]{G0253}
Federrath, C., Rathborne, J.~M., Longmore, S.~N., {et~al.} 2016, The
  Astrophysical Journal, 832, 143

\bibitem[{{Ferri{\`e}re}(2009)}]{FerriereMagneticField2009}
{Ferri{\`e}re}, K. 2009, Astron. \& Astroph., 505, 1183

\bibitem[{{Ferri{\`e}re}(2012)}]{Inner10pc}
{Ferri{\`e}re}, K. 2012, Astron. \& Astroph., 540, A50

\bibitem[{Ferri{\`e}re {et~al.}(2007)Ferri{\`e}re, Gillard, \&
  Jean}]{MassGalacticCenter2}
Ferri{\`e}re, K., Gillard, W., \& Jean, P. 2007, Astron. Astrophys., 467, 611

\bibitem[{{Ferri{\`e}re} \& {Terral}(2014)}]{X-ShapeModel}
{Ferri{\`e}re}, K. \& {Terral}, P. 2014, Astron. \& Astroph., 561, A100

\bibitem[{Finkbeiner(2004)}]{WmapHaze}
Finkbeiner, D.~P. 2004, Astrophys. J., 614, 186

\bibitem[{{Goldsmith} {et~al.}(1990){Goldsmith}, {Lis}, {Hills}, \&
  {Lasenby}}]{SgrB2}
{Goldsmith}, P.~F., {Lis}, D.~C., {Hills}, R., \& {Lasenby}, J. 1990,
  Astroph.~J., 350, 186

\bibitem[{{Heywood} {et~al.}(2019){Heywood}, {Camilo}, {Cotton}, {Yusef-Zadeh},
  \& {Abbott}}]{Heywood2019}
{Heywood}, I., {Camilo}, F., {Cotton}, W.~D., {Yusef-Zadeh}, F., \& {Abbott},
  T.~D. 2019, Nature, 573, 235

\bibitem[{{Hildebrand} {et~al.}(1990{\natexlab{a}}){Hildebrand}, {Gonatas},
  {Platt}, {Wu}, {Davidson}, {Werner}, {Novak}, \& {Morris}}]{CNDObservation}
{Hildebrand}, R.~H., {Gonatas}, D.~P., {Platt}, S.~R., {et~al.}
  1990{\natexlab{a}}, The Astrophysical Journal, 362, 114

\bibitem[{{Hildebrand} {et~al.}(1990{\natexlab{b}}){Hildebrand}, {Gonatas},
  {Platt}, {Wu}, {Davidson}, {Werner}, {Novak}, \& {Morris}}]{Hildebrand1990}
{Hildebrand}, R.~H., {Gonatas}, D.~P., {Platt}, S.~R., {et~al.}
  1990{\natexlab{b}}, Astroph.~J., 362, 114

\bibitem[{{Immer} {et~al.}(2012){Immer}, {Menten}, {Schuller}, \&
  {Lis}}]{Immer2012}
{Immer}, K., {Menten}, K.~M., {Schuller}, F., \& {Lis}, D.~C. 2012, Astronomy
  and Astrophysics, 548, A120

\bibitem[{Jaffe {et~al.}(2013)Jaffe, Ferri{\`e}re, Banday, Strong, Orlando,
  Macias-Perez, Fauvet, Combet, \& Falgarone}]{MagneticFieldModelJaffe2013}
Jaffe, T.~R., Ferri{\`e}re, K.~M., Banday, A.~J., {et~al.} 2013, Monthly
  Notices of the Royal Astronomical Society, 431, 683

\bibitem[{Jansson \& Farrar(2012)}]{Farrar2012}
Jansson, R. \& Farrar, G.~R. 2012, The Astrophysical Journal, 757, 14

\bibitem[{{Johnson} {et~al.}(2015){Johnson}, {Fish}, {Doeleman}, {Marrone},
  {Plambeck}, {Wardle}, {Akiyama}, {Asada}, {Beaudoin}, {Blackburn},
  {Blundell}, {Bower}, {Brinkerink}, {Broderick}, {Cappallo}, {Chael}, {Crew},
  {Dexter}, {Dexter}, {Freund}, {Friberg}, {Gold}, {Gurwell}, {Ho}, {Honma},
  {Inoue}, {Kosowsky}, {Krichbaum}, {Lamb}, {Loeb}, {Lu}, {MacMahon},
  {McKinney}, {Moran}, {Narayan}, {Primiani}, {Psaltis}, {Rogers}, {Rosenfeld},
  {SooHoo}, {Tilanus}, {Titus}, {Vertatschitsch}, {Weintroub}, {Wright},
  {Young}, {Zensus}, \& {Ziurys}}]{Johnson2015}
{Johnson}, M.~D., {Fish}, V.~L., {Doeleman}, S.~S., {et~al.} 2015, Science,
  350, 1242

\bibitem[{{Kauffmann}(2017)}]{StarFormationGC}
{Kauffmann}, J. 2017, ArXiv e-prints [\eprint[arXiv]{1712.01453}]

\bibitem[{{Kauffmann} {et~al.}(2017{\natexlab{a}}){Kauffmann}, {Pillai},
  {Zhang}, {Menten}, {Goldsmith}, {Lu}, \& {Guzman}}]{CMZMC}
{Kauffmann}, J., {Pillai}, T., {Zhang}, Q., {et~al.} 2017{\natexlab{a}},
  Astron. \& Astroph., 603, A89

\bibitem[{{Kauffmann} {et~al.}(2017{\natexlab{b}}){Kauffmann}, {Pillai},
  {Zhang}, {Menten}, {Goldsmith}, {Lu}, {Guzm{\'a}n}, \&
  {Schmiedeke}}]{CMZCloud2}
{Kauffmann}, J., {Pillai}, T., {Zhang}, Q., {et~al.} 2017{\natexlab{b}},
  Astron. \& Astroph., 603, A90

\bibitem[{{Kleimann} {et~al.}(2019){Kleimann}, {Schorlepp}, {Merten}, \&
  {Becker Tjus}}]{Kleimann2019}
{Kleimann}, J., {Schorlepp}, T., {Merten}, L., \& {Becker Tjus}, J. 2019,
  Astroph.~J., 877, 76

\bibitem[{Lang {et~al.}({1999a})Lang, Anantharamaiah, Kassim, \&
  Lazio}]{Cornelia1999}
Lang, C.~C., Anantharamaiah, K.~R., Kassim, N.~E., \& Lazio, T. J.~W. {1999a},
  The Astrophysical Journal Letters, 521, L41

\bibitem[{{Lang} {et~al.}({1999b}){Lang}, {Morris}, \& {Echevarria}}]{Lang1999}
{Lang}, C.~C., {Morris}, M., \& {Echevarria}, L. {1999b}, The Astrophysical
  Journal, 526, 727

\bibitem[{LaRosa {et~al.}(2005)LaRosa, Brogan, Shore, Lazio, Kassim, \&
  Nord}]{LaRosa330MHz}
LaRosa, T.~N., Brogan, C.~L., Shore, S.~N., {et~al.} 2005, Astrophys. J., 626,
  L23

\bibitem[{{LaRosa} {et~al.}(2000){LaRosa}, {Kassim}, {Lazio}, \&
  {Hyman}}]{LaRosaFilaments}
{LaRosa}, T.~N., {Kassim}, N.~E., {Lazio}, T.~J.~W., \& {Hyman}, S.~D. 2000,
  \aj, 119, 207

\bibitem[{LaRosa {et~al.}(2001)LaRosa, Lazio, \& Kassim}]{LaRosaFilaments2001}
LaRosa, T.~N., Lazio, T. J.~W., \& Kassim, N.~E. 2001, The Astrophysical
  Journal, 563, 163

\bibitem[{{Law} {et~al.}(2008){Law}, {Yusef-Zadeh}, \& {Cotton}}]{Law2008}
{Law}, C.~J., {Yusef-Zadeh}, F., \& {Cotton}, W.~D. 2008,
  Astroph.~J.~Suppl.~Series, 177, 515

\bibitem[{{Lesch} \& {Reich}(1992)}]{Lesch1992}
{Lesch}, H. \& {Reich}, W. 1992, Astron. \& Astroph, 264, 493

\bibitem[{Lin {et~al.}(2000)Lin, Penn, \& Tomczyk}]{Lin2000}
Lin, H., Penn, M.~J., \& Tomczyk, S. 2000, Astroph.~J., 541, L83

\bibitem[{{Lis} {et~al.}(1991){Lis}, {Carlstrom}, \& {Keene}}]{MassSgrD}
{Lis}, D.~C., {Carlstrom}, J.~E., \& {Keene}, J. 1991, The Astronomical
  Journal, 380, 429

\bibitem[{{Mangilli} {et~al.}(2019){Mangilli}, {Aumont}, {Bernard}, {Buzzelli},
  {de Gasperis}, {Durrive}, {Ferri{\`e}re}, {Fo{\"e}nard}, {Hughes}, {Lacourt},
  {Misawa}, {Montier}, {Mot}, {Ristorcelli}, {Roussel}, {Ade}, {Alina}, {de
  Bernardis}, {de Gouveia Dal Pino}, {Dubois}, {Engel}, {Hargrave}, {Laureijs},
  {Longval}, {Maffei}, {Magalh{\~a}es}, {Marty}, {Masi}, {Montel}, {Pajot},
  {Rodriguez}, {Salatino}, {Saccoccio}, {Stever}, {Tauber}, {Tibbs}, \&
  {Tucker}}]{Mangilli2019}
{Mangilli}, A., {Aumont}, J., {Bernard}, J.~P., {et~al.} 2019, arXiv e-prints,
  arXiv:1901.06196

\bibitem[{{Merten}(2015)}]{MertenMaster}
{Merten}, L. 2015, Master's thesis, Ruhr-University Bochum, Germany, Germany

\bibitem[{Merten {et~al.}(2017)Merten, Becker~Tjus, Fichtner, Eichmann, \&
  Sigl}]{CRpropa2017}
Merten, L., Becker~Tjus, J., Fichtner, H., Eichmann, B., \& Sigl, G. 2017,
  JCAP, 1706, 046

\bibitem[{{Michelson} {et~al.}(2010){Michelson}, {Atwood}, \&
  {Ritz}}]{FermiBubble2}
{Michelson}, P.~F., {Atwood}, W.~B., \& {Ritz}, S. 2010, Reports on Progress in
  Physics, 73, 074901

\bibitem[{{Morris}(1990)}]{Morris1990}
{Morris}, M. 1990, in IAU Symposium, Vol. 140, Galactic and Intergalactic
  Magnetic Fields, ed. R.~{Beck}, P.~P. {Kronberg}, \& R.~{Wielebinski},
  361--367

\bibitem[{{Morris}(2015)}]{Morris2015}
{Morris}, M.~R. 2015, {Manifestations of the Galactic Center Magnetic Field}
  (Springer), 391

\bibitem[{{Mo{\'s}cibrodzka} {et~al.}(2009){Mo{\'s}cibrodzka}, {Gammie},
  {Dolence}, {Shiokawa}, \& {Leung}}]{Moscibrodzka2009}
{Mo{\'s}cibrodzka}, M., {Gammie}, C.~F., {Dolence}, J.~C., {Shiokawa}, H., \&
  {Leung}, P.~K. 2009, The Astrophysical Journal, 706, 497

\bibitem[{{Muno} {et~al.}(2004){Muno}, {Baganoff}, {Bautz}, {Feigelson},
  {Garmire}, {Morris}, {Park}, {Ricker}, \& {Townsley}}]{muno2004}
{Muno}, M.~P., {Baganoff}, F.~K., {Bautz}, M.~W., {et~al.} 2004, Astroph.~J.,
  613, 326

\bibitem[{{Nakashima} {et~al.}(2019){Nakashima}, {Koyama}, {Wang}, \&
  {Enokiya}}]{Nakashima2019}
{Nakashima}, S., {Koyama}, K., {Wang}, Q.~D., \& {Enokiya}, R. 2019,
  Astroph.~J., 875, 32

\bibitem[{Nishiyama {et~al.}(2010)Nishiyama, Hatano, Tamura, Matsunaga,
  Yoshikawa, Suenaga, Hough, Sugitani, Nagayama, Kato, \& Nagata}]{Nishiyama}
Nishiyama, S., Hatano, H., Tamura, M., {et~al.} 2010, The Astrophysical Journal
  Letters, 722, L23

\bibitem[{{Nishiyama} {et~al.}(2009){Nishiyama}, {Tamura}, {Hatano}, {Kanai},
  {Kurita}, {Sato}, {Matsunaga}, {Nagata}, {Nagayama}, {Kandori}, {Nakajima},
  {Kusakabe}, {Sato}, {Hough}, {Sugitani}, \& {Okuda}}]{Nishiyama2009}
{Nishiyama}, S., {Tamura}, M., {Hatano}, H., {et~al.} 2009, The Astrophysical
  Journal, 690, 1648

\bibitem[{{Nishiyama} {et~al.}(2013){Nishiyama}, {Yasui}, {Nagata},
  {Yoshikawa}, {Uchiyama}, {Sch{\"o}del}, {Hatano}, {Sato}, {Sugitani},
  {Suenaga}, {Kwon}, \& {Tamura}}]{Nishiyama2}
{Nishiyama}, S., {Yasui}, K., {Nagata}, T., {et~al.} 2013, Astroph.~J.~Lett.,
  769, L28

\bibitem[{Novak {et~al.}(2003)Novak, Chuss, Renbarger, Griffin, Newcomb,
  Peterson, Loewenstein, Pernic, \& Dotson}]{Polarization2}
Novak, G., Chuss, D.~T., Renbarger, T., {et~al.} 2003, The Astrophysical
  Journal Letters, 583, L83

\bibitem[{Oka {et~al.}(2019)Oka, Geballe, Goto, Usuda, Benjamin, \&
  Indriolo}]{Oka2019}
Oka, T., Geballe, T.~R., Goto, M., {et~al.} 2019, The Astrophysical Journal,
  883, 54

\bibitem[{{Pedlar} {et~al.}(1989){Pedlar}, {Anantharamaiah}, {Ekers}, {Goss},
  {van Gorkom}, {Schwarz}, \& {Zhao}}]{RadioHalo}
{Pedlar}, A., {Anantharamaiah}, K.~R., {Ekers}, R.~D., {et~al.} 1989, APJ, 342,
  769

\bibitem[{{Peratt}(1984)}]{Peratt1984}
{Peratt}, A.~L. 1984, Sky and Telescope, 68, 118

\bibitem[{Pillai {et~al.}(2015)Pillai, Kauffmann, Tan, Goldsmith, Carey, \&
  Menten}]{Pillai2015}
Pillai, T., Kauffmann, J., Tan, J.~C., {et~al.} 2015, The Astrophysical
  Journal, 799, 74

\bibitem[{{Plante} {et~al.}(1994){Plante}, {Lo}, {Crutcher}, \&
  {Killeen}}]{CNDMF2}
{Plante}, R., {Lo}, K., {Crutcher}, R., \& {Killeen}, N. 1994, Springer, 445,
  205

\bibitem[{{Plante} {et~al.}(1995){Plante}, {Lo}, \& {Crutcher}}]{Plante1995}
{Plante}, R.~L., {Lo}, K.~Y., \& {Crutcher}, R.~M. 1995, The Astrophysical
  Journal Letters, 445, L113

\bibitem[{{Ponti} {et~al.}(2019){Ponti}, {Hofmann}, {Churazov}, {Morris},
  {Haberl}, {Nandra}, {Terrier}, {Clavel}, \& {Goldwurm}}]{Ponti2019}
{Ponti}, G., {Hofmann}, F., {Churazov}, E., {et~al.} 2019, Nature, 567,
  pages347–350

\bibitem[{{Reich} {et~al.}(1988){Reich}, {Sofue}, {Wielebinski}, \&
  {Seiradakis}}]{FilamentSpectralIndexSgrA}
{Reich}, W., {Sofue}, Y., {Wielebinski}, R., \& {Seiradakis}, J.~H. 1988,
  Astron. \& Astroph., 191, 303

\bibitem[{{Serabyn} \& {Morris}(1994)}]{Serabyn1994}
{Serabyn}, E. \& {Morris}, M. 1994, Astroph.~J.~Lett., 424, L91

\bibitem[{{Shukurov} {et~al.}(2019){Shukurov}, {Rodrigues, Luiz Felippe S.},
  {Bushby, Paul J.}, {Hollins, James}, \& {Rachen, J\"org P.}}]{ShukorovGMF}
{Shukurov}, A., {Rodrigues, Luiz Felippe S.}, {Bushby, Paul J.}, {Hollins,
  James}, \& {Rachen, J\"org P.} 2019, A\&A, 623, A113

\bibitem[{{Staguhn} {et~al.}(1998){Staguhn}, {Stutzki}, {Uchida}, \&
  {Yusef-Zadeh}}]{Staguhn1998}
{Staguhn}, J., {Stutzki}, J., {Uchida}, K.~I., \& {Yusef-Zadeh}, F. 1998,
  Astron. \& Astroph, 336, 290

\bibitem[{{Su} {et~al.}(2010){Su}, {Slatyer}, \& {Finkbeiner}}]{FermiBubble3}
{Su}, M., {Slatyer}, T.~R., \& {Finkbeiner}, D.~P. 2010, Astrophys. J., 724,
  1044

\bibitem[{{Sun, X. H.} {et~al.}(2008){Sun, X. H.}, {Reich, W.}, {Waelkens, A.},
  \& {En\ss{}lin, T. A.}}]{MagneticFieldModelSun2008}
{Sun, X. H.}, {Reich, W.}, {Waelkens, A.}, \& {En\ss{}lin, T. A.} 2008, A\&A,
  477, 573

\bibitem[{{Terral} \& {Ferri{\`e}re}(2017)}]{Ferriere2017}
{Terral}, P. \& {Ferri{\`e}re}, K. 2017, Astron. \& Astroph., 600, A29

\bibitem[{{Uchida} {et~al.}(1996){Uchida}, {Morris}, {Serabyn}, \&
  {G{\"u}sten}}]{Uchida1996}
{Uchida}, K.~I., {Morris}, M., {Serabyn}, E., \& {G{\"u}sten}, R. 1996,
  Astroph.~J., 462, 768

\bibitem[{Unger \& Farrar(2019)}]{Unger2019}
Unger, M. \& Farrar, G. 2019, in {Ultra High Energy Cosmic Rays (UHECR 2018)
  Paris, France, October 8-12, 2018}

\bibitem[{{Walker} {et~al.}(2015){Walker}, {Longmore}, {Bastian}, {},
  {Rathborne}, {Jackson}, {Foster}, \& {Contreras}}]{Walker2015}
{Walker}, D.~L., {Longmore}, S.~N., {Bastian}, N., {et~al.} 2015,
  Mon.~Not.~Roy.~Astron.~Soc., 449, 715

\bibitem[{{Walker} {et~al.}(2018){Walker}, {Longmore}, {Zhang}, {Battersby},
  {Keto}, {Kruijssen}, {Ginsburg}, {Lu}, {Henshaw}, {Kauffmann}, {Pillai},
  {Mills}, {Walsh}, {Bally}, {Ho}, {Immer}, \& {Johnston}}]{Walker2018}
{Walker}, D.~L., {Longmore}, S.~N., {Zhang}, Q., {et~al.} 2018,
  Mon.~Not.~Roy.~Astron.~Soc., 474, 2373

\bibitem[{{Wardle} \& {K\"onigl}(1990)}]{CNRMagneticField}
{Wardle}, M. \& {K\"onigl}, A. 1990, The Astrophysical Journal, 362, 120

\bibitem[{{Wink} {et~al.}(1982){Wink}, {Altenhoff}, \& {Mezger}}]{Wink1982}
{Wink}, J.~E., {Altenhoff}, W.~J., \& {Mezger}, P.~G. 1982, \aap, 108, 227

\bibitem[{{Yusef-Zadeh}(2003{\natexlab{a}})}]{FilamentsYusufZadeh}
{Yusef-Zadeh}, F. 2003{\natexlab{a}}, The Astronomical Journal, 598, 325

\bibitem[{{Yusef-Zadeh}(2003{\natexlab{b}})}]{YusufZadehOriginNTF}
{Yusef-Zadeh}, F. 2003{\natexlab{b}}, Astroph.~J., 598, 325

\bibitem[{{Yusef-Zadeh} \& {Morris}(1987)}]{Yusef-ZadehArc1987}
{Yusef-Zadeh}, F. \& {Morris}, M. 1987, Astron.~J., 94, 1178

\end{thebibliography}
\end{document}